\documentclass[12pt]{article}

\usepackage{cancel,slashed}
\usepackage{bbm}
\usepackage{amssymb}
\usepackage{amsfonts}
\usepackage{amsmath}
\usepackage{graphicx}
\usepackage{cite}

\usepackage{latexsym}
\usepackage{color}
\usepackage{xypic}
\usepackage{cancel,slashed}
\usepackage[linktocpage]{hyperref}
\usepackage{color}
\usepackage{transparent}

\newcommand{\bmat}{\left(\begin{array}}
\newcommand{\emat}{\end{array}\right)}

\def\p{\partial}
\def\a{\alpha}

\def\b{\beta}
\def\g{\gamma}
\def\d{\delta}
\def\th{\theta}

\def\vphi{\varphi}
\def\-{\hphantom{-}}

\def\s2{\frac{1}{\sqrt2}}

\def\oh{\frac{1}{2}}
\def\beq{\begin{equation}}
\def\eeq{\end{equation}}
\def\beqa{\begin{eqnarray}}
\def\eeqa{\end{eqnarray}}

\def\tr{{\rm tr \,}}

\def\Dsl{\,\raise.15ex\hbox{/}\mkern-13.5mu D} %this one can be subscripted

\def\e{\epsilon}

\def\CR {{\cal R}}

\def\CO {{\cal O}}

\def\tr{\mbox{Tr}}
\def\str{\mbox{STr}}

\def\be{\begin{equation}}
\def\ee{\end{equation}}
\def\bea{\begin{eqnarray}}
\def\eea{\end{eqnarray}}
\def\raw{\rightarrow}

\def\IC{\mathbb{C}}

\def\IZ{\mathbb{Z}}

\def\oh{\frac{1}{2}}
\def\a{{\alpha}}
\def\b{{\beta}}
\def\d{{\delta}}
\def\eps{{\epsilon}}
\def\th{{\theta}}
\def\Lam{{\Lambda}}
\def\lam{{\lambda}}

\def\g{{\gamma}}

\def\vphi{{\varphi}}
\def\p{{\partial}}

\def\vec#1{{\overrightarrow{#1}}}
\def\str{\mbox{STr}}

\def\w{{\wedge}}

\def\sm2{{\mbox{\small 2}}}

\begin{document}
\pagestyle{plain}

%----------------------------------------------------------------------%
%  numbering equations with section number
%----------------------------------------------------------------------%
\makeatletter
\@addtoreset{equation}{section}
\makeatother
\renewcommand{\theequation}{\thesection.\arabic{equation}}
%----------------------------------------------------------------------%
%  title page
%----------------------------------------------------------------------%
\pagestyle{empty}
%\vspace*{1.0in}
\rightline{ IFT-UAM/CSIC-15-023}
\rightline{ MPP-2015-40}
%\rightline{\tt hep-th/yymmnnn}
\vspace{0.5cm}
\begin{center}
%\LARGE{{Yukawa hierarchies at the $E_8$ point in F-theory}
\LARGE{{Yukawa hierarchies at the point of $E_8$ in F-theory}
\\[10mm]}
\large{Fernando Marchesano,$^1$ Diego Regalado$^{2}$ and Gianluca Zoccarato$^{1,3}$ \\[10mm]}
\small{
${}^1$ Instituto de F\'{\i}sica Te\'orica UAM-CSIC, Cantoblanco, 28049 Madrid, Spain \\[2mm] 
${}^2$ Max-Planck-Institut f\"ur Physik, F\"ohringer Ring 6, 80805 Munich, Germany\\[2mm]
${}^3$ Departamento de F\'{\i}sica Te\'orica, 
Universidad Aut\'onoma de Madrid, %Cantoblanco, 
28049 Madrid, Spain
\\[8mm]} 
\small{\bf Abstract} \\[5mm]
\end{center}
\begin{center}
\begin{minipage}[h]{15.0cm} 

We analyse the structure of Yukawa couplings in local SU(5) F-theory models with $E_8$ enhancement. In this setting the $E_8$ symmetry is broken down to SU(5) by a 7-brane configuration described by T-branes, all the Yukawa couplings are generated in the vicinity of a point and only one family of quarks and leptons is massive at tree-level. The other two families obtain their masses when non-perturbative effects are taken into account, being hierarchically lighter than the third family. However, and contrary to previous results, we find that this hierarchy of fermion masses is not always appropriate to reproduce measured data. We find instead that different T-brane configurations breaking $E_8$ to SU(5) give rise to distinct hierarchical patterns for the holomorphic Yukawa couplings. Only some of these patterns allow to fit the observed fermion masses with reasonable local model parameter values, adding further constraints to the construction of F-theory GUTs. We consider an $E_8$ model where such appropriate hierarchy is realised and compute its physical Yukawas, showing that realistic charged fermions masses %and mixing angles 
can indeed be obtained in  this case.

\end{minipage}
\end{center}
\newpage
%----------------------------------------------------------------------%
%  Resetting of counters
%----------------------------------------------------------------------%
\setcounter{page}{1}
\pagestyle{plain}
\renewcommand{\thefootnote}{\arabic{footnote}}
\setcounter{footnote}{0}
%----------------------------------------------------------------------%
%  Paper begins
%----------------------------------------------------------------------%

%\end{document}

\tableofcontents

%\newpage

\section{Introduction}
\label{s:intro}

The proposal made in \cite{dw1,bhv1,bhv2,dw2} to build realistic 4d vacua by means of GUT constructions in F-theory has undoubtedly generated a wealth of activity in the past few years. While from all the classes of models considered in the string phenomenology literature \cite{thebook,reviews,ftheoryreviews} none is a priori preferred to achieve a fully realistic model of Particle Physics, F-theory vacua present a number of conceptual and technical advantages that has allowed to make substantial recent progress on this front. 

For instance, type IIB/F-theory flux compactifications contain the class of vacua where at present moduli stabilisation is best understood \cite{Grana:2005jc,Douglas:2006es,Quevedo:2014xia}, and where most of the statistical analysis of string vacua has been applied \cite{Denef:2007pq}. In addition, the construction of F-theory compactifications relies on complex geometry, and so allows to implement powerful results in algebraic geometry. In practice, this translates into very useful techniques that can be used to construct explicit examples of F-theory models, as well as to gain a global perspective of the set of vacua as a whole. 

But perhaps the key ingredient that highlights F-theory GUTs as a promising avenue to construct realistic vacua is the localisation properties of 7-branes and their consequences. Indeed, the fact that 7-branes localise gauge and chiral degrees of freedom lets us formulate the construction of F-theory GUTs in a bottom-up fashion \cite{Aldazabal:2000sa}. This in turn permits to express the basic features that the gauge sector of realistic model should contain in terms of a small internal region where the GUT fields are localised, and to compute many quantities of physical interest in terms of such local data. 

A good example of the latter is the computation of Yukawa couplings in F-theory GUTs. Through a series of works \cite{hv08,Hayashi:2009ge,Randall:2009dw,Font:2009gq,cchv09,Conlon:2009qq,hktw2,mm09,Leontaris:2010zd,cchv10,Chiou:2011js,afim11,fimr12,fmrz13} it has been realised that to extract the flavour structure of an F-theory model one may implement an ultra-local approach and compute its Yukawas by analysing small regions of the four-cycle $S_{\rm GUT}$ where the GUT degrees of freedom are localised. More precisely one finds that, if gauge fields are localised in $S_{\rm GUT}$ and chiral fields at complex matter curves $\Sigma_i$ inside $S_{\rm GUT}$, then holomorphic Yukawas can be computed by looking at the points of intersection of such matter curves. Physical Yukawas, on the other hand, can be computed ultra-locally if the internal wavefunctions for the chiral fields are sufficiently localised in a region near such point of intersection.\footnote{For a more precise statement in terms of the notion of local chirality see \cite{Palti:2012aa}. For other applications of seven-brane wavefunctions see \cite{Marchesano:2008rg,Marchesano:2010bs,Camara:2011nj,Ibanez:2012zg,Camara:2013fta,Camara:2014tba}.}

An important outcome of the analysis of Yukawas in F-theory is that one may easily engineer GUT models where the Yukawa matrices are of rank one, by simply imposing a topological condition on the matter curves \cite{cchv10}. This will automatically give a mass hierarchy between one family and the rest. The masses of the two lightest families can then be generated when taking into account the effect of an Euclidean D3-brane instanton on a different four-cycle, along the lines of \cite{mm09}.\footnote{For different approaches to the generation of Yukawa hierarchies in F-theory GUTs see e.g.  \cite{fi08,DudasPalti,Leontaris:2010zd,Krippendorf,Krippendorf:2014xba,Bizet:2014uua,Klevers:2014bqa,Garcia-Etxebarria:2014qua,Mayrhofer:2014haa,Cvetic:2015moa}.} Notice that  rank one Yukawas are not exclusive of F-theory models (see e.g. \cite{Cremades:2003qj,Cremades:2004wa} for type II examples). However, the F-theory framework does allow to compute them systematically for a wide class of models by means of the ultra-local approach. Quite remarkably, this remains true even after we include the non-perturbative corrections of \cite{mm09}. 

In this spirit, the computation of Yukawa couplings in the presence of non-perturbative effects has been carried out in \cite{afim11,fimr12,fmrz13}. In particular, refs.\cite{fimr12} and \cite{fmrz13} respectively analysed the fermion mass hierarchy  developed for down-type $\mathbf{10 \times \bar{5} \times \bar{5}}$ and up-type $\mathbf{10 \times 10 \times 5}$ couplings in SU(5) models, which become MSSM Yukawas once that hypercharge flux effects breaking $SU(5) \raw SU(3) \times SU(2) \times U(1)_Y$ are taken into account. In both cases it was found that a family hierarchy of the form $(1, \eps, \eps^2)$ is generated for fermion masses, with $\eps$ a small parameter measuring the strength of the non-perturbative effect. Such hierarchy is already present at the level of the holomorphic Yukawas, and allows to fit empirical data upon taking $\eps \sim 10^{-4}$ and including the dependence of physical Yukawas on the worldvolume fluxes threading $S_{\rm GUT}$. 

The computations carried out in \cite{fimr12} and \cite{fmrz13} are independent from each other because, in principe, down-type and up-type Yukawas can be generated at very different points of the GUT four-cycle $S_{\rm GUT}$. However, as pointed out in \cite{Heckman:2009mn}  experimentally the CKM matrix describes non-trivial correlations between U and D-quark Yukawas, and this  strongly suggests that in a realistic model these two points should be in the same neighbourhood of $S_{\rm GUT}$, so that they experience the same local geometry and worldvolume flux densities. A model in which the two Yukawa points are very close to each other or even coincide is very attractive from the bottom-up perspective, as one is then able to compute the whole set of Yukawas with the mere knowledge of the local patch of $S_{\rm GUT}$. In this sense, even more appealing is the case where further matter curves  intersect at a single point $p_{E_8} \in S_{\rm GUT}$, such that the singularity of the F-theory elliptic fibre enhances to $E_8$ at that point. One would then be able to compute ultra-locally the whole set of couplings which are most relevant for the GUT gauge theory by analysing a local patch around $p_{E_8}$, which is usually dubbed the {\em point of $E_8$ in F-theory} \cite{Heckman:2009mn}.

Roughly speaking, the aim of this work is to apply the scenario of \cite{mm09} to F-theory models of SU(5) unification with a point of $E_8$ where all Yukawa couplings are generated.  More precisely, we consider F-theory models that can be locally described in terms of an $E_8$ symmetry higgsed down to SU(5) by the 7-brane position background. This region of $E_8$ symmetry contains both Yukawa points $p_{\rm up}$ and $p_{\rm down}$, which may be coincident or not, and the matter curves intersecting at them are chosen in such a way that we obtain rank one Yukawas at tree level. The question we address is then if, by taking into account non-perturbative effects, a realistic hierarchical pattern of Yukawa couplings is generated such that it allows to fit experimental data, in the same spirit of \cite{afim11,fimr12,fmrz13}.

As mentioned, the hierarchy that has so far allowed to fit empirical data is of the form $(1, \eps , \eps^2)$, at least for reasonable values of $\eps$ and local model parameters like flux densities. Therefore we take this hierarchical pattern as the guiding principle to achieve a realistic fermion mass spectrum in the context at hand. Remarkably, this hierarchical structure can already be seen at the level of holomorphic Yukawas, which depend on very few parameters of the local model. As a result, the above criterion is very robust and can be applied even without specifying the oftentimes complicated 7-brane flux background. 

As we will see, while one may construct several $E_8$ models with rank one Yukawas at tree level, the addition of non-perturbative effects does not always yield a hierarchy of the form $(1, \eps , \eps^2)$. In fact, from all the models that we have analysed only one choice generates the desired hierarchy for both types of Yukawas and at the same time contains interesting mechanisms for realistic $\mu$-term and neutrino masses. Of course having this hierarchy at the holomorphic level does not guarantee that one can reproduce the whole set of empirical data related to Yukawa couplings. Hence, once selected the most promising $E_8$ model we proceed to describe it in full detail and to compute its physical Yukawa couplings. We find that, similarly to the results in \cite{afim11,fimr12,fmrz13}, within the MSSM scheme fermion masses at the GUT scale can be fit for $\eps \sim 10^{-4}$ and large values of tan $\b$. This result is valid for the families of quark and leptons over which we have good control given the approximations taken in our analysis, namely the two heaviest families. This not only applies to the fermion mass spectrum but also to the quark mixing angles. We determine the latter in terms of the separation of the two Yukawa points $p_{\rm up}$ and $p_{\rm down}$, providing precise formulas that illustrate previous statements in the literature.

The paper is organised as follows. In section \ref{s:yukawas} we review the construction of local F-theory GUTs and how hierarchies of Yukawa couplings arise via non-perturbative effects. In section \ref{s:models} we analyse several local $E_8$ models and compute their Yukawas at the holomorphic level, selecting one model based on the criterion described above. In section \ref{s:E8model} we describe this particular model in detail and in section \ref{s:realwave} we solve for its chiral zero modes wavefunctions and compute the normalisation factors that render their kinetic terms canonical. Such normalisations are the missing ingredient to compute the physical Yukawas, whose hierarchies are analysed in section \ref{s:physical}. We conclude in  section \ref{s:conclu}. %We draw our conclusions in section \ref{s:conclu}.

Several technical details have been relegated to the appendices. Appendix \ref{ap:e8} contains the details and notation regarding the $E_8$ Lie algebra used throughout the rest of the paper. Appendix \ref{ap:details} contains the details of the computation of holomorphic Yukawas for the models of section \ref{s:models}. Appendix \ref{ap:chiral} discusses the notion of local chirality applied to the $E_8$ model of section \ref{s:E8model}, while appendix  \ref{ap:holoyuk} spells out the computation of its holomorphic Yukawas, and appendix  \ref{ap:wave} its zero mode wavefunctions in a real gauge.

\section{Yukawa hierarchies in F-theory GUTs}
\label{s:yukawas}

The standard description of F-theory GUT models \cite{bhv1,bhv2,dw1,dw2} (see \cite{ftheoryreviews} for reviews) is done in terms of a Calabi-Yau fourfold which is elliptically fibered over a three-fold base $B$. The fibration is such that the fibre degenerates over a 4-cycle $S_{\rm GUT}$, with a fibre singularity whose Dynkin diagram corresponds to $G_{\rm GUT}$. At certain 2-cycles $\Sigma$ within $S_{\rm GUT}$ the fibre may display a higher singularity type, signalling the presence of chiral multiplets charged under $G_{\rm GUT}$ and localised at such matter curve. The precise 4d chiral matter content of the model depends on the four-form flux $G_4$ threading $S_{\rm GUT}$ which, if chosen appropriately, can break $G_{\rm GUT}$ to the subgroup $SU(3) \times SU(2) \times U(1)_Y$. Finally, the couplings among different chiral multiplets will depend on the intersection pattern of the corresponding matter curves. Hence just by knowing the local geometry around $S_{\rm GUT}$ one may see which Yukawa couplings may be generated in the effective 4d GUT theory. 

While such a picture is rather compelling it is not the appropriate one for the actual computation of Yukawa couplings. Instead, it proves more useful to work with an alternative description of the degrees of freedom localised at $S_{\rm GUT}$. Namely, one can use a 8d action related to the 7-branes wrapping $S_{\rm GUT}$ and those intersecting them. Such an action is defined on a 4-cycle $S$ and in terms of a non-Abelian symmetry group $G \supset G_{\rm GUT}$, and is such that the 4d gauge theory described above can be obtained upon dimensional reduction. In particular, Yukawa couplings can be obtained from the following superpotential
\be
W\, =\, m_*^4 \int_{S} \tr \left( F \wedge \Phi \right)
\label{supo7}
\ee
where $m_*$ is the F-theory characteristic scale, $F = dA - i A \wedge A$ is the field strength of the 7-branes gauge boson $A$, and $\Phi$ is the so-called Higgs field: a (2,0)-form on the 4-cycle $S$ describing the 7-branes transverse geometrical deformations. Both $A$ and $\Phi$ transform in the adjoint of the initial gauge group $G$, which is nevertheless broken to a subgroup due to their non-trivial profile. In particular, the profile $\langle \Phi \rangle$ is such that it only commutes with the generators of $G_{\rm GUT}$ in the bulk of $S_{\rm GUT}$, while on top of the matter curves of $S_{\rm GUT}$ it also commutes with further roots of $G$. The  background profiles $\langle \Phi \rangle$ and $\langle A\rangle$ cannot be arbitrary, but they must solve for the equations of motion that arise from (\ref{supo7}) and the D-term
\be
D\, =\, \int_S \omega \wedge F + \frac{1}{2}  [\Phi, \Phi^\dagger]
\label{FI7}
\ee
where $\omega$ stands for the fundamental form of $S$. Given the background profiles for $\Phi$ and $A$ one can use (\ref{supo7}) and (\ref{FI7}) to solve for the internal wavefunction of their zero mode fluctuations representing the 4d matter fields. Finally,  these wavefunctions can be plugged back into (\ref{supo7}) to compute the precise value of the Yukawa couplings of the model. 

In this setting, two important results that apply to the holomorphic part of the Yukawa couplings are {\it i)} they do not depend on the profile of the worldvolume flux $F$ and {\it ii)} they only depend on the local geometry around the point $p$ where the involved matter curves intersect  \cite{cchv09}. Thanks to this, in order to compute holomorphic Yukawa couplings it suffices to describe the profile $\langle \Phi \rangle$ around a neighbourhood $U_p \subset S_{\rm GUT}$ of the intersection point $p$, and take $G = G_p$ to be a symmetry group just large enough to contain $G_{\rm GUT}$ and describe the matter curves intersecting at $p$. This ultra-local approach is also valid to compute physical Yukawa couplings if one assumes that the wavefunction profile for the corresponding 4d chiral fields is peaked within $U_p$, something usually achievable due to the localisation properties of matter curves and of the worldvolume flux $F$ threading them.

This approach to compute Yukawa couplings has been mostly developed in the case where $G_{\rm GUT}=SU(5)$. There we have that the up-type Yukawa couplings $\mathbf{10 \times 10 \times 5}$ can be described by taking $G_p = E_6$ or larger, while the down-type couplings $\mathbf{10 \times \bar{5} \times \bar{5}}$ require at least of $G_p = SO(12)$. While in principle one may consider several intersection points of each kind, it was proposed in \cite{hv08} a scenario where all up-type Yukawas are generated from a single Yukawa point $p_{\rm up}$, and all down-type Yukawas from $p_{\rm down}$, on the grounds that it is then natural to obtain that one fermion family much heavier than the other two. In fact, what one finds for the down-type couplings is that the Yukawa matrix has rank one, and hence only one family of quarks and leptons becomes massive \cite{cchv09}. This one-rank result is rather robust in the sense that it only depends on the topological intersection pattern of matter curves, and so deforming the divisor $S_{\rm GUT}$ or its worldvolume flux $F$ will not change it. Nevertheless, non-perturbative effects associated to other 4-cycles $S_{\rm np} \subset B$ will affect the result and increase the Yukawa rank from one to three \cite{mm09}. 

Indeed following the discussion in \cite{mm09}, if $S_{\rm np}$ hosts a 7-brane with a gaugino condensate or a 3-brane instanton with the appropriate number of zero modes then the $S_{\rm GUT}$ 7-brane superpotential (\ref{supo7}) will be modified, obtaining
\be
W\, =\, m_*^4\,  \int_S \tr \left( F \wedge \Phi \right) + \eps\, \frac{\theta_0}{2} \tr \left( F \wedge F\right)
\label{suponp}
\ee
where $\eps$ measures the strength of the non-perturbative effect, and $\theta_0$ is a function that depends on the embedding of the 4-cycle $S_{\rm np}$ (we have that $\theta_0\, =\,  (4\pi^2 m_*)^{-1} [{\rm log\, } h/h_0]_{z=0}$, with $h$ the divisor function such that $S_{\rm np} = \{h = 0\}$ and $h_0 = \int_S h$). Further corrections that depend on the derivatives of $h$ normal to $S$ do also appear, as well as corrections at higher powers of $\eps$, see \cite{mm09,afim11,fimr12} for explicit expressions. Nevertheless, for realistic models the increase of the Yukawa rank and the generation of hierarchies can already be seen from the leading correction shown in (\ref{suponp}), with which we will work in the following, while the remaining contributions are rather suppressed.

The analysis of down-type Yukawa couplings for $G_{\rm GUT} = SU(5)$, $G_p = SO(12)$ and with the superpotential (\ref{suponp}) was carried out in ref.\cite{fimr12}. It was obtained a fermion mass hierarchy of the form $(\CO(1), \CO(\eps), \CO(\eps^2))$ that is already present at the level of the holomorphic Yukawas. Such robust hierarchy allows to fit the experimental values of quark and lepton mass ratios once run to the unification scale.  For this one needs  $\eps \sim 10^{-4}$ and to take into account the worldvolume flux dependence of the physical Yukawas. In particular, their dependence on the hypercharge flux $F_Y$ allows to obtain a realistic ratio for the $\tau$ and $b$-quark Yukawas and for the different mass quotients between the second and third families of $D$-quarks and leptons. Being allowed to fit all these data is a clear improvement with respect to classical 4d field theory models of SU(5) unification.

In principle one can implement the same strategy to obtain a realistic spectrum of up-type Yukawa couplings. However, because the coupling $\mathbf{10 \times 10 \times 5}$ involves two fields with the same quantum numbers one cannot achieve a tree-level rank-one result for these Yukawas with the simple intersection of three matter curves. Instead, one needs to consider more involved matter curve geometries which, in terms of the 7-brane position field $\Phi$ can be described by a non-Abelian background profile for $\langle \Phi \rangle$, usually dubbed T-brane \cite{hktw2,cchv10} (see also \cite{Donagi:2011jy,Anderson:2013rka,Collinucci:2014qfa}).

The presence of T-branes does complicate the 7-brane wavefunction equations but, as shown in \cite{fmrz13}, one can still analyse the case of up-type Yukawas by taking $G_{\rm GUT} = SU(5)$, $G_p = E_6$ and the superpotential (\ref{suponp}). It was found in this reference that the inclusion of non-perturbative effects also modifies the tree-level rank result to a hierarchical structure of the form $(\CO(1), \CO(\eps), \CO(\eps^2))$ for the holomorphic Yukawa eigenvalues. Finally, realistic values for the top-quark and for the quotients of $U$-quarks can also be achieved by again taking $\eps \sim 10^{-4}$ and values for the worldvolume flux densities very similar to those needed around $p_{\rm down}$ to fit the down-type Yukawa data.

These previous results suggest that one may naturally obtain a realistic mass spectrum by considering $p_{\rm up}$ and $p_{\rm down}$ in the same neighbourhood of $S_{\rm GUT}$, as this would explain why local parameters like flux densities are similar around both points. In practice, this amounts to consider a symmetry group large enough to describe the whole set of matter curves containing the MSSM chiral content, which selects either $G_p = E_7$ or $G_p = E_8$. In fact, it was proposed in \cite{Heckman:2009mn} that considering $p_{\rm up}$ and $p_{\rm down}$ to coincide in a point of $E_8$ enhancement would account for all the flavour hierarchies observable in the Standard Model. In this paper we would like to investigate if that is indeed possible in the scheme discussed above, namely where the hierarchies are obtained from perturbing a tree-level rank-one result via non-perturbative effects, as encoded by the corrected superpotential (\ref{suponp}).\footnote{To be precise, our scheme only assumes that the hierarchies present in the up-type and down-type Yukawa matrices are all due to a rank-one tree-level superpotential corrected by non-perturbative effects, while the hierarchies in the neutrino sector could be due to a different mechanism. In this sense it is equally interesting to explore the presence of hierarchies in points of $E_7$ enhancement, which will be discussed in a separate publication.}

The set of $E_8$ models that can accommodate the matter spectrum of the MSSM is rather rich, and it is usually classified in terms of the matter curves present in $S_{\rm GUT}$. This classification is particularly powerful whenever the matter curves content admits a spectral cover description, which in particular means that there is an underlying $E_8$ structure globally defined over $S_{\rm GUT}$ \cite{Donagi:2009ra,Marsano:2009gv,Marsano:2009wr,Dudas:2010zb,Dolan:2011iu} or even through the whole threefold base \cite{Baume:2015wia}. In our ultra-local approach such spectral cover description is not necessary, in the sense that the Yukawa couplings will depend on the profiles of $\Phi$ and $F$ near the Yukawa point $p=p_{\rm up} = p_{\rm down}$. In particular, the holomorphic Yukawas will only depend on the profile of $\Phi$ around $p$. As the hierarchy of couplings is already captured at the holomorphic level, we will proceed to classify our $E_8$ models based on their local profile $\langle \Phi \rangle$. Finally, just as in the $E_6$ case we need the presence of T-branes in order to have a tree-level rank-one up-type Yukawa matrix, so the models to consider will be T-brane profiles for the field $\Phi$ around the Yukawa point $p$ of $E_8$ enhancement.\footnote{Strictly speaking due to the T-brane profile $\langle \Phi \rangle$ does not vanish and so the symmetry is not enhanced to $E_8$ at any point of the local model. In particular at the point $p$ where all matter curves meet and all Yukawas are generated we will get a fibre singularity enhancement but we do not expect to recover a full $E_8$ singularity. For the sake of simplicity, we will abuse of language and still refer to this $p$ as a point of $E_8$ in the model.}

To summarise, the findings of \cite{fimr12,fmrz13} suggest one can describe the flavour hierarchies of the Standard Model via an F-theory SU(5) local model with a $E_8$ point where all the Yukawas originate from. The geometry near such point should be described by an $E_8$ T-brane profile for $\Phi$, and non-perturbative effects should be taken into account to increase the rank of the Yukawa matrix from one to three.  Since the holomorphic Yukawas already detect the fermion mass hierarchies, one can already see at this level whether a specific T-brane model is promising for reproducing empirical data. In the following section we will consider different classes of T-brane backgrounds and analyse if they give rise to the appropriate flavour hierarchy of fermion masses.

\section{$SU(5)$ models with $E_8$ enhancement}
\label{s:models}

In this section we will present a set of local SU(5) F-theory models that can be described as an $E_8$ theory higgsed by a T-brane background. Each of these models has the appropriate structure of matter curves so that they can embed the full content of the MSSM chiral spectrum, with only one massive family at tree-level. The remaining families of quark and leptons will become massive due to non-perturbative corrections, but then we find that one may get a hierarchy of masses either of the form $(1, \eps, \eps^2)$ or of the form $(1,\eps^2, \eps^2)$. Since $\eps$ is a very small number that measures the strength of a non-perturbative effect, the latter hierarchical pattern is very unlikely to reproduce empirical data, while the former has already been shown to be adequate in simpler local F-theory models \cite{fimr12,fmrz13}. 

As discussed in the previous section, to discover the hierarchical pattern that non-perturbative effects give rise to it suffices to compute the holomorphic piece of the Yukawa couplings. This will greatly simplify the analysis of this section, as these couplings can be computed via a residue formula \cite{cchv09,cchv10,fimr12,fmrz13}. Finally, for the sake of clarity we will only display the basic features of each model, their structure of matter curves and the final result for the holomorphic Yukawas, leaving the computational details to appendix  \ref{ap:details}.

\subsection{T-branes and matter curves}

One crucial feature of an F-theory local model with respect to the computation of Yukawa couplings is the profile for the 7-brane Higgs field $\Phi$ in the vicinity of the Yukawa point. As mentioned above, obtaining a third family much heavier than the other two naturally selects a T-brane profile for $\Phi$, which then specifies an appropriate local structure of matter curves. In the following we will briefly discuss the relation between T-brane profiles and matter curves, as they will be used when discussing each model. For a more thorough discussion on this subject we refer the reader to \cite{cchv10,fmrz13}. 

A T-brane background is specified by a particular configuration for the Higgs background $\langle \Phi \rangle$ that does not commute with its adjoint, namely $[\langle\Phi \rangle, \langle \Phi \rangle^\dag] \neq 0$. In this class of backgrounds the identification of the matter curves can be subtle because, unlike in the case of commuting Higgs field, there will not be an enhancement of the symmetry group in a complex curve within $S$. Nevertheless, it is still true that some additional roots of the algebra of $G_p$ will commute at specific complex codimension one loci, and this allows us to identify the matter curves as these particular loci. 

In order to detect the structure of matter curves it proves useful to work with matrix representations of the Higgs field in the algebra $\mathfrak{g}_\perp$ defined such that $\mathfrak{g}_{\rm GUT} \oplus \mathfrak{g}_\perp$ is a maximal subalgebra of $\mathfrak{g}_p = {\rm Lie} (G_p)$. Let us consider the case of interest in this paper, namely $\mathfrak{g}_p = \mathfrak{e}_8$ and $\mathfrak{g}_{\rm GUT}=\mathfrak{su}_{5}$. Then we have the well-known maximal decomposition
\bea
\label{maximal}
\mathfrak{e}_8 & \supset & \mathfrak{su}_{5}^{\rm GUT} \otimes \mathfrak{su}_{5}^\perp \\ \nonumber
\textbf{248} & \rightarrow & (\textbf{24},\textbf{1}) \oplus (\textbf{1},\textbf{24}) \oplus ((\textbf{10},\textbf{5}) \oplus c.c.)\oplus ((\overline{\textbf{5}},\textbf{10})\oplus c.c.)
\eea
Since the Higgs profile $\langle \Phi \rangle$ belongs to the adjoint of $\mathfrak{e}_8$ and by construction commutes with $\mathfrak{su}_{5}^{\rm GUT}$, it will only act non-trivially on the each of the representations $\CR$ of $\mathfrak{g}_\perp =\mathfrak{su}_{5}^\perp$ that appear in (\ref{maximal}). This action can be expressed in terms of  a matrix $\Phi_{\mathcal{R}}$ such that $[\langle \Phi\rangle, \CR] = \Phi_\CR \CR$ (see e.g. \cite{fmrz13}, section 3) so whenever the determinant of $\Phi_\CR$ vanishes 
 an element of $\CR$ will be commuting with $\langle \Phi\rangle$. Finally, given that $(\CR_{\rm GUT}, \CR) \subset \textbf{248}$, this will indicate that we have a zero mode transforming as $\CR_{\rm GUT}$ in the locus where ${\rm det\, } \Phi_\CR = 0$, and so the corresponding matter curve.

Interestingly, these facts allow to express the structure of matter curves in terms of the spectral surface of the Higgs field, which is defined as\footnote{The following
expression for the spectral surface holds if $\Phi$ takes values in a $\mathfrak{u}_n$ subalgebra of $\mathfrak{g}_p$.}
\be
P_{\Phi_{\mathcal{R}}} (x,y,z) = \text{det} (\Phi_{\mathcal R} - z I) =0
\label{ssurface}
\ee
for each of the matrices $\Phi_\CR$ associated to $\langle \Phi \rangle$. Following \cite{cchv10}, we say that $\Phi_{\mathcal R}$ is reconstructible if its spectral surface is a non singular algebraic variety, and that it is block reconstructible if it has the structure of a block diagonal matrix such that every block is reconstructible. As the property of reconstructibility is independent of the representation $\CR$ we then say that the Higgs field is block reconstructible, and in this case the whole information of $\langle \Phi \rangle$ is carried by its spectral surfaces.\footnote{As shown in \cite{cchv10}, $SU(k)$ reconstructible T-branes correspond to spectral covers with monodromy $\IZ_k$.} 

Now it is easy to see how the pattern of matter curves can be encoded in the spectral surface (\ref{ssurface}): when the Higgs field is block reconstructible its spectral surfaces will be the product of polynomials whose zero locus is a non-singular algebraic variety, and there will be a one to one correspondence between these varieties and the matter curves in a specific representation. Hence, the presence of several matter curves will induce a splitting of the spectral surface into irreducible polynomials, the number of factors of this splitting matching with the number of matter curves.

In the following we will present a number of local $E_8$ models whose local spectrum of matter curves can be detected by means of the above  considerations. For the sake of simplicity, we will focus on models in which the Higgs field background $\langle \Phi \rangle $ is block reconstructible, since then we can classify our models by the number of matter curves near the Yukawa point. It would however be interesting to extend our set of examples to more general, non-reconstructible backgrounds.

\subsection{Catalogue of models}

We now proceed to describe several kinds of local $E_8$ models with only one massive family at tree level. Such models are candidates to yield a realistic hierarchical fermion mass pattern after non-perturbative effects have been taken into account although, as already advertised, this will not always be the case. To find out we will compute the holomorphic Yukawa couplings, which depend on the profile for $\Phi$ in the holomorphic gauge \cite{cchv09}. For this purpose we only need to specify $\langle \Phi \rangle$ as a linear combination of holomorphic functions multiplying the $E_8$ roots, following the notation of appendix  \ref{ap:e8}. As explained above we may also describe this background as a matrix $\Phi_{\bf 5}$ acting on the representation ${\bf 5}$ of $\mathfrak{su}_{5}^\perp$, which allows to find the local set of {\bf 10} matter curves via eq.(\ref{ssurface}). Since we are considering reconstructible backgrounds, the $5\times 5$ matrix $\Phi_{\bf 5}$ will be automatically block diagonal, so we can classify our local models by the different dimension of each of these blocks.

For simplicity we will only provide the basic data for each of the models that we have studied, leaving the details for appendix  \ref{ap:details}. That is, we will describe the local set of matter curves that arise from the profile for $\Phi$ and the different possible assignments of the MSSM fields within them. Recall that in the absence of hypercharge flux the matter spectrum is organised in SU(5) multiplets, so for the purpose of computing holomorphic Yukawas we can consider that SU(5) is unbroken. Then to achieve rank one Yukawas at tree-level we need to have three copies of the matter representation ${\bf 10}_M$ within the same {\bf 10}-curve and three copies of ${\bf \bar{5}}_M$ in the same {\bf 5}-curve. Finally the Higgs multiplets $\mathbf{5}_U$ and $\mathbf{\bar{5}}_D$ should be in {\bf 5}-curves different from the one of ${\bf \bar{5}}_M$ and such that the couplings ${\bf 10}_M \times {\bf 10}_M \times {\bf 5}_U$ and ${\bf 10}_M \times {\bf \bar{5}}_M \times {\bf \bar5}_D$ are allowed. 

For each assignment we will present the structure of holomorphic Yukawa couplings that arise from the superpotential (\ref{suponp}) with 
\be
\theta_0 \, =\,  i(x \theta_x + y \theta_y)
\label{th0}
\ee
and $(x,y)$ parametrising the complex coordinates of the 4-cycle $S$. We will then discuss whether such structure accommodates favourable hierarchies to fit empirical data. In the next section we will provide a more detailed description of one of the models with such favourable structure, providing all the details that allow to compute its physical Yukawas. 

\subsubsection*{4+1 models}

Let us first consider a holomorphic background for $\Phi = \Phi_{xy} dx \wedge dy$ of the form 
\be
\langle \Phi_{xy} \rangle \, = \, \lambda (\hat H_1 +2 \hat H_2 +3 \hat H_3 +4 \hat H_4) +m( E_1^++E_2^+ + E_5^++mx E_3^-)
\ee
where the notation and definitions that are used for the $E_8$ roots are given in appendix  \ref{ap:e8}. Here $\lambda = \mu^2 (bx - y)$ is a holomorphic linear function of $(x,y)$ vanishing at the origin, which is where the Yukawa point $p$ will be located. By acting on the fundamental representation of $\mathfrak{su}_{5}^\perp$ we obtain the matrix representation 
\be
\Phi_{\bf 5}\, =\, \left(\begin{array}{c c c c c}\lambda & m & 0 & 0 &0\\0& \lambda & m & 0 & 0\\0&0&\lambda&m&0\\m^2 x& 0 & 0 &\lambda & 0\\0&0&0&0&-4\lambda\end{array}\right)\,,
\ee
which displays a $4+1$ block structure. The various matter representations and their matter curves are then the following ones
\begin{itemize}
\item[-] $\mathbf{10}$ sector
\be\nonumber
\mathbf{10}_a : \,  \lambda^4=m^5x\,, \qquad \mathbf{10}_b : \, \lambda = 0\,,
\ee
\item[-] $ \mathbf{5}$ sector
\be\nonumber
\mathbf{5}_a : \,  (3\lambda)^4 =m^5x \,, \qquad \mathbf{5}_b : \,  \lambda^2 (m^5 x + 4 \lambda^4 )=0\,,
\ee
\end{itemize}
and it is easy to see that all the curves meet at the origin.

This local model has already been considered in \cite{Chiou:2011js}, where it was found a rank one structure for the holomorphic Yukawas by using the tree-level superpotential  (\ref{supo7}). In the following we would like to extend this result by considering the superpotential (\ref{suponp}) corrected by non-perturbative effects and providing the resulting holomorphic Yukawas up to order $\CO(\eps^2)$. 

As pointed out in \cite{Chiou:2011js} (see also appendix  \ref{ap:details}) in order to generate an up-type Yukawa coupling ${\bf 10}_M \times {\bf 10}_M \times {\bf 5}_U$ it is necessary to assign the representation $\mathbf{10}_M$ to the $\mathbf{10}_a$ curve and $\mathbf{5}_U$ to the curve $\mathbf{5}_b$. Because there are only two {\bf 5}-curves, we will also consider that $\mathbf{\bar 5}_D$ is also localised in $\mathbf{5}_b$ while the three copies of the $\mathbf{\bar 5}_M$ are in $ \mathbf{5}_a$. 

With this setup one can compute the Yukawa matrices via a residue calculation. Schematically we find that
\be 
Y_U = \left(\begin{array}{c c c}0 & 0 & \epsilon\, y _{13}\\0& \epsilon\, y _{22}& 0\\ \epsilon\, y _{31} & 0 & y_{33}\end{array}\right) + \mathcal{O}(\epsilon^2)\,,
\ee
\be
Y_{D/L} = \left(\begin{array}{c c c}0 & 0 & \epsilon\, y _{13}\\0& \epsilon\, y _{22} & 0\\ \epsilon\, y _{31} & 0 & y_{33}\end{array}\right) + \mathcal{O}(\epsilon^2)\,,
\ee
where $y_{ij}$ are order one numbers (detailed expressions for $y_{ij}$ are found in appendix  \ref{ap:details}). We then reproduce the results of \cite{Chiou:2011js} in the limit $\eps \raw 0$, while we see that for $\eps\neq0$ the rank of both matrices is increased to three. Finally, both matrices will have a hierarchy of eigenvalues of the form $(\CO(1), \CO(\eps), \CO(\eps^2))$ so this model has a Yukawa structure which is favourable to reproduce the empirical data. 

Despite this favourable hierarchy, this model has the less attractive feature of having both up and down Higgses ${\bf 5}_U$, $\mathbf{\bar 5}_D$ in the same curve. Hence some particular mechanism should be invoked to prevent a large $\mu$-term to be generated. Because of this potential drawback we will not consider this model in the following. 

\subsubsection*{3+2 models}

We next consider a Higgs background of the form 
\be
\langle \Phi_{xy} \rangle = -\lambda \left(\frac{2}{3} \hat H_1 + \frac{4}{3} \hat H_2 +2 \hat H_3+\hat H_4\right)+\tilde m(E_1^+ + E_5^++\tilde m y E_8^-)+m(E_{10}^+ +m x E_{10}^-)\,.
\ee
where again $\lambda = \mu^2 (bx - y)$. Its action on the fundamental of $\mathfrak{su}_{5}^\perp$ is given by
\be
\Phi_{\bf 5} \, =\,  \left(\begin{array}{c c c c c}-\frac{2}{3}\lambda & \tilde m & 0 & 0 &0\\0& -\frac{2}{3}\lambda & \tilde m & 0 & 0\\\tilde m^2 y&0&-\frac{2}{3}\lambda&0&0\\0& 0 & 0 &\lambda & m\\
0&0&0&m^2x&\lambda\end{array}\right)\,,
\ee
showing a $3+2$ block structure. The various matter representations and curves are now
\begin{itemize}
\item[-] $\mathbf{10}$ sector
\be\nonumber
\mathbf{10}_a : \,  -\frac{8}{27}\lambda^3+\tilde m^4y=0\,, \qquad \mathbf{10}_b : \, \lambda^2-m^3x=0\,,
\ee
\item[-] $ \mathbf{5}$ sector
\be\begin{split}\nonumber
\mathbf{5}_a : \,\tilde m^4 y+\frac{64}{27}\lambda^3=0 \,, \ \ &\mathbf{5}_b : \,m^9 x^3=\frac{\lambda ^2 m^6x^2}{3}+m^3x \left(2 \lambda 
   \tilde m^4y-\frac{\lambda ^4}{27}\right)+\frac{1}{729}
   \left(\lambda ^3+27\tilde m^4 y\right)^2 ,\\ &\mathbf{5}_c : \lambda =0\,.
\end{split}\ee
\end{itemize}
In this class of models we can assign the three copies of $\mathbf{10}_M$ to either the $\mathbf{10}_a$ or the $\mathbf{10}_b$ sector. If we assign the $\mathbf{10}_M$ to the $\mathbf{10}_a$ sector then we need to assign $\mathbf{5}_U$ to the $\mathbf{5}_a$ sector in order to have ${\bf 10}_M \times {\bf 10}_M \times {\bf 5}_U$ Yukawas which are singlets under $SU(5)_\perp$. Nevertheless, an explicit computation shows that the holomorphic up-type Yukawa couplings vanish for this arrangement. This vanishing result is analogous to the one found in \cite{Chiou:2011js} for the $E_7$ model studied in there, with the matter curves involved in the up-type Yukawas having a similar structure. We therefore see that this assignment of chiral matter to curves does not yield realistic Yukawas.  

The other possibility in this model is to assign $\mathbf{10}_M$ to the $\mathbf{10}_b$ sector, which requires that $\mathbf{5}_U$ corresponds to the $\mathbf{5}_c$ sector. In addition one has to choose how to assign the representations $\mathbf{\bar 5}_M$ and $\mathbf{\bar 5}_D$ to the sectors $\mathbf{5}_a$ and $\mathbf{5}_b$, having two possibilities. The final structure of the Yukawa matrices does however not depend on this choice. In both cases we find that
\be
Y_U = \left(\begin{array}{c c c}0 & 0 & \epsilon\, y _{13}\\0& \epsilon\, y _{22}& \epsilon\, y _{23}\\ \epsilon\, y _{31} & \epsilon\, y _{32}& y_{33}\end{array}\right) + \mathcal{O}(\epsilon^2)\,,
\ee
\be
Y_{D/L} = \left(\begin{array}{c c c}0 & 0 & 0\\0& 0 & \epsilon\, y _{23}\\ 0& \epsilon\, y _{32} & y_{33}\end{array}\right) + \mathcal{O}(\epsilon^2)\,,
\ee
where again $y_{ij}$ are order one numbers whose explicit expression is given in appendix  \ref{ap:details}.
We see therefore that for this class of models the down-type Yukawa matrix does not have a favourable hierarchical structure.

\subsubsection*{2+2+1 models}

We finally consider a Higgs background of the form
\be
\langle \Phi_{xy} \rangle = \lam_1 (\hat H_1 +2 \hat H_2 -2\hat H_4)  - \lam_2 (\hat H_3+2 \hat H_4) + m (E_1^++mx E_1^-)+\tilde m(E_2^++\tilde my E_2^-)
\label{prephi}
\ee
whose action on the fundamental of $\mathfrak{su}_{5}^\perp$ is
\be
\Phi_{\bf 5} = \left(\begin{array}{c c c c c}\lambda_1 & m & 0 & 0 &0\\m^2 x& \lambda_1 & 0 & 0 & 0\\0&0& -2\lam_1-\lambda _2 &\tilde m&0\\0& 0 &\tilde m^2 y & -2\lam_1-\lambda _2  & 0\\0&0&0&0&2( \lambda_1+ \lambda_2)\end{array}\right)\,,
\label{prephi5}
\ee
and where now $\lambda_1$ and $\lambda_2$ are two different polynomials of $x$, $y$ which we shall take as $\lambda_1 = \mu_1^2 (b x - y)$ and $ \lambda _2 = \mu_2^2 (b x - y)$. As we will see in the next section, taking $\lam_1 \neq \lam_2$ with a slightly more general Ansatz will allow us to separate the two Yukawa points $p_{\rm up}$ and $p_{\rm down}$ from each other, introducing an interesting source of family mixing. 

The matter representations and matter curves are in this case
\begin{itemize}
\item[-] $\mathbf{10}$ sector
\be\nonumber
\mathbf{10}_a : \,  \lambda_1^2 -m^3 x =0\,, \qquad \mathbf{10}_b : \, (2\lambda_1+\lambda_2)^2-\tilde m^3 y=0\,,\qquad \mathbf{10}_c : \, \lambda_1+\lambda_2 =0\,,
\ee
\item[-] $ \mathbf{5}$ sector
\be\nonumber
\begin{split}
&\mathbf{ 5}_a : \, \lambda_1 =0  \,, \quad \mathbf{5}_b : \,  2\lambda_1+\lambda_2 = 0 \,,\quad \mathbf{5}_c : (3\lambda_1 +2\lambda_2)^2 - m^3 x=0\,,\\& \mathbf{5}_d : \lambda_2^2 - \tilde m^3y=0\,,
\quad \mathbf{5}_e : (\lambda_1+\lambda_2)^4 -2(\lambda_1+\lambda_2)^2 ( m^3x+ \tilde m^3y) +(m^3x-\tilde m^3y)^2=0\,.
\end{split}\ee
\end{itemize}
so the amount of matter curves increases considerably with respect to previous models. 

In this case we can assign the representation $\mathbf{10}_M$ to either the $\mathbf{10}_a$ or the $\mathbf{10}_b$ sectors. Since both choices end up leading to the same results we will choose the first option, which fixes the $\mathbf{5}_U$ representation within the $\mathbf{5}_a$ sector. The up-type Yukawas then have the following structure
\be
Y_U =\left(\begin{array}{c c c}0 & 0 & \epsilon\, y _{13} \\ 0& \epsilon\, y _{22} & 0\\ \epsilon\, y _{31}& 0 & y_{33}\end{array}\right) + \mathcal{O} (\epsilon^2)\,.
\label{preup}
\ee
We then find an eigenvalue hierarchy of the form $(\CO(1), \ \CO(\eps), \CO(\eps^2))$ and therefore  a suitable hierarchical structure to fit empirical data.

There are some possibilities now on how to associate the representations $\mathbf{\bar 5}_M$ and $\mathbf{ \bar 5}_D$ to the remaining matter curves, and this choice affects the down-type Yukawa matrix. We list here the possible choices and the resulting Yukawa matrices:
\begin{itemize}
\item[-] Either $\mathbf{\bar 5}_M$ is associated to $\mathbf{5}_d$ and $\mathbf{\bar 5}_D$ is associated to $\mathbf{5}_e$ or the other way round. In the first case we find that the down-type Yukawa matrix has the form
\be
Y_{D/L} = \left(\begin{array}{c c c}0 & 0 & 0 \\ 0& 0 & \epsilon\, y _{23} \\ 0& \epsilon\, y _{32} & y_{33}\end{array}\right) + \mathcal{O} (\epsilon^2)
\ee
whose eigenvalues are $(\CO(1), \ \CO(\eps^2), \CO(\eps^2))$.  Therefore this assignment for the matter fields does not lead to a good hierarchical structure for the down-type Yukawas. On the other hand, if we identify the $\mathbf{\bar 5}_M$ with $\mathbf{5}_e$, then it is not clear how to perform the analysis due to the fact that the matter curve is singular at the Yukawa point.

\item[-] Either $\mathbf{\bar 5}_M$ is associated to $\mathbf{5}_b$ and $\mathbf{\bar 5}_D$ to $\mathbf{5}_c$ or the other way round. In both cases we find that the down-type Yukawa matrix has the structure
\be
Y_{D/L} = \left(\begin{array}{c c c}0 & 0 & \epsilon\, y _{13} \\ 0& \epsilon\, y _{22} &0 \\ \epsilon\, y _{31}&0& y_{33}\end{array}\right) + \mathcal{O} (\epsilon^2)\,.
\label{predown}
\ee
that has the favourable eigenvalue hierarchy $(\CO(1), \ \CO(\eps), \CO(\eps^2))$. 

\end{itemize}

We then see that in the present  $2+2+1$ model there are two particular assignments of matter fields that yield a promising hierarchical structure for both up and down-type Yukawas. However, as we discuss next, one of the two has a more attractive structure for the $\mu$-term and neutrino masses, namely the choice of assigning $\mathbf{\bar 5}_M$ to $\mathbf{5}_b$ and $\mathbf{\bar 5}_D$ to $\mathbf{5}_c$.

\subsection{Comments on $\mu$-term and neutrino masses}\label{sec:neutrino}

One of the most attractive features of the $E_8$ models is that it is possible to describe the masses of the neutrinos and the $\mu$-term for the MSSM Higgs sector at the same time as the Yukawa couplings \cite{Heckman:2009mn}. Although in the following sections will not deal with them, let us briefly analyse here the structure of neutrino masses and $\mu$-term in the case of the two $2+2+1$ models with good hierarchical structures for the Yukawa matrices, in order to select one of them.

In the case in which we assign 
the $\mathbf{\bar 5}_M$ to $\mathbf{5}_c$ and the $\mathbf{\bar 5}_D$ to $\mathbf{5}_b$ we find that there is a singlet under $SU(5)_{GUT}$ that can give a coupling of the form $\mathbf{1} \times \mathbf{5}_U \times \mathbf{
\bar 5}_M$ and, after breaking $SU(5)_{GUT}$ down to the standard model gauge group this will imply the presence of the following coupling in the superpotential
\be
W \supset \lambda \, H_u L  \, S\,,
\ee
where we called the singlet $S$ and $L$ the lepton doublet superfield. This coupling, as analysed in \cite{Bouchard:2009bu}, corresponds to a Dirac mass for the neutrinos if we identify the singlet $S$ with the right handed neutrino $N_R$.
With this assignment of matter curves however it is not possible to have a renormalisable $\mu$-term for the Higgs fields. It is possible to generate a non-renormalisable $\mu$-term nonetheless if we consider the 
interactions of the Higgs fields with modes coming from other matter curves. In particular when the fields in the $\mathbf{5}_e$ come in vector-like pairs the following couplings will be allowed in the superpotential
\be
W \supset \lam_1 H_u \tilde S \phi + \lam_2 H_d \tilde S \phi^c+ \Lam \, \phi \phi^c\,,
\ee
where we called  $\phi$ any field in the $\mathbf{5}_e$ sector and $\phi^c$ its conjugate, $\Lambda$ is a mass term for $\phi$ and $\tilde S$ is a singlet. After integrating out $\phi$ and $\phi^c$ using their F-term
equations we find in the superpotential the following term
\be
W \supset \frac{\lam_1 \lam_2}{\Lam} \tilde S^2 H_u H_d\,,
\ee
which becomes an effective $\mu$-term  for the Higgs fields if the singlet $\tilde S$ gets a non-vanishing vev. Note that this kind of non-renormalisable effective $\mu$-term has already been considered in \cite{Kim:1983dt} and 
can provide a solution to the $\mu$-problem in the MSSM.

In the second case, namely in the case we assign
the $\mathbf{\bar 5}_M$ to $\mathbf{5}_b$ and the $\mathbf{\bar 5}_D$ to $\mathbf{5}_c$, we find that it is possible to have the following coupling in the superpotential 
\be
W \supset S H_u H_d 
\ee
which becomes an effective $\mu$-term if the singlet $S$ gets a non-vanishing vev. This class of effective $\mu$-term is particularly interesting because it can provide a mechanism for solving the $\mu$-term problem in the MSSM
\cite{Giudice:1988yz}.
However for this assignment of matter curves we find the feature that no masses for the neutrinos are possible if they are localised at the intersection of two 7-branes. Since one of the major motivations for studying Yukawa couplings at the point of $E_8$ is 
the generations of all couplings in the MSSM, including $\mu$-term and masses for neutrinos, we will henceforth focus our attention on the first model discussed in this subsection and start analysing it in detail in the next section.

\subsubsection*{Summary}

To sum up, from all the models discussed in this section, we have found that the $2+2+1$ model specified by (\ref{prephi}) is the most interesting phenomenologically, in the sense that it yields a hierarchical structure of Yukawa couplings of the form $(\CO(1), \CO(\eps), \CO(\eps^2))$ for a specific assignment of $SU(5)$ representations to matter curves. Such assignment is
\be
{\bf 10}_M \, \raw \, {\bf 10}_a \quad \quad \mathbf{\bar 5}_M \, \raw \, \mathbf{5}_c \quad \quad \mathbf{5}_U \, \raw \, \mathbf{5}_a \quad \quad \mathbf{\bar 5}_D \, \raw \, \mathbf{5}_b 
\ee
which also exhibits interesting mechanisms to generate realistic neutrino masses and $\mu$-term. In the next sections we will analyse this model in detail, specifying a background that includes the appropriate worldvolume fluxes on each of the above matter curves (section \ref{s:E8model}). We will then compute the wavefunctions (section \ref{s:realwave}) and the physical Yukawas (section \ref{s:physical}) for this local $E_8$ model, showing how empirical fermion masses and mixings can be fit upon an appropriate choice of parameters.

\section{An $E_8$ model with hierarchical Yukawas} \label{s:E8model}

Let us now consider in some detail the most promising of the $E_8$ local models discussed above. In particular we will describe the 7-brane background for the last 2+2+1 model and show that it indeed incorporates a realistic hierarchy of Yukawa couplings. 

Describing the background of a 7-brane local model with $E_8$ symmetry group entails specifying a Higgs field $\Phi$ and gauge connection $A$, both valued in the algebra $\mathfrak{su}_{5}^\perp$. If we want to preserve supersymmetry at the GUT scale, they must obey the F-term equations
\begin{subequations}
\label{Fterm7}
\begin{align}
\bar \partial_{A} \Phi = &\, 0\\
F^{(0,2)} =&\, 0
\end{align}
\end{subequations}
that arise from minimising the superpotential (\ref{supo7}). Also, the D-term (\ref{FI7}) should vanish
\be\label{Dterm}
\omega \wedge F + \frac{1}{2}  [\Phi, \Phi^\dagger]=0.
\ee
In order to find a solution to the above, one usually exploits the fact that the F-terms are invariant under the complexified gauge group, as opposed to the D-term which is only invariant under the real group. More explicitly, any holomorphic Higgs together with $A^{(0,1)}=0$ automatically satisfies the F-terms. This is referred to as a solution in \emph{holomorphic gauge} \cite{Font:2009gq,cchv09} which is not physical since it does not obey the full set of equations of motion and the gauge field is not real. However, one can still extract useful information at the holomorphic level, such as the structure of matter curves and rank of the Yukawa couplings by solving a relatively simple algebraic problem. Finally, by performing a complex gauge transformation, we can bring the fields in a real gauge that satisfy the D-term. This last step is the most challenging as it requires solving a set of partial differential equations that become particularly complicated in models including T-branes. However, it is unavoidable if we want to obtain the kinetic terms and hence the magnitude of the Yukawa couplings.

Following the above approach, we will first introduce the background for the Higgs field in holomorphic gauge and discuss the structure of matter curves and their intersections. We will then consider the background in a real gauge by imposing the D-term equation, which forces the introduction of non-primitive fluxes. We will complete the description of the background by introducing additional fluxes to achieve chirality for the MSSM fields as well as GUT symmetry breaking. This choice of background will yield an MSSM spectrum whose holomorphic Yukawa couplings are of the form (\ref{preup}) and (\ref{predown}), as we show by giving explicit expressions computed by means of a residue formula. The computation of physical Yukawa couplings and mixing angles is left for sections \ref{s:realwave} and \ref{s:physical}. 

\subsection{Higgs background}\label{sec:higgs}

\subsubsection*{Holomorphic gauge}

The first ingredient necessary to define our local model model is the background Higgs field $\langle \Phi \rangle = \langle \Phi_{xy}\rangle\, dx \wedge dy$ that triggers the breaking of $E_8$ to $SU(5)_{GUT}$. In holomorphic gauge we choose
\be\label{hback}
\langle \Phi_{xy}\rangle \,  =\,  \lam Q_1 + d(\lam+\kappa) Q_2 + m ( E_1^+ + m x  E_1^-)+\tilde m (E_2^+ + \tilde m y E_2^-),
\ee
where $Q_i, E^\pm_i$ are $E_8$ generators whose definition and all other details involving the $E_8$ Lie algebra are given in appendix  \ref{ap:e8}. Here $\lam= \mu^2 (bx-y)$, $m,\tilde m,\mu$ and $ \kappa$ are constants with dimensions of mass and $b,d$ are dimensionless constants. Notice that in terms of the background (\ref{prephi}) we have chosen $\lam_1 = \lam$ and $ \lambda_2 = - (d+2)\lambda - d\kappa$, with $\kappa$ being the distance between the two zeroes of these polynomials. As we will see, $\kappa$ will control the distance between the two Yukawa points $p_{\rm up}$ and $p_{\rm down}$ of this model.

As discussed in appendix  \ref{ap:details} this background breaks $SU(5)_\perp$ to $S(U(2) \times U(2) \times U(1))_\perp$, the representations of $SU(5)_\perp$ decomposing as
\bea
SU(5)_\perp &\longrightarrow& S(U(2) \times U(2)\times U(1))_\perp \\\nonumber
\mathbf{5} &\longrightarrow& (\mathbf{2},\mathbf{1})_{\frac{3}{5},-\frac{2}{5}}\oplus(\mathbf{1},\mathbf{2})_{-\frac{2}{5},\frac{3}{5}}\oplus(\mathbf{1},\mathbf{1})_{-\frac{2}{5},-\frac{2}{5}} \\\nonumber
\mathbf{10} &\longrightarrow& (\mathbf{1},\mathbf{1})_{\frac{6}{5},-\frac{4}{5}}\oplus(\mathbf{1},\mathbf{1})_{-\frac{4}{5},\frac{6}{5}}\oplus(\mathbf{2},\mathbf{1})_{\frac{1}{5},-\frac{4}{5}}\oplus(\mathbf{1},\mathbf{2})_{-\frac{4}{5},\frac{1}{5}}\oplus(\mathbf{2},\mathbf{2})_{\frac{1}{5},\frac{1}{5}}
\eea
where the subscripts denote the charges under the traces of the two $U(2)$ factors. Each of these subsectors corresponds to a different {\bf 5} or {\bf 10} matter curve, and some of them will be associated to the $SU(5)_{\rm GUT}$ fields $\{ \mathbf{10}_M$, $\mathbf{\bar 5}_M$, $\mathbf{5}_U$, $\mathbf{\bar 5}_D$\}.

In order to describe the different assignments of fields to matter curves and representations of $S(U(2) \times U(2) \times U(1))_\perp$ it proves useful to introduce a basis of the fundamental representation of $\mathfrak{su}_{5}^\perp$, that we denote by $e_1,\dots,e_5$. The action of the background Higgs field on this basis reads 
\be
\Phi_{\bf 5} = \left(
\begin{array}{ccccc}
 \lam  & m & 0 & 0 & 0 \\
 m^2 x & \lam  & 0 & 0 & 0 \\
 0 & 0 & d(\lam+\kappa) & \tilde m & 0 \\
 0 & 0 &  \tilde m^2 y  & d(\lam+\kappa) & 0 \\
 0 & 0 & 0 & 0 & -2(1+d)\lam - 2d\kappa
   \\
\end{array}
\right)\,,
\ee
in agreement with (\ref{prephi5}). A similar matrix $\Phi_{\bf 10}$ can be built for the antisymmetric representation of $\mathfrak{su}_{5}^\perp$, with elements of the form $e_i \wedge e_j$ with $i \neq j$. From $\Phi_{\bf 5}$ and $\Phi_{\bf 10}$ and using eq.(\ref{ssurface}) one can detect the ${\bf 10}$ and ${\bf 5}$ matter curves, respectively. For this model one then finds three {\bf 10}-curves and five {\bf 5}-curves, each of them corresponding to a factor in the above decomposition of the {\bf 5} and {\bf 10} of $SU(5)_\perp$, respectively.
 
Assigning matter fields to matter curves as in the previous section is then equivalent to specifying their $G_\perp = S(U(2) \times U(2)\times U(1))_\perp$ quantum numbers. One finds that for the model with a satisfactory hierarchy of Yukawa couplings the assignment is
\be\begin{array}{ccc}
\mathbf{10}_M : \, \left( \begin{array}{c} e_1 \\ e_2\end{array}\right ) \sim (\mathbf{2},\mathbf{1})_{\frac{3}{5},-\frac{2}{5}},
&\quad  \quad& \mathbf{\bar 5}_M  : \, \left (\begin{array}{c} e_1 \wedge e_5 \\ e_2\wedge e_5 \end{array}\right ) \sim (\mathbf{2},\mathbf{1})_{\frac{1}{5},-\frac{4}{5}},\\
\mathbf{5}_U :\, \left ( e^*_1 \wedge e^*_2 \right) \sim (\mathbf{1},\mathbf{1})_{-\frac{6}{5},\frac{4}{5}}, 
& \quad \quad & \mathbf{\bar 5}_D : \, \left ( e_3 \wedge e_4 \right ) \sim (\mathbf{1},\mathbf{1})_{-\frac{4}{5},\frac{6}{5}},
\end{array}
\ee
where we have also expressed these quantum numbers in terms of the basis $e_1,\dots,e_5$, see appendix  \ref{ap:details} for further details.

To cross-check this assignment let us read off each matter curve from the action of $\langle \Phi \rangle$ on the $G_\perp$ quantum numbers of each matter field. More precisely, we define $\Phi|_{{\mathcal R}_{\rm GUT}}$ as the action of $\langle \Phi_{xy} \rangle$ on the $\mathfrak{g}^\perp$ part of $(\CR_{\rm GUT}, \CR) \subset \textbf{248}$. We have that 
\be\begin{array}{ll}\label{phisector}
\Phi|_{\mathbf{10}_M}=\left(\begin{array}{cc}
\lambda  & m\\
m^2 x & \lam \end{array}\right ),& \Phi|_{\mathbf{\bar 5}_M}=\left(\begin{array}{cc}
-(1+2d)\lam-2d\kappa & m\\
m^2 x & -(1+2d)\lam-2d\kappa \end{array}\right )\\
\Phi|_{\mathbf{5}_U}=-2\lam,& \Phi|_{\mathbf{\bar 5}_D}=2d(\lam+\kappa).
\end{array}
\ee
The matter curves are then given by the vanishing of $\det \Phi|_{\mathcal R_{\rm GUT}}$, namely
\be
\begin{array}{ll}
\Sigma_{\mathbf{10}_M}\,:\, \lam^2-m^3x=0,\qquad&\Sigma_{\mathbf{\bar 5}_M}\,:\, \left ((1+2d)\lam+2d\kappa)\right )^2-m^3x=0\\
\Sigma_{\mathbf{5}_U}\,:\, \lam=0, \qquad& \Sigma_{\mathbf{\bar 5}_D}\,:\, d(\lam+\kappa)=0,
\end{array}
\ee
in agreement with the discussion of the previous section and that of appendix  \ref{ap:details}. 

The two types of Yukawa couplings are generated when these curves meet. In particular, the up-type Yukawa ${\bf 10}_M \times {\bf 10}_M \times {\bf 5}_U$ is developed at the intersection between $\Sigma_{\mathbf{10}_M}$ and $\Sigma_{\mathbf{5}_U}$. On the other hand, the down-type coupling ${\bf 10}_M \times {\bf 10}_M \times {\bf 5}_U$ appears at the point where $\Sigma_{\mathbf{10}_M}$, $\Sigma_{\mathbf{\bar 5}_M}$ and $\Sigma_{\mathbf{\bar 5}_D}$ coincide.\footnote{We are demanding that three curves in a surface meet, which does not look generic. However, due to gauge invariance, one of these equations is a linear combination of the others and the coupling is in fact generic.} These points are
\bea
Y_U&:&  \Sigma_{\mathbf{10}_M}\cap\Sigma_{\mathbf{5}_U}=\{x=y=0\} = p_{\rm up} \\\nonumber
Y_{D/L}&:&  \Sigma_{\mathbf{10}_M}\cap\Sigma_{\mathbf{\bar 5}_M}\cap \Sigma_{\mathbf{\bar 5}_D}=\left \{x=x_0,y=y_0\right \} = p_{\rm down} ,
\eea

\noindent with
\be\label{pdown}
x_0=\frac{\kappa^2}{m^3},\qquad y_0=\frac{\kappa}{\mu^2 }\left(1+\frac{\kappa b\mu^2}{m^3}\right).
\ee

\noindent We see that for each type of Yukawa there is a single intersection point and that, in general, these are not the same. It is the parameter $\kappa$ that controls the separation between them. For $\kappa=0$ both couplings are developed at the origin, whereas the two Yukawa points separate as $|\kappa|$ increases.

\subsubsection*{Real gauge}

The previous background fields are in holomorphic gauge and we would now like to find a physical solution, namely one that satisfies the D-term equation. As explained earlier, this is achieved by performing an arbitrary complex gauge transformation and imposing (\ref{Dterm}), which translates into a set of differential equations for such transformation.

More explicitly, consider the following transformation
\be
\Phi \rightarrow g\,  \Phi\,  g^{-1}\,, \quad A_{0,1} \rightarrow A_{0,1} + i g\, \bar \p\, g^{-1},
\ee
where $g$ is an element of $SU(5)^\perp_{\mathbb C}$. We propose the following Ansatz
\be\label{cgt}
g = e^{\frac{1}{2}(f P_1 +\tilde f P_2)}
\ee
where $P_K=[E^+_K,E^-_K]$ and $f,\tilde f$ are real functions of $(x,y)$. After the transformation, the Higgs and gauge fields are
\begin{subequations}\label{back}
\begin{align}
\Phi =&\, \lam Q_1 + d(\lam+\kappa) Q_2 + m (e^{f } E_1^+ + m x e^{-f} E_1^-)+\tilde m (e^{\tilde f } E_2^+ +\tilde m ye^{-\tilde f} E_2^-)\\
A_{0,1} =& \,-\frac{i}{2} \,(\bar \p f P_1 +\bar \p \tilde f P_2).
\end{align}
\end{subequations}
We can now plug these fields into the D-term (\ref{Dterm}) which yields equations for $f$ and $\tilde f$. Taking the K\"ahler form as
\be
\omega = \frac{i}{2} (d x \wedge d \bar x + d y \wedge d \bar y),
\label{omega}
\ee
these become
\begin{subequations}
\begin{align}
(\p_x \bar \p_{\bar x}+\p_y \bar \p_{\bar y}) f =&\, m^2 (e^{2 f}-m^2 |x|^2 e^{-2 f})\label{pain1}\\
(\p_x \bar \p_{\bar x}+\p_y \bar \p_{\bar y}) \tilde f =&\, \tilde m^2 (e^{2 \tilde f}-\tilde m^2 |y|^2 e^{-2 \tilde f}).
\end{align}
\end{subequations}
Following \cite{cchv10,fmrz13}, we take $f$ to depend only on $r_{x}$, the radial coordinate in the $(x,\bar x)$ plane. Then, eq.(\ref{pain1}) reduces to
\be\label{Pain}
\left(\frac{d^2}{ds^2}+\frac{1}{s}\frac{d}{ds}\right) h = \frac{1}{2} \sinh (2 h)
\ee
where $s= \frac{8}{3}( m r_x)^{\frac{3}{2}}$, and the function $h$ is defined as
\be
e^{2 f} = m r_x e^{2 h}%,\qquad\qquad  e^{2 f_2} = \tilde m r_2 e^{2 h_2}.
\ee
This is a particular case of the Painlev\'e III equation whose solution over the whole complex plane $\IC$ can be found in \cite{mtw77}. However, since we are working in a patch of $S_{\rm GUT}$ that contains the origin, we just need an approximate solution,\footnote{Far away from the origin the equations themselves receive corrections so it does not make sense to insist on solving the Painlev\'e equation for the whole of $\mathbb C$.} which is \cite{fmrz13}
\be
f (r_x) =  \log c + c ^2 m ^2 r_x^2 +m^4 r_x ^4 \left( \frac{c^4}{2}-\frac{1}{4 c^2}\right)+\dots
\ee
and similarly for $\tilde f$, replacing $x\rightarrow y$. In the previous equation the constant $c$ needs to be fixed to the values
\be
c = 3^{1/3}\frac{\Gamma\left[\frac{2}{3}\right]}{\Gamma\left[\frac{1}{3}\right]}\sim 0.73\,,
\ee
if we ask for a regular solution for all values of $r_x$. However, as mentioned above, we will not restrict to this particular choice since the actual value will be fixed only when all the global details of the background are specified. %Thus, in the spirit of a local model we treat it as a free parameter with values around $0.73$.

\subsection{Primitive fluxes}

The fields (\ref{back}) define a consistent background that solves both the F and D-term equations. We can still find a more general background by turning on additional gauge fluxes, however, these cannot be generic since that would require modifying $\Phi$. The most general flux that respects the Higgs field in (\ref{back}) has to commute with it and be primitive on $S_{\rm GUT}$. If it also keeps the gauge group $SU(5)_{GUT}$ unbroken such flux is of the form
\be
F_Q = i (dx \wedge d \bar x-dy \wedge d\bar y ) (M_1 Q_1+M_2Q_2) + i (dx \wedge d \bar y +dy \wedge d \bar x)( N_1 Q_1+N_2Q_2)\,.
\label{Qflux}
\ee
As usual, the presence of such worldvolume flux is necessary to induce 4d chirality in the matter curves. The modes of opposite chirality {\bf 5}, ${\bf \bar{5}}$ and {\bf 10}, ${\bf \overline{10}}$ feel the background (\ref{back}) in a similar way, and so whenever there is a zero mode solution for one chirality there will also be a solution for the opposite chirality. This is no longer true for the background flux (\ref{Qflux}), that will locally select modes of one chirality or the other depending on the signs of $M_i$, $N_i$, $i =1,2$. This chirality selection can be characterised in terms of a local chirality index \cite{Palti:2012aa}, as discussed in more detail in appendix  \ref{ap:chiral}.

Finally, following the standard strategy in F-theory GUTs, we break $SU(5)_{GUT}$ down to the SM gauge group by turning on a flux along the hypercharge generator. Keeping the associated gauge boson massless amounts to imposing a global condition, which is invisible at the local level of our discussion. We then parametrise such flux as
\be
F_Y = i \left[ \tilde N_Y (dy \wedge d\bar y-dx \wedge d\bar x)+N_Y (dx \wedge d \bar y+dy \wedge d\bar x)\right] Q_Y\,,
\ee
where the hypercharge generator is defined as follows
\be
Q_Y = \frac{1}{3} \left(\tilde H_1 +2 \tilde H_2 +3 \tilde H_3\right)+\frac{1}{2} \tilde H_4\,.
\ee
The total primitive flux is then
\be
F_p=iQ_R(dy\wedge d\bar y-dx\wedge d\bar x)+iQ_S(dx\wedge d\bar y+dy\wedge d\bar x)
\ee
with 
\be
Q_R=\tilde N_YQ_Y-M_1Q_1-M_2Q_2,\qquad Q_S= N_YQ_Y+N_1Q_1+N_2Q_2.
\label{qrqs}
\ee
These fluxes will enter into the Dirac equation for the zero modes of our model. As a result, each of the MSSM chiral zero modes will feel a different flux depending on their quantum numbers (and in particular its hypercharge) and will then develop a different wavefunction profile. As mentioned before, these flux differences will not affect the Yukawas at the holomorphic level, but they will enter into the final expression for the physical Yukawas. We have then gathered the flux felt by the different MSSM sectors of the present $E_8$ model in table \ref{t:sectors}, and in particular the effective combination of fluxes $q_R$ and $q_S$ that will be crucial for the computations of section \ref{s:realwave}.

\begin{table}[htb]
%\footnotesize
\renewcommand{\arraystretch}{1.2}
\setlength{\tabcolsep}{2pt}
\begin{center}
\begin{tabular}{|c||c|c|l||c|c|}
\hline
MSSM & Sector & $S(U(2)\times U(2)\times U(1))_\perp$ & $G_{\rm MSSM}$  & $q_R$ & $q_S$\\
\hline
\hline
$Q$&$\mathbf{10}_M$&$(\mathbf{2},\mathbf{1})_{\frac{3}{5},-\frac{2}{5}}$&$(\mathbf{3},\mathbf{2})_{-\frac{1}{6}}$&$-\frac{1}{6}\tilde N_Y-M_1$&$-\frac{1}{6} N_Y+N_1$\\
\hline
$U$&$\mathbf{10}_M$&$(\mathbf{2},\mathbf{1})_{\frac{3}{5},-\frac{2}{5}}$&$(\mathbf{\bar 3},\mathbf{1})_{\frac{2}{3}}$&$\frac{2}{3}\tilde N_Y-M_1$&$\frac{2}{3} N_Y+N_1$\\
\hline 
$E$&$\mathbf{10}_M$&$(\mathbf{2},\mathbf{1})_{\frac{3}{5},-\frac{2}{5}}$&$(\mathbf{1},\mathbf{1})_{-1}$&$-\tilde N_Y-M_1$&$- N_Y+N_1$\\
\hline
$D$&$\mathbf{\bar 5}_M$&$(\mathbf{2},\mathbf{1})_{\frac{1}{5},-\frac{4}{5}}$&$(\mathbf{\bar 3},\mathbf{1})_{-\frac{1}{3}}$&$-\frac{1}{3}\tilde N_Y+M_1+2M_2$&$-\frac{1}{3} N_Y+N_1-2N_2$\\
\hline
$L$&$\mathbf{\bar 5}_M$&$(\mathbf{2},\mathbf{1})_{\frac{1}{5},-\frac{4}{5}}$&$(\mathbf{1},\mathbf{2})_{\frac{1}{2}}$&$\frac{1}{2}\tilde N_Y+M_1+2M_2$&$\frac{1}{2} N_Y+N_1-2N_2$\\
\hline
$H_U$&$\mathbf{5}_U$&$(\mathbf{1},\mathbf{1})_{-\frac{6}{5},\frac{4}{5}}$&$(\mathbf{1},\mathbf{2})_{-\frac{1}{2}}$&$-\frac{1}{2}\tilde N_Y+2M_1$&$-\frac{1}{2} N_Y-2N_1$\\
\hline
$H_D$&$\mathbf{\bar 5}_D$&$(\mathbf{1},\mathbf{1})_{-\frac{4}{5},\frac{6}{5}}$&$(\mathbf{1},\mathbf{2})_{\frac{1}{2}}$&$\frac{1}{2}\tilde N_Y-2M_2$&$\frac{1}{2} N_Y+2N_2$\\
\hline
%$T_U$&$\mathbf{5}_U$&$(\mathbf{1},\mathbf{1})_{-\frac{6}{5},\frac{4}{5}}$&$(\mathbf{3},\mathbf{1})_{\frac{1}{3}}$&$\frac{1}{3}\tilde N_Y+2M_1$&$\frac{1}{3} N_Y-2N_1$\\
%\hline
%$T_D$&$\mathbf{\bar 5}_D$&$(\mathbf{1},\mathbf{1})_{-\frac{4}{5},\frac{6}{5}}$&$(\mathbf{\bar 3},\mathbf{1})_{-\frac{1}{3}}$&$-\frac{1}{3}\tilde N_Y-2M_2$&$-\frac{1}{3} N_Y+2N_2$\\
%\hline
\end{tabular}
\end{center}
\caption{Different sectors and charges for the $E_8$ model of this section. Here $q_R$ and $q_S$ are the $E_8$ operators (\ref{qrqs}) evaluated at each different sector. All the multiplets in the table have the same chirality.}
\label{t:sectors}
\end{table}

\subsection{Residue formula for Yukawa couplings}\label{sec:residuemain}

The computation of the holomorphic Yukawa couplings can be performed via dimensional reduction of the 7-brane superpotential
\be\label{eq:superp}
W = m_*^4\int_S \text{Tr}\left(\Phi \wedge F\right) +\frac{\eps}{2}\theta_0 \text{Tr}\left(F \wedge F\right)\,.
\ee
As discussed in section \ref{s:yukawas}, the second term in (\ref{eq:superp}) is due to the presence of non-perturbative effects in the compactification. In particular we have that $\theta_0$ is a holomorphic section on $S$ and $\epsilon$ is a parameter that measures the strength of the non-perturbative effect. We note that the presence of this additional term will eventually change the BPS equations for the background that we previously solved, and so the background profiles for $\Phi$ and $F$ will have $\CO(\eps)$ corrections. One can however show that these corrections do not affect the computation of holomorphic Yukawas \cite{fimr12}, and so they can be ignored in what follows.\footnote{The full superpotential expression involves terms of the form $\theta_k \str (\Phi_{xy}^k F^2)$, $k\geq2$, but suppressed by higher powers in $m_*$ \cite{mm09,afim11,fimr12}. Hence their contributions will be less relevant than those from $\theta_0$.} 

The zero mode equations can be derived by expanding the 7-brane fields into background and fluctuations
\be\label{expan}
\Phi = \langle \Phi \rangle + \varphi\,, \quad A= \langle A \rangle + a\,,
\ee
and linearising in fluctuations the F-term equations derived from (\ref{eq:superp}). We obtain 
\be\label{eq:fluctF}\begin{split}
\bar \p_{\langle A\rangle} a &=0\,,\\
\bar \p_{\langle A\rangle} \varphi &=  i [a, \langle \Phi\rangle ]- \eps \p \theta_0 \wedge (\p_{\langle A\rangle} a +\bar \p_{\langle A\rangle} a^\dagger)\,.
\end{split}
\ee
A similar procedure can be applied to the D-term equation, but since we are simply looking at the holomorphic part of the Yukawa couplings in this section we will postpone that discussion to the following section. It is possible to solve explicitly for the system (\ref{eq:fluctF}), the solution being
\be\label{eq:Fsol}\begin{split}
a &= \bar \p _{\langle A \rangle} \xi \,,\\
\varphi  &=h-i [\langle \Phi\rangle,\xi] +\eps \p \theta_0 \wedge (a^\dag - \p_{\langle A\rangle} \xi)\,,
\end{split}
\ee
where $\xi$ is a section of $\Omega^{(0,0)}(S) \otimes \text{ad}(E_8)$ and $h$ is a holomorphic section of $\Omega^{(2,0)}(S) \otimes \text{ad}(E_8)$. We stress that while the solution (\ref{eq:Fsol}) contains dependence on $a^\dag$ which may in principle introduce some non-holomorphic terms in the 4d superpotential, these terms will appear only in total derivatives and so they will not eventually appear in the resulting 4d superpotential \cite{fimr12}. Using this solution it is possible to prove \cite{cchv09,fimr12,cchv10,fmrz13} that the Yukawa couplings are
\be
Y = -i \frac{m_*^4}{3} \int_S \text{Tr} (h \wedge \bar \p_{\langle A \rangle} \xi \wedge \bar \p_{\langle A \rangle} \xi)\,.
\ee
It is also possible to write the Yukawa couplings as a residue evaluated at the Yukawa point. We simply quote here the result referring to  \cite{fimr12,fmrz13} for the general proof:
\be
Y =  m_*^4 \pi^2 f_{abc}\,\text{Res}_p \left[\eta^a \eta^b h_{xy}\right] =  m_*^4 \pi^2 f_{abc} \int_{\mathcal{C}}\eta^a \eta^b h_{xy}d x \wedge dy\,,
\label{yukres}
\ee
where $\mathcal{C}$ can be continuously contracted to a product of unit circles surrounding the Yukawa point $p$ without encountering singularities in the integrand
and we defined the function $\eta$ as
\be
\eta = - i \Phi^{-1}\left[h_{xy}+i \epsilon\p_x \theta_0 \p_y \left (\Phi^{-1} h_{xy}\right)-i \epsilon\p_y \theta_0 \p_x \left (\Phi^{-1} h_{xy}\right)\right]\,.
\label{eta}
\ee

\subsection{Holomorphic Yukawa couplings for the $E_8$ model}

Let us finally discuss the structure of the Yukawa couplings that arise in the present local $E_8$ model. 
We focus our attention on the sector involving only the MSSM fields so we will have only two Yukawa matrices, namely the $\mathbf{10}_M \times \mathbf{10}_M \times \mathbf{5}_H$ and the $\mathbf{10}_M \times \mathbf{\bar 5}_M \times \mathbf{\bar 5}_H$. The functions $h_{xy}$ for the different fields are
\be\label{haches}
\begin{array}{ll}
h_{\mathbf{10}_M} = \g_{10,i}\,m_*^{3-i}(bx-y)^{3-i}& h_{\mathbf{ \bar 5}_M}=  \g_{5,i}\,m_*^{3-i}(b(x-x_0)-(y-y_0))^{3-i}\\
h_{\mathbf{5}_H} = \gamma_U & h_{\mathbf{\bar 5}_H} = \gamma_D,
\end{array}
\ee
where $(x_0,y_0)$ corresponds to the coordinates (\ref{pdown}) of the down-type Yukawa point $p_{\rm down}$, the constants $\g_{10,i}, \g_{5,i},\g_U,\g_D$ are normalisation factors to be fixed in the next section and $i=1,2,3$ is a family index. Using these we can compute the functions $\eta$ in (\ref{eta}) which in turn are needed to compute the holomorphic couplings via the residue formula (\ref{yukres}). Such $\eta$'s are computed in appendix  \ref{ap:holoyuk}, where the following Yukawa couplings are found:
\be\label{hyuk}
\begin{split}
Y_U =  &\frac{\pi^2\,\g_U\,\g_{10,3}^2}{2\rho_m\rho_\mu}\left (\begin{array}{ccc}
0&0&\tilde \eps\frac{\g_{10,1}}{2\rho_\mu\g_{10,3}}\\
0&\tilde \eps\frac{\g_{10,2}^2}{2\rho_\mu\g_{10,3}^2}&0\\
\tilde \eps\frac{\g_{10,1}}{2\rho_\mu\g_{10,3}}&0&1\end{array}\right )\\ 
Y_{D/L} = &-\frac{\pi^2\,\g_D\,\g_{10,3}\,\g_{5,3}}{2d\,\rho_m\rho_\mu}\left (\begin{array}{ccc}
-\tilde\eps\tilde\kappa^2\frac{\g_{10,1}\g_{5,1}}{2d\rho_\mu^3\g_{10,3}\,\g_{5,3}}&\tilde\eps\tilde\kappa\frac{\g_{10,1}\g_{5,2}}{d\rho_\mu^2\g_{10,3}\g_{5,3}}&\left (\frac{2\tilde \kappa^2}{\rho_\mu}-\frac{\tilde \eps}{d} \right )\frac{\g_{10,1}}{2\rho_\mu\g_{10,3}}\\
\tilde\eps\tilde\kappa\frac{\g_{10,2}\g_{5,1}}{2d\rho_\mu^2\g_{10,3}\g_{5,3}} &-\tilde\eps \frac{\g_{10,2}\g_{5,2}}{2d\rho_\mu\g_{10,3}\g_{5,3}}&-\tilde \kappa\frac{\g_{10,2}}{\rho_\mu\g_{10,3}}\\
-\tilde\eps \frac{\g_{5,1}}{2d\rho_\mu\g_{5,3}}&0&1\end{array}\right )
\end{split}
\ee
\vspace*{-.5cm}
with
\be
\tilde \eps=\eps(\th_x+b\th_y),\qquad \tilde\kappa=\frac{\kappa}{m_*},\qquad \rho_m=\frac{m^2}{m_*^2},\qquad \rho_\mu=\frac{\mu^2}{m_*^2}
\ee
and where we have kept terms linear in $\epsilon$.
Notice that $Y_U$ is of the form (\ref{preup}) and $Y_{D/L}$ reduces to (\ref{predown}) in the limit $\kappa \raw 0$. A non-vanishing $\kappa$ distorts the form of $Y_{D/L}$, but it does not spoil its hierarchical structure of eigenvalues. In fact, as we will see in section \ref{s:physical}, $\kappa$ will only enter in the CKM matrix describing quark mixing angles.% with an approximate value of $\tilde \kappa \sim \sqrt{\tilde \eps}$. 

\section{Zero mode wavefunctions}\label{s:realwave}

So far our discussion of the Yukawa couplings has been restricted to the holomorphic level, namely we have been discussing the cubic couplings that appear in the four dimensional superpotential. The missing ingredient in the computation of the physical couplings involves the normalisation of the wavefunctions as well as their kinetic mixing. Indeed, recall that the Yukawa couplings that are measured experimentally can be compared to those that appear in the Lagrangian once the kinetic terms for the chiral fields are canonical. In order to compute such kinetic terms, we necessarily need to solve the equations of motion for the zero modes in a real gauge, which introduces the dependence on the worldvolume flux densities.

In this section we will solve for such zero mode wavefunctions of the $E_8$ model of section \ref{s:E8model}, and compute the kinetic terms for them. Solving analytically for real gauge wavefunctions is a much more complicated problem than doing it at the holomorphic level, especially for T-brane models like ours, and the problem becomes particularly involved when non-perturbative corrections are taken into account. However, similarly to \cite{fmrz13} these zero mode equations can be solved  for a certain region of parameters of the local model, allowing to see how such wavefunctions depend on flux densities. Our approach will be to first consider the perturbative case and compute the kinetic terms, where no kinetic mixing arises for the choice of wavefunction for each family that we make. Second, we include the non-perturbative corrections and argue that they do not change the result. Many computational details will be relegated to appendix \ref{ap:wave} (see also \cite{afim11,fimr12,fmrz13}).

That is, in this section we will see that {\it i)} there is no kinetic mixing between families and {\it ii)} non-perturbative corrections do not affect their kinetic terms, at least at the level of approximation that we are working. As a result, to compute physical Yukawas one may combine the residue computation of the non-perturbative holomorphic couplings with the computation of the normalisation factors $\g_{10}$, $\g_{5}$ at tree-level that we do in the following, and which should be inserted in eq.(\ref{hyuk}) to obtain the physical Yukawas.\footnote{Alternatively, one may directly compute the physical Yukawas by performing the triple overlap of real gauge zero modes. Similarly to \cite{fimr12,fmrz13} one can show that both approaches give the same result. }

\subsection{Wavefunctions in the perturbative limit}

Before including non-perturbative corrections, the equations for the zero modes are obtained from (\ref{Fterm7}) and (\ref{Dterm}) by expanding to linear order around a given background as in (\ref{expan}) which yields
\begin{subequations}\label{zeropert}
\begin{align}
\bar \p_{\langle A\rangle} a=&\,0\,,\label{treeF1}\\
\bar \p_{\langle A\rangle} \varphi =&\,i [a , \langle \Phi\rangle]\,,\label{treeF2}\\
\omega \wedge  \p_{\langle A\rangle} a=&\,\frac{1}{2}[\langle \bar \Phi \rangle,\varphi]\,.\label{treeD}
\end{align}
\end{subequations}
where $\langle \Phi\rangle$ and $\langle A\rangle$ correspond to the background in real gauge. These can be solved for every particular sector using the techniques in \cite{afim11,fimr12,fmrz13} and in the following we just quote the result for the relevant sectors, namely those that appear in table \ref{t:sectors}. The details of the computation can be found in appendix  \ref{ap:wave}.

We use the following notation for the zero modes,
\be
\vec \varphi_\rho=  \left(\begin{array}{c}
a^s_{\bar x}\\
a^s_{\bar y}\\
\varphi^s_{xy}\end{array}\right) E_{\rho,s}
\ee
where $E_{\rho,s}$ denotes the particular set of roots, labeled by $s$, for each sector $\rho$. In the case of the up and down-type Higgses, $s$ only takes one value since these are only charged under an Abelian subgroup of $S(U(2)\times U(2)\times U(1))$. On the other hand, the matter sectors transform as doublets of the first $U(2)$ factor so $s$ takes two different values in that case.

\subsubsection*{Higgs wavefunctions}

The solution for the $\mathbf{5}_U$ sector is
\be
\vec \varphi_U = \gamma_U \left(\begin{array}{c}-i \frac{\zeta_U}{2\mu^2}\\-i\frac{\zeta_U-\lambda_U}{2\mu^2}\\1\end{array}\right) \chi_U\,E_U
\ee
where
\be
\chi_U(x,y)=e^{\frac{q_R}{2}(|x|^2-|y|^2)-q_S (x \bar y + y \bar x)+(x-y)(\zeta_U\bar x-(\lambda_U-\zeta_U)\bar y)}
\ee
and $\lambda_U$ is a function of the flux densities and intersection parameters given as the lowest solution to the cubic equation (\ref{cubU}). Also, $\zeta_U= \frac{\lambda_U(\lambda_U-q_R-q_S)}{2(\lambda_U-q_S)}$.

Similarly, the solution for the $ \mathbf{\bar 5}_D$ is 
\be
\vec \varphi_D = \gamma_D \left(\begin{array}{c}i \frac{\zeta_D}{2d\mu^2}\\i\frac{\zeta_D-\lambda_D}{2d\mu^2}\\1\end{array}\right) e^{-i\psi}\chi_D(x-x_0,y-y_0)\,E_D
\ee
with 
\be
\chi_D(x,y)=e^{\frac{q_R}{2}(|x|^2-|y|^2)-q_S (x \bar y + y \bar x)+(x-y)(\zeta_D\bar x-(\lambda_D-\zeta_D)\bar y)}.
\ee
and where $\psi$ is defined in (\ref{gaugepar}). Finally, $\lambda_D$ is the lowest solution to (\ref{cubD}) and $\zeta_D= \frac{\lambda_D(\lambda_D-q_R-q_S)}{2(\lambda_D-q_S)}$.

\subsubsection*{Matter wavefunctions}

These sectors are a bit more involved because the fields are charged under the T-brane background. Given our choice of background, both the $\mathbf{10}_M$ and $\mathbf{\bar5}_M$ transform as doublets of the first $SU(2)$ factor in the decomposition of $SU(5)^\perp$. Thus, we write the solution as
\be
\vec \varphi = \left(\begin{array}{c} a^+_{\bar x}\\ a^+_{\bar y}\\\varphi^+ _{xy}\end{array}\right) E_1^++\left(\begin{array}{c} a^-_{\bar x}\\ a^-_{\bar y}\\\varphi^- _{xy}\end{array}\right) E_1^-=
\vec \varphi_+E_1^+ +\vec \varphi_- E_1^-\,,
\ee
where we use a + to denote the upper component of the $U(2)_1$ doublet and $-$ to denote the lower one. The zero mode equations in these two sectors turn out to be rather complicated to solve in general, but it is still possible to find approximate solutions in the limit $\mu,\kappa \ll m$. The real wavefunction for the ${\mathbf {10}}_M$ is
\be
\vec \varphi_{10}^i = \gamma_{10}^i \left(\begin{array}{c}\frac{i \lambda_{10}}{m^2}\\-i\frac{\lambda_{10} \zeta_{10}}{m^2}\\0\end{array}\right)e^{f/2} \chi_{10}^i E^+_1+\gamma_{10}^i \left(\begin{array}{c}0\\0\\1\end{array}\right)e^{-f/2} \chi_{10}^i E^-_1
\ee
where $\lambda_{10}$ is the negative solution to the cubic (\ref{cub10}) and $\zeta_{10} = -q_S / (\lambda_{10} -q_R)$. Finally the wavefunctions $\chi_{10}^i$ are 
\be
\chi_{10}^i = e^{\frac{q_R}{2}(|x|^2-|y|^2)-q_S (x\bar y+y\bar x)+\lambda_{10} x(\bar x- \zeta_{10} \bar y)} g_{10}^i(y+\zeta_{10} x)\,,
\ee
where $g_{10}^i$ are holomorphic functions of $y+\zeta_{10} x$ and $i=1,2,3$ is a generation index. As in \cite{fimr12,fmrz13} we choose these holomorphic functions in the following way
\be\label{fam10}
g_{10}^i (y+\zeta_{10} x) = m_*^{3-i} (y+\zeta_{10} x)^{3-i}\,.
\ee
The solution to the $\bar{\mathbf 5}_M$ is very similar and reads
\be
\vec \varphi_{5}^i = \gamma_{5}^i \left(\begin{array}{c}\frac{i \lambda_{5}}{m^2}\\-i\frac{\lambda_{5} \zeta_{5}}{m^2}\\0\end{array}\right)e^{i\tilde\psi +f/2} \chi_{5}^i(x,y-\nu/a) E^+_1+\gamma_5^i \left(\begin{array}{c}0\\0\\1\end{array}\right)e^{i\tilde\psi-f/2} \chi_{5}^i(x,y-\nu/a) E^-_1
\ee
with $\tilde\psi$ defined in (\ref{gaugepar2}). Also, $\lambda_{5}$ is a function of the fluxes and intersection parameters defined as the lowest solution to (\ref{cub5}) and $\zeta_{5} = -q_S / (\lambda_{5} -q_R)$. Finally the wavefunctions $\chi_{5}^i$ are 
\be
\chi_{5}^i(x,y) = e^{\frac{q_R}{2}(|x|^2-|y|^2)-q_S (x\bar y+y\bar x)+\lambda_{5} x(\bar x- \zeta_{5} \bar y)} g_{5}^i(y+\zeta_{5} x)\,,
\ee
where $g_{5}^i$ are holomorphic functions of $y+\zeta_{5} x$ and $i=1,2,3$ is a generation index. Analogously, the family functions are
\be\label{fam5}
g_{5}^i (y+\zeta_{5} x) = m_*^{3-i} (y+\zeta_{5} x)^{3-i}\,.
\ee

\subsection{Normalisation factors}

Once we have the perturbative wavefunctions we can compute the normalisation factors and kinetic mixing. The appropriate scalar product is given by
\be
K_\rho^{ij}\, =\, \langle \vec{\vphi}_{\rho}^{i} | \vec{\vphi}_{\rho}^{j} \rangle \, = \,
m_*^4 \int_S \tr \,( \vec{\vphi}_{\rho}^{i}{}^\dag \cdot \vec{\vphi}_{\rho}^{j})\, {\rm d vol}_S
\label{4dnorm}
\ee
as can be seen by performing the dimensional reduction.

Given the choice of family functions (\ref{fam10}) and (\ref{fam5}) we find that the kinetic terms (\ref{4dnorm}) are diagonal, so we only need to compute the corresponding normalisation factors. We find
\begin{subequations}\label{gammas}
\begin{align}
|\gamma_U|^2=&\,-\frac{4}{\pi^2}\left (\frac{\mu}{m_*} \right )^4\frac{(2\zeta_{U}+q_R)(q_R+2\zeta_{U}-2\lam_{U})+(q_S+\lam_{U})^2}{4\mu^4+\zeta_{U}^2+(\zeta_{U}-\lam_{U})^2}\\
|\gamma_D|^2=&\,-\frac{4d^2}{\pi^2}\left (\frac{\mu}{m_*} \right )^4\frac{(2\zeta_{D}+q_R)(q_R+2\zeta_{D}-2\lam_{D})+(q_S+\lam_{D})^2}{4d^2\mu^4+\zeta_{D}^2+(\zeta_{D}-\lam_{D})^2}\\
|\gamma_{10, j}|^2=&\,-\frac{c}{m_*^2\pi^2(3-j)!}\frac{1}{\frac{1}{2\lam_{10}+q_R(1+\zeta_{10}^2)-m^2 c^2}+\frac{c^2\lam_{10}^2}{m^4}\frac{1}{2\lam_{10}+q_R(1+\zeta_{10}^2)+m^2 c^2}}\left (\frac{q_R}{m_*^2}\right )^{4-j}\\
|\gamma_{5, j}|^2=&\,-\frac{c}{m_*^2\pi^2(3-j)!}\frac{1}{\frac{1}{2\lam_{5}+q_R(1+\zeta_{5}^2)-m^2 c^2}+\frac{c^2\lam_{5}^2}{m^4}\frac{1}{2\lam_{5}+q_R(1+\zeta_{5}^2)+m^2 c^2}}\left (\frac{q_R}{m_*^2}\right )^{4-j}
\end{align}
\end{subequations}
Recall that the parameters $\lam_\rho$ and $\zeta_\rho$ for a given sector $\rho$, depend on the flux densities felt by such a sector and, in particular, depend on the hypercharge flux. Thus, each MSSM multiplet in a given GUT multiplet will have different normalisation factors.

\subsection{Non-perturbative corrections to the wavefunctions}

The computation of the kinetic terms performed above allows to obtain the physical Yukawa couplings at tree level. However, since we are interested in the leading corrections induced by the non-perturbative effects, we need to compute the normalisation factors and mixings at $\CO(\eps)$. To do so, we will solve for the zero mode equations in real gauge including all $\CO(\eps)$ corrections. As we will see, it turns out that no mixing is generated and the normalisation factors are not corrected at this order, so we may use the ones obtained earlier.

Following section \ref{sec:residuemain}, the F-term equations for the zero modes at $\CO(\eps)$ read
\be\begin{split}
\bar \p_{\langle A\rangle} a &=0\,,\\
\bar \p_{\langle A\rangle} \varphi &=  i [a, \langle \Phi\rangle ]- \eps \p \theta_0 \wedge (\p_{\langle A\rangle} a +\bar \p_{\langle A\rangle} a^\dagger)\,.
\end{split}
\ee
which have to be solved together with the D-term equation (\ref{treeD}) that remains unchanged \cite{fimr12}. As in the tree-level case, we quote the relevant results for each sector and relegate the computations to appendix \ref{ap:wave}. We start with the Higgs sectors that are not charged under the T-brane and then consider the more involved case of the matter sectors.

\subsubsection*{Higgs sectors}

The solution to the non-perturbative zero mode equations for the sector $\mathbf 5_U$ is
\be
\vec\vphi_{U}\, =\, \gamma_{U}\left ( \begin{array}{c}
i\frac{\zeta_{U}}{2\mu^2}\\
i\frac{(\zeta_{U}-\lam_{U})}{2\mu^2}\\
1\end{array}\right )\, \chi_{U}^{\rm np}, \
\qquad
\chi_{U}^{\rm np} = 
e^{\frac{q_R}{2}(|x|^2-|y|^2)-q_S (x \bar y +y\bar x)+(x-y)(\zeta_{U}\bar x-(\lam_{U}-\zeta_{U})\bar y))}(1+\eps\Upsilon_{U}).
\ee
The $\mathcal O(\eps)$ non-perturbative correction is
\be\label{corU}
\Upsilon_{U}=-\frac{1}{4\mu^2}(\zeta_{U} \bar{x} - (\lam_{U} - \zeta_{U})\bar{y})^2(\th_x+\th_y)+\frac{\delta_1}{2}(x-y)^2+\frac{\delta_2}{\zeta_{U}}(x-y)(\zeta_{U} y+(\lam_{U}-\zeta_{U})x)
\ee
with the constants $\delta_1$, $\delta_2$ given by (\ref{d1}) and (\ref{d2}) respectively. 
The solution to the $\mathbf 5_D$ is essentially the same and can be obtained by performing the replacements explained in appendix \ref{ap:wave} so we do not write it explicitly. 

Now one can see that this particular correction to the wavefunction will not generate a correction to the normalisation factor at order $\eps$.
The reason is that the extra terms that appear in the integrand of (\ref{4dnorm}) will be those in (\ref{corU}) and its complex conjugate which are not invariant under the rotation $(x,y) \raw e^{i\a}(x,y)$.

\subsubsection*{Matter sector}

As shown in the appendix, the structure of the solution for the $\mathbf{10}_M$ sector is 
\be
\label{phys10np} 
\vec{\vphi}_{{10}^+} \, =\, 
\left(\begin{array}{c} \bullet \\ \bullet \\ 0\end{array}\right)
 + \eps
\left(\begin{array}{c} 0 \\ 0 \\ \bullet\end{array}\right) +\CO(\eps^2)
 \qquad 
\vec{\vphi}_{{10}^-} \, =\,
\left(\begin{array}{c} 0 \\ 0 \\ \bullet\end{array}\right)
 + \eps
 \left(\begin{array}{c} \bullet \\ \bullet \\ 0\end{array}\right)+\CO(\eps^2).
\ee
and similarly for the $\mathbf{\bar 5}_M$. This structure already shows that the $\CO(\eps)$ corrections to the kinetic terms of the matter sectors vanish, even without specifying their explicit form. Indeed, from (\ref{4dnorm}) such corrections will be proportional to the scalar products $\vec{\vphi}_{{10}^+}^{(0)}\cdot\vec{\vphi}_{{10}^-}^{(1)}$ and $\vec{\vphi}_{{10}^-}^{(0)}\cdot\vec{\vphi}_{{10}^+}^{(1)}$, where the superscript $(0)$ denotes the tree-level term and $(1)$ the $\CO(\eps)$ correction. Given the solution (\ref{phys10np}), we see that those products are trivially zero.

\section{Physical Yukawas and hierarchies}\label{s:physical}

Combining the results of the last two sections one finds the following physical Yukawas for quarks and charged leptons in our local $E_8$ model
\begin{subequations}\label{matrices}
\begin{align}\label{matrixU}
Y_U= &\,\frac{\pi^2\,\g_U\,\g^Q_{10,3}\g^U_{10,3}}{2\rho_m\rho_\mu}\left (\begin{array}{ccc}
0&0&\tilde \eps\frac{\g^Q_{10,1}}{2\rho_\mu\g^Q_{10,3}}\\
0&\tilde \eps\frac{\g^Q_{10,2}\g^U_{10,2}}{2\rho_\mu\g^Q_{10,3}\g^U_{10,3}}&0\\
\tilde \eps\frac{\g^U_{10,1}}{2\rho_\mu\g^U_{10,3}}&0&1\end{array}\right )  + \CO(\tilde \eps^2)\\
\label{matrixD}
Y_{D}=&\,-\frac{\pi^2\,\g_D\,\g^Q_{10,3}\,\g^D_{5,3}}{2d\,\rho_m\rho_\mu}\left (\begin{array}{ccc}
0&\tilde\eps\tilde\kappa\frac{\g^Q_{10,1}\g^D_{5,2}}{d\rho_\mu^2\g^Q_{10,3}\g^D_{5,3}}&\left (\frac{2\tilde \kappa^2}{\rho_\mu}-\frac{\tilde \eps}{ d}\right )\frac{\g^Q_{10,1}}{2\rho_\mu\g^Q_{10,3}}\\
\tilde\eps\tilde\kappa\frac{\g^Q_{10,2}\g^D_{5,1}}{2d\rho_\mu^2\g^Q_{10,3}\g^D_{5,3}} &-\tilde\eps \frac{\g^Q_{10,2}\g^D_{5,2}}{2d\rho_\mu\g^Q_{10,3}\g^D_{5,3}}&-\tilde \kappa\frac{\g^Q_{10,2}}{\rho_\mu\g^Q_{10,3}}\\
-\tilde\eps \frac{\g^D_{5,1}}{2d\rho_\mu\g^D_{5,3}}&0&1\end{array}\right ) + \CO(\tilde \eps^2)\\
Y_{L}=&\,-\frac{\pi^2\,\g_D\,\g^E_{10,3}\,\g^L_{5,3}}{2d\,\rho_m\rho_\mu}\left (\begin{array}{ccc}
0&\tilde\eps\tilde\kappa\frac{\g^E_{10,1}\g^L_{5,2}}{d\rho_\mu^2\g^E_{10,3}\g^L_{5,3}}&\left (\frac{2\tilde \kappa^2}{\rho_\mu}-\frac{\tilde \eps}{ d}\right )\frac{\g^E_{10,1}}{2\rho_\mu\g^E_{10,3}}\\
\tilde\eps\tilde\kappa\frac{\g^E_{10,2}\g^L_{5,1}}{2d\rho_\mu^2\g^E_{10,3}\g^L_{5,3}} &-\tilde\eps \frac{\g^E_{10,2}\g^L_{5,2}}{2d\rho_\mu\g^E_{10,3}\g^L_{5,3}}&-\tilde \kappa\frac{\g^E_{10,2}}{\rho_\mu\g^E_{10,3}}\\
-\tilde\eps \frac{\g^L_{5,1}}{2d\rho_\mu\g^L_{5,3}}&0&1\end{array}\right ) + \CO(\tilde \eps^2)
\end{align}
\end{subequations}
with the dimensionless complex parameters defined by
\be
\tilde \eps=\eps(\th_x+b\th_y),\qquad \tilde\kappa=\frac{\kappa}{m_*},\qquad \rho_m=\frac{m^2}{m_*^2},\qquad \rho_\mu=\frac{\mu^2}{m_*^2}
\ee
and the normalisation factors $\g_{10,j}^\a$, $\g_{5,j}^\a$  given by (\ref{gammas}). The superscript $\a$ denotes the particular MSSM chiral multiplet within ${\bf 10}_M$ or ${\bf 5}_M$, namely the first column in table \ref{t:sectors}.

We would like to see if this structure can reproduce empirical data for charged fermion masses. Since our results apply at the GUT scale, presumably at around $10^{16}$ GeV, experimental values at weak scale need to be run using the renormalisation group equations. Table \ref{tab:massas} shows the extrapolation to the unification scale of such observed quantities taken from \cite{Ross:2007az} in the context of the MSSM. These depend on the parameter $\tan\beta$ that controls the relative magnitude of the vevs of $H_U$ and $H_D$. In particular, we have that $m_{\tau,b}=Y_{\tau,b}V\cos\beta$ and $m_t=Y_t V\sin\beta$ with $V=\sqrt{V_u^2+V_d^2}\approx 174$ GeV. In the following we will discuss the comparison between the experimental results with the predictions from our local $E_8$ model.
\begin{table}[htb] 
%\footnotesize
\renewcommand{\arraystretch}{1.25}
\begin{center}
\begin{tabular}{|c||c|c|c|}
\hline
tan$\beta$  &  10&   38  &  50 \\
\hline\hline
$m_u/m_c$ &   $2.7\pm 0.6\times 10^{-3}$   &  $2.7\pm 0.6\times 10^{-3}$&$2.7\pm 0.6\times 10^{-3}$  \\
\hline
$m_c/m_t$ &   $2.5\pm 0.2\times 10^{-3}$ &$2.4\pm 0.2\times 10^{-3}$&$2.3\pm 0.2\times 10^{-3}$ \\
\hline\hline
$m_d/m_s$ &   $5.1\pm 0.7\times 10^{-2}$   &  $5.1\pm 0.7\times 10^{-2}$  & $5.1\pm 0.7\times 10^{-2}$  \\
\hline
$m_s/m_b$ &    $1.9\pm 0.2\times 10^{-2}$   &  $1.7\pm 0.2\times 10^{-2}$  & $1.6\pm 0.2\times 10^{-2}$  \\
\hline\hline
$m_e/m_\mu$  &   $4.8\pm 0.2\times 10^{-3}$   &  $4.8\pm 0.2\times 10^{-3}$  & $4.8\pm 0.2\times 10^{-3}$   \\
\hline
$m_\mu/m_\tau$  &    $5.9\pm 0.2\times 10^{-2}$   &  $5.4\pm 0.2\times 10^{-2}$  & $5.0\pm 0.2\times 10^{-2}$  \\
\hline\hline
%$m_b/m_\tau$  &    $0.73\pm0.03 $   &  $0.73\pm0.03 $ &    $0.73\pm0.04 $  \\
%\hline\hline
$Y_\tau $  &    $0.070\pm0.003 $   &  $0.32\pm0.02 $ &    $0.51\pm0.04 $ \\
\hline
$Y_b $  &    $0.051\pm0.002 $   &  $0.23\pm0.01 $ &    $0.37\pm0.02 $ \\
\hline
$Y_t $  &    $0.48\pm0.02 $   &  $0.49\pm0.02 $ &    $0.51\pm0.04 $ \\
\hline
\end{tabular}
\end{center}
\caption{\small Running mass ratios of leptons and quarks at the unification scale from ref.\cite{Ross:2007az}.}
\label{tab:massas}
\end{table}

\subsection{Fermion masses}

The masses for quarks and charged leptons will directly depend on the eigenvalues of the physical Yukawa matrices. From  (\ref{matrices}) we see that such eigenvalues are
\be\label{Yeigen}
\begin{array}{ccc}
Y_t=\frac{\pi^2\,\g_U\,\g^Q_{10,3}\g^U_{10,3}}{2\rho_m\rho_\mu}, &  \qquad Y_c=\tilde\eps \,\frac{\pi^2\,\g_U\,\g^Q_{10,2}\g^U_{10,2}}{4\rho_m\rho^2_\mu}, &\qquad Y_u=\mathcal O(\tilde\eps^2)\\
Y_b=\frac{\pi^2\,\g_D\,\g^Q_{10,3}\g^D_{5,3}}{2d\rho_m\rho_\mu}, &\qquad  Y_s=\tilde\eps \,\frac{\pi^2\,\g_D\,\g^Q_{10,2}\g^D_{5,2}}{4d^2\rho_m\rho^2_\mu}, & \qquad Y_d=\mathcal O(\tilde\eps^2) \\
Y_\tau=\frac{\pi^2\,\g_D\,\g^E_{10,3}\g^L_{5,3}}{2d\rho_m\rho_\mu}, &\qquad  Y_\mu=\tilde\eps\, \frac{\pi^2\,\g_D\,\g^E_{10,2}\g^L_{5,2}}{4d^2\rho_m\rho^2_\mu}, &\qquad Y_e=\mathcal O(\tilde\eps^2),
\end{array}
\ee
which makes manifest the mass hierarchy $(\CO(1), \CO(\tilde \eps), \CO(\tilde \eps^2))$ between families. However, it still remains to see if the data of table \ref{tab:massas} can be reproduced via these expressions, and if affirmative for which range of values for $\tan{\beta}$. This question is non-trivial in the sense that the normalisation factors $\g_{10}$, $\g_5$ are complicated functions of the multiple flux densities present in the model, which makes it hard to proceed analytically.\footnote{Notice that when embedded into a global model, this large number of flux densities should depend on a few K\"ahler moduli of the compactification, and in this sense many of the free parameters that are present in the local approach become constrained.}

\subsubsection*{Mass ratios}

\begin{figure}[ht]
\center\includegraphics[width=8.5cm]{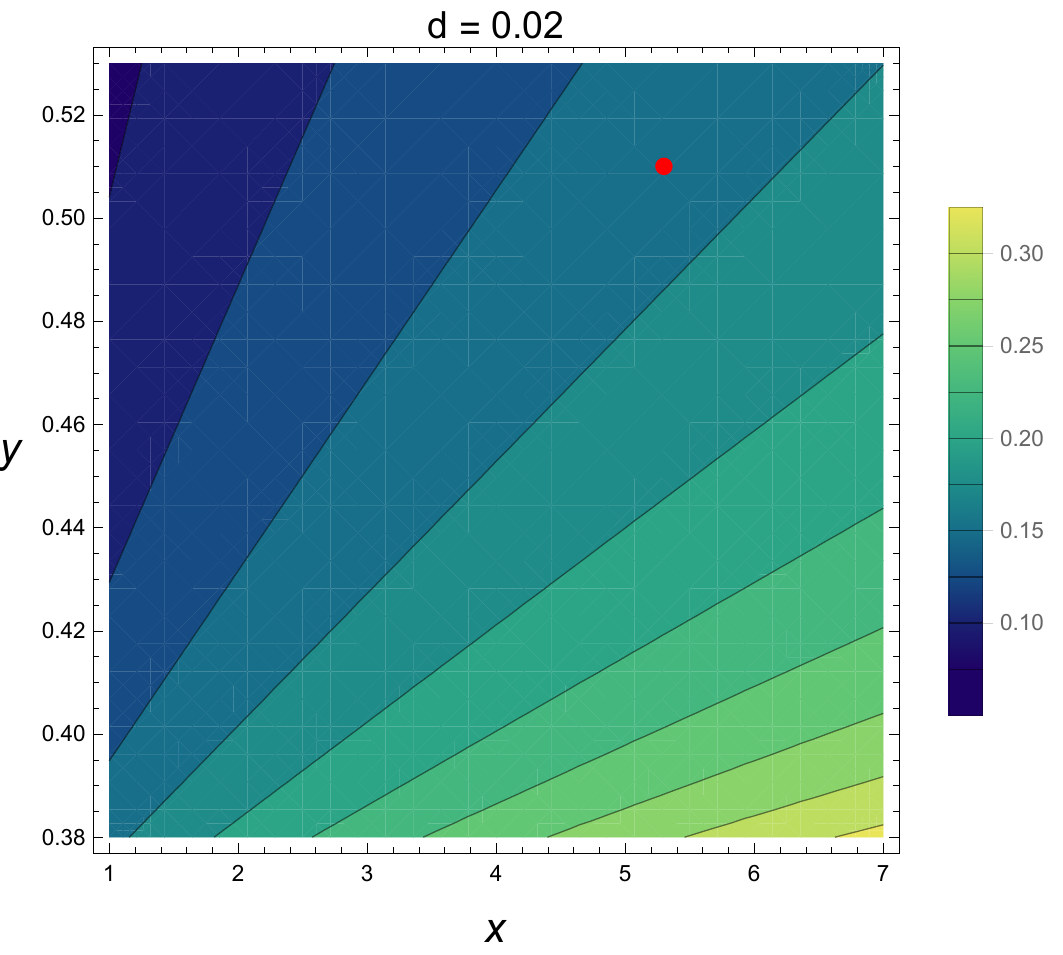}\includegraphics[width=8.5cm]{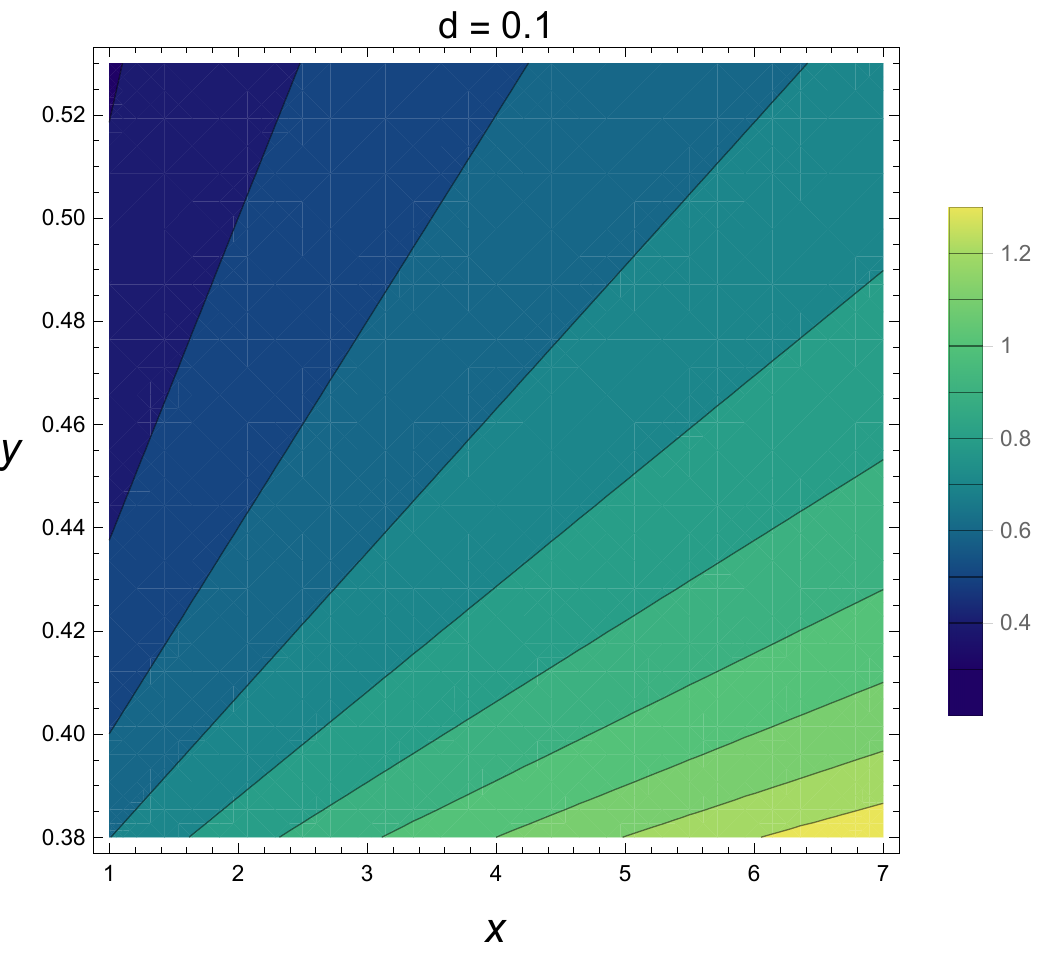}
 \caption{\small{Ratio of ratios (\ref{ratiosa}) where the horizontal axis is $x =-M_1/\tilde N_Y$ and the vertical is $y=(M_1+2M_2)/\tilde N_Y$. The right figure displays the value for $|d| = 0.1$ and the left one for $|d| = 0.02$. In the latter, the red dot corresponds to the values chosen in (\ref{values2}).}}
 \label{fig:ratiosa}
\end{figure}

\begin{figure}[ht]
\center\includegraphics[width=8.5cm]{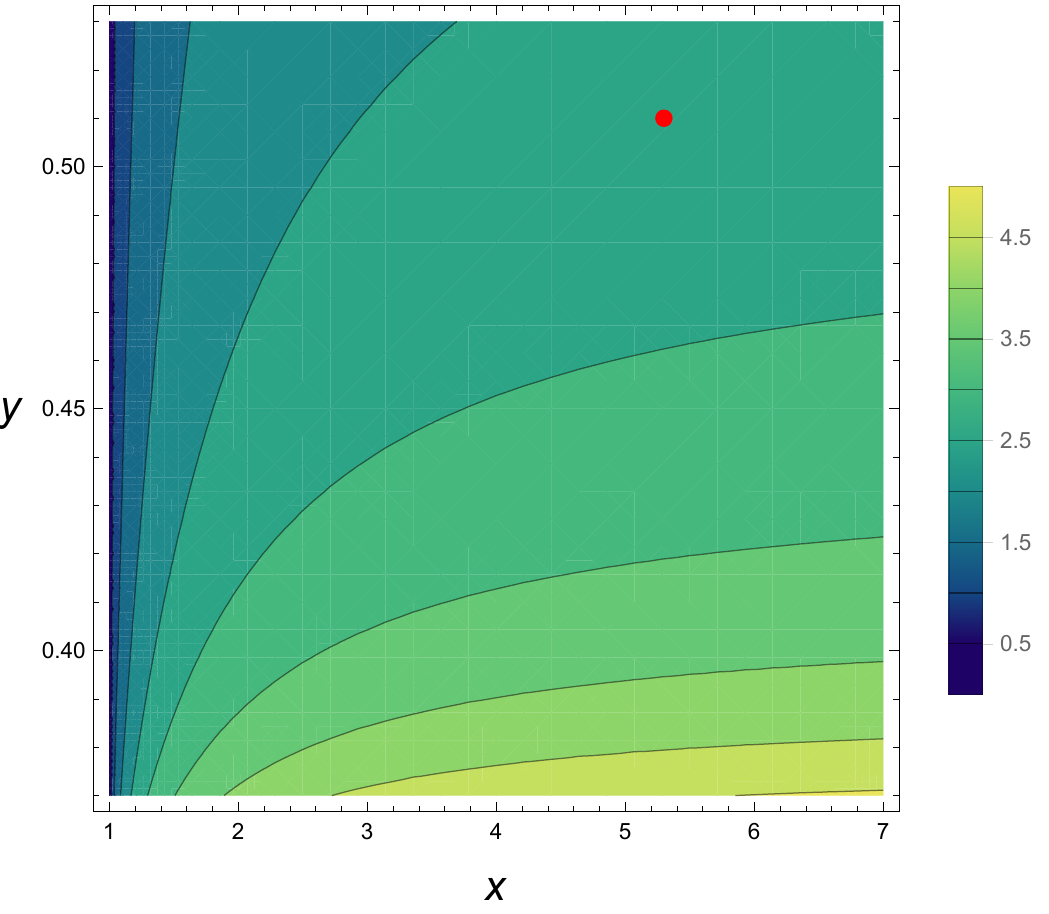}
 \caption{\small{Ratio (\ref{ratiosb}) with $x =-M_1/\tilde N_Y$, $y=(M_1+2M_2)/\tilde N_Y$. The red dot represents (\ref{values2}).}}
 \label{fig:ratiosb}
\end{figure}

\begin{figure}[h!]
\center\includegraphics[width=10cm]{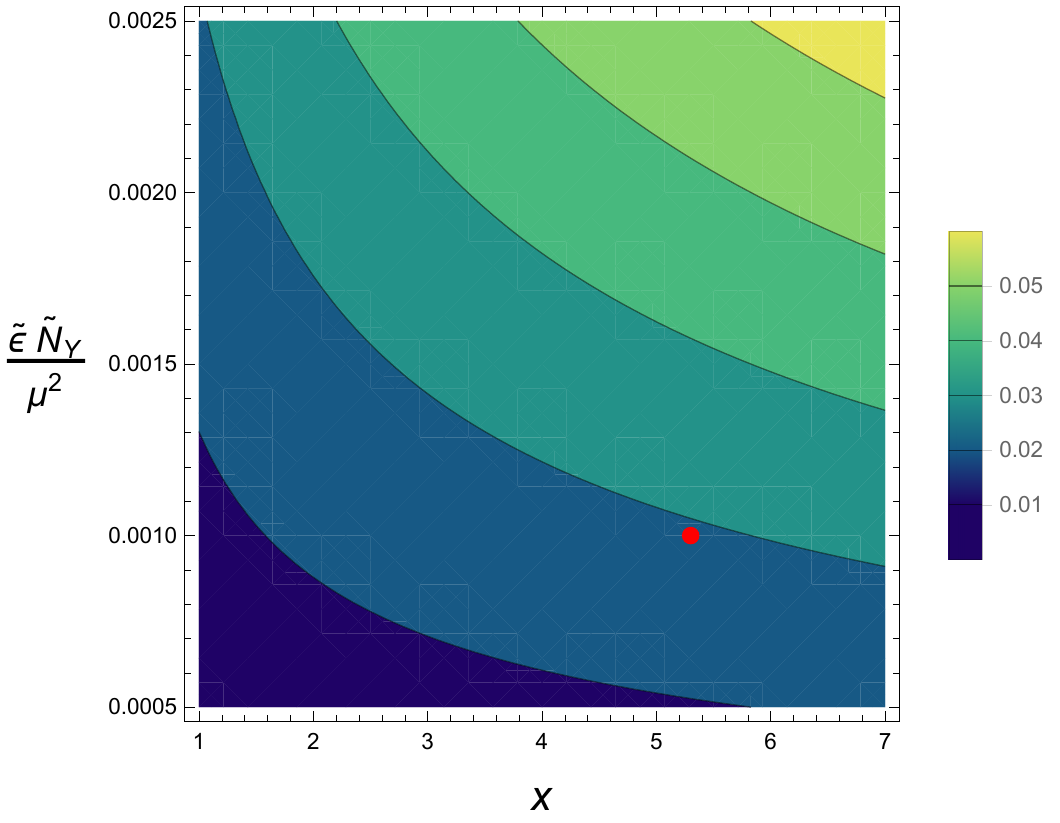}
 \caption{\small{Ratio $m_s/m_b$ that fixes the order of magnitude of  $|\tilde\eps \tilde N_Y/2\rho_\mu|$. %The red dot corresponds to the values chosen in (\ref{values2}).
 }}
 \label{fig:ratioseps}
\end{figure}

While the Yukawa eigenvalues are complicated functions of the local flux densities, fermion mass ratios have a much simpler dependence on them, as already noticed in \cite{fimr12}. In particular, let us consider those mass ratios between the second and third generation that are independent of tan $\beta$. These are
\begin{subequations}\label{ratioss}
\begin{align}
\frac{m_c}{m_t}\,=&\,\,\left| \frac{\tilde\eps}{2\rho_\mu}\right|  \sqrt{ q_{R}^Q\,q_{R}^{U} }\,=\,  \left|\frac{\tilde\eps \,\tilde N_Y}{2 \rho_\mu}\right| \sqrt{\left (x-\frac{1}{6}\right )\left (x + \frac{2}{3}\right )} \\
\frac{m_s}{m_b}\,=&\,\, \left| \frac{\tilde\eps }{2d\rho_\mu}\right|   \sqrt{ q_{R}^{Q}\,q_{R}^{D} }\,=\, \left|\frac{\tilde\eps\,\tilde N_Y }{2 d\rho_\mu}\right| \sqrt{\left (x-\frac{1}{6}\right )\left (y -\frac{1}{3}\right )} \\
\frac{m_\mu}{m_\tau}\,=&\,\, \left|\frac{\tilde\eps }{2d\rho_\mu}\right|  \sqrt{ q_{R}^{E}\,q_{R}^{L} }\,=\,\left|\frac{\tilde\eps\,\tilde N_Y }{2 d\rho_\mu}\right| \sqrt{\left (x-1\right )\left (y +\frac{1}{2}\right )}
\end{align}
\end{subequations}
where $q_{R}^Q$, $q_{R}^{U}$ are the linear combinations of flux densities that appear in table \ref{t:sectors}, and we have defined the quotients
\be\label{xy}
x \, =\, - \frac{M_1}{\tilde N_Y} \quad \quad \quad y\, =\, \frac{M_1+2M_2}{\tilde N_Y}
\ee
We see that for these mass quotients the dependence on the normalisation factors for the Higgses drops and we obtain fairly simple formulae in terms of a few flux densities. More precisely, besides $x$ and $y$ these three ratios depend on two more parameters, namely $|d|$ and $|\tilde\eps \tilde N_Y/2\rho_\mu|$. Furthermore, by considering quotients of ratios we can eliminate the dependence on the last parameter, since
\begin{subequations}\label{ratrat}
\begin{align}
\frac{m_c/m_t}{m_s/m_b}=&\,|d| \left ( \frac{x +\frac{2}{3}}{y -\frac13} \right )^{1/2}
\label{ratiosa}\\
\frac{m_\mu/m_\tau}{m_s/m_b}=&\,3 \left (\frac{\left (x-1\right) \left (y+\frac12\right )}{\left (3x-\frac12\right )\left (3y-1\right )}  \right )^{1/2}.
\label{ratiosb}
\end{align}
\end{subequations}
So one may proceed to constrain these three parameters of the local model in terms of two empirical quantities, which from the data of table \ref{tab:massas} read
\begin{subequations}\label{exper}
\begin{align}
\left. \frac{m_c/m_t}{m_s/m_b}\right |_{\rm exp.}=&\ 0.13\pm {0.03}\\
\left. \frac{m_\mu/m_\tau}{m_s/m_b}\right |_{\rm exp.}=&\ 3.3\pm1.
\end{align}
\end{subequations}
Finally, recall that the values of $x$, $y$ are constrained from the results of appendix  \ref{ap:chiral}. For $\tilde N_Y <0$ we have that  $x < -2/3$, $y < -1/2$, while for $\tilde N_Y > 0$ we have that $x > 1$, $y > 1/3$. We find that it is easier to fit the above empirical values for the latter case and by taking a small value for $|d|$, as illustrated in figures \ref{fig:ratiosa} and \ref{fig:ratiosb}. For instance, taking $d\sim 0.02$ we can use (\ref{ratrat}) together with (\ref{exper}) to estimate $x,y$ defined above, namely
\bea
\begin{split}
x=&\, 5\pm 3  \\
y=&\, 0.45\pm0.05.
\end{split}
\eea
Then, using these approximate values for the fluxes we can find the order of magnitude of $|\tilde\eps\tilde N_Y/\rho_\mu|$ by fitting one of the mass ratios in (\ref{ratioss}), which yields (see figure \ref{fig:ratioseps})
\be\label{epseps}
\left |\frac{\tilde\eps\tilde N_Y}{\rho_\mu}\right |=(1.3\pm0.7)\cdot 10^{-3}.
\ee
Notice that, up to now, we did not specify the value of any flux density but only quotients of fluxes. In the following we discuss the absolute value of Yukawa couplings for which we actually need the flux density values.

%%%%%%%%%%%%%%%%%%%%%%%
%%%%%%%%%%%%%%%%%%%%%%%%
%
%%%%%%%%%%%%%%%%%%%%%%%
%%%%%%%%%%%%%%%%%%%%%%%%
%
%%%%%%%%%%%%%%%%%%%%%%%
%%%%%%%%%%%%%%%%%%%%%%%%

\subsubsection*{Yukawas for the third generation}

So far we have discussed ratios of Yukawa couplings for which we have simple expressions. However, for the Yukawa couplings themselves, we do not have such simple results and it is much harder to understand how these depend on the parameters of the model. In particular, these depend on flux densities and not just on quotients so let us estimate their order of magnitude.

The fluxes that are not along the hypercharge generator, i.e. $M_1,M_2, N_1,N_2$, determine the number of chiral multiplets for the matter sectors. On the other hand, the hypercharge fluxes $N_Y$ and $\tilde N_Y$ are also subject to quantisation conditions since they are responsible for the doublet-triplet splitting. Thus, all of the flux densities satisfy that $|\int_{\Sigma} F|\simeq 2\pi$, where $\Sigma$ is a matter curve and if we take $F$ to be approximately constant we find that $ |F|\simeq 2\pi/V_{\Sigma}$, with $V_{\Sigma}$ the volume of $\Sigma$. Furthermore, we know that the volume of the GUT divisor, $V_{GUT}$, is related to the coupling constant $\a_{GUT}$ as (see e.g. \cite{thebook})
\be
\a_{GUT}\simeq \frac{2\pi^2g_s}{m_{st}^4V_{GUT}},
\ee
where the string scale $m_{st}$ is related to $m_*$ by $m^4_{st}=(2\pi)^3g_sm_*^4$. From unification of the gauge couplings, we expect $\a_{GUT}\simeq 1/24$, so assuming $V_{\Sigma}\simeq V_{GUT}^{1/2}$, we find that 
\be\label{esti}
\frac{|F|}{m_{st}^2}\simeq \left (\frac{2\a_{GUT}}{g_s}\right )^{1/2}\simeq \frac{0.3}{g_s^{1/2}},
\ee
which tells us the order of magnitude for the the flux densities. Notice that our approach relies on having diluted fluxes (as well as intersection slopes) so the coupling constant should not be arbitrarily small. 

Now one can perform a scan for flux densities in the ballpark of (\ref{esti}) and see whether there is a region that allows to fit the Yukawa couplings for the third generation of U and D-type quarks as well as charged leptons. We find that this is indeed possible and, for instance, taking the following flux densities (in units of $m^2_{st}$)
\be\label{values2}
(M_1,M_2,N_1,N_2,\tilde N_Y, N_Y) = (-0.16,0.09,-0.501,0.501,0.03,-1.0)\,,
\ee
together with intersection angles
\be
(\rho_m,\rho_\mu,c,d) = (0.09,0.004,0.53,0.025)\,,
\ee
%\be
%(m,\mu,c,d) = (0.28,0.09,0.014,0.02)\,,
%\ee
we find
\be
Y_t= 0.46  \qquad Y_b= 0.08 \qquad Y_\tau= 0.15,
\ee
consistent with the experimental values in table \ref{tab:massas} for $\tan\beta\simeq 10-20$. Also, using these parameters we get from (\ref{epseps}) that
\be
\tilde \eps=1.3\cdot 10^{-4}
\ee
which is small enough to be consistent with its non-perturbative nature.

%%%%%%%%%%%%%%%%%%%%%%%
\begin{figure}[htb]
\center\includegraphics[width=9.5cm]{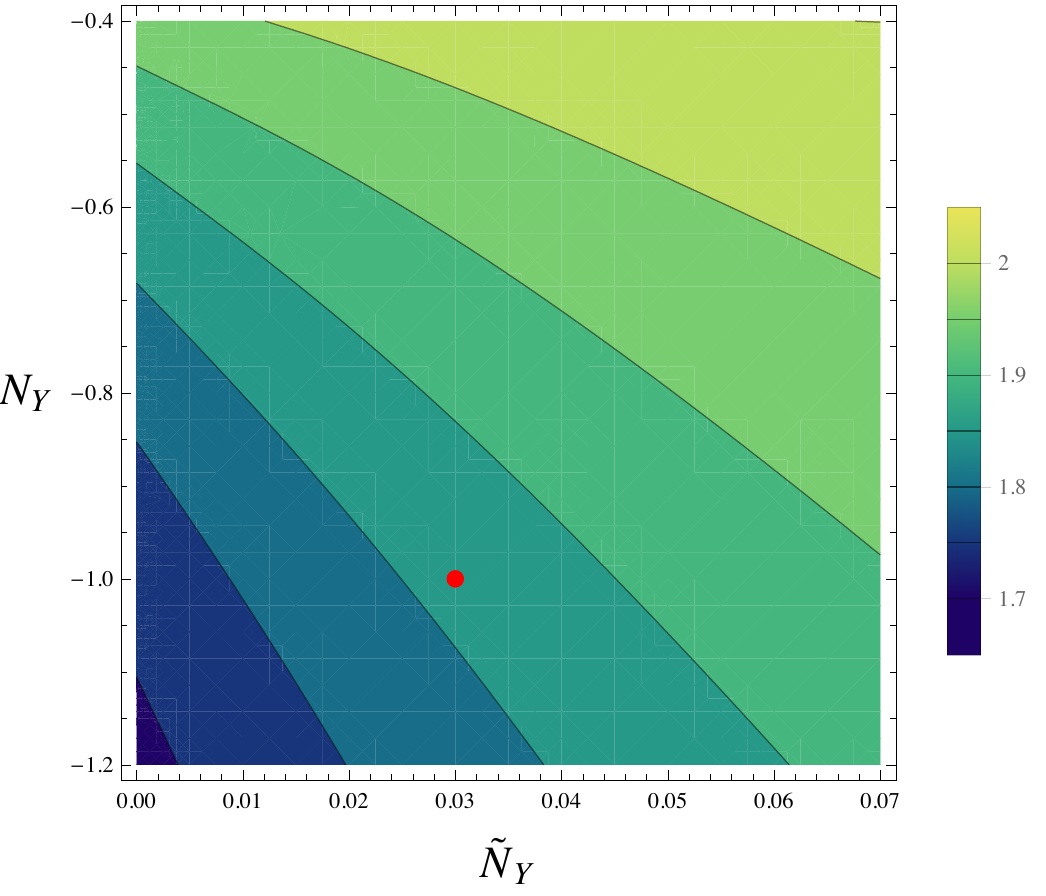}
 \caption{\small{Ratio of $\tau$ and $b$ mass. The red dot corresponds to the values chosen in (\ref{values2}).}}
 \label{fig:ratios2}
\end{figure}
%%%%%%%%%%%%%%%%%%%%%%%

As we can see, our model allows to accommodate a large top Yukawa coupling, which is usually troublesome in perturbative type II GUTs.\footnote{Notice that the expression for the top Yukawa in our model is exactly like in the $E_6$ model of \cite{fmrz13}.} Also, the hypercharge flux may induce the correct difference between $Y_b$ and $Y_\tau$. Recall that these come from the same Yukawa in the $SU(5)$ GUT but, due to the hypercharge breaking, they are different even at $M_{GUT}$. From (\ref{Yeigen}) we have that
\be
\frac{Y_\tau}{Y_b}=\frac{\g^E_{10,3}\g^L_{5,3}}{\g^Q_{10,3}\g^D_{5,3}},
\ee
which, for vanishing hypercharge fluxes, is exactly one. However, for our particular choice of parameters we find that
\be
\frac{Y_\tau}{Y_b}=1.81,
\ee
consistent with the observed ratio.

%\subsubsection*{Fitting fermion masses}

%Choosing the following values for the flux densities and intersection parameters (in units of $m_*$)
%
%\be\label{values2}
%\left(m,\,\mu,\,\frac{M_1}{\tilde N_Y},\,\frac{\tilde M}{\tilde N_Y},\, \frac{N_1}{N_Y},\,\frac{N_2}{N_Y},\,N_Y,\,\tilde N_Y,\, d,\,c\right) =(1.05,\,0.33,\, -5,\, 0.48,\,1.4 ,\-3 ,\,-2.44,\,0.19,\, 0.024 ,0.014) \,.
%\ee
%
%together with the non-perturbative parameter $\tilde \eps = 1.5\cdot 10^{-4}$, we find
%
%\be
%\left(\frac{m_\mu/m_\tau}{m_s/m_b},\frac{m_c}{m_t},\frac{Y_\tau}{Y_b},\frac{Y_s}{Y_b},Y_b,Y_t\right) =(2.35, 2.4 \times 10^{-3},1.38,1.6 \times 10^{-2},0.13,0.5)
%\ee
%
%in good agreement with the experimental data (see table {\ref{tab:massas}}) for low $\tan\b$.

\subsection{Quark mixing angles}

Let us now analyse the quark mixing angles for this model. Recall that the CKM matrix is defined in terms of the unitary matrices $V_U$ and $V_D$ such that they diagonalise the Hermitian product of quark Yukawa matrices $YY^\dagger$. More precisely we have that
\begin{subequations}\label{uves}
\begin{align}
M_U=&\,V_U Y_UY_U^\dagger V_U^\dagger\\
M_D=&\,V_D Y_DY_D^\dagger V_D^\dagger
\end{align}
\end{subequations}
with $M_U$ and $M_D$ diagonal. We then define the CKM matrix as
\be
V_{CKM}=V_UV_D^\dagger.
\ee

Directly applying these definitions to (\ref{matrixU}) and (\ref{matrixD}) and not taking into account their $\CO(\eps^2)$ corrections, one finds that the following matrices satisfy (\ref{uves})
\begin{subequations}
\begin{align}
\hat V_U=&\,\left (\begin{array}{ccc}
1&0&-\frac{\tilde\eps\g^Q_{10,1}}{2\rho_\mu\g^Q_{10,3}}\\
0&1&0\\
\frac{\tilde\eps^*\g^Q_{10,1}}{2\rho^*_\mu\g^Q_{10,3}}&0&1 \end{array}\right )\\
\hat V_D=&\,\left( \begin{array}{ccc}
1&  \frac{i\tilde \eps{\rm Im}(\tilde \kappa\rho_\mu)\g^Q_{10,1}\g^Q_{10,2}}{d\rho_\mu |\rho_\mu|^2(\g^Q_{10,3})^2}  &\left (\frac{\tilde\eps}{ d}-\frac{2\tilde\kappa^2}{\rho_\mu}\right )\frac{\g^Q_{10,1}}{2\rho_\mu\g^Q_{10,3}}\\
-\tilde\eps^*\tilde\kappa^*\frac{\g^Q_{10,1}\g^L_{10,2}}{2d^*\rho_\mu^{*2}(\g^Q_{10,3})^2}&1-\frac{|\tilde\kappa|^2(\g^Q_{10,2})^2}{2|\rho_\mu|^2(\g^Q_{10,3})^2}&\frac{\tilde\kappa\g^Q_{10,2}}{\rho_\mu\g^Q_{10,3}}\\
-\left (\frac{\tilde\eps^*}{ d^*}-\frac{2\tilde\kappa^{*2}}{\rho^*_\mu}\right )\frac{\g^Q_{10,1}}{2\rho^*_\mu\g^L_{10,3}}&-\frac{\tilde\kappa^*\g^Q_{10,2}}{\rho^*_\mu\g^Q_{10,3}}&1-\frac{|\tilde\kappa|^2(\g^Q_{10,2})^2}{2|\rho_\mu|^2(\g^Q_{10,3})^2}\end{array}\right )
\end{align}
\end{subequations}
Taking into account $\CO(\eps^2)$ corrections would modify these expressions, in particular those related to the rotation angles for the first family. In particular, one expects that the final rotation matrices are to a good approximation of the form
\be
V_U=R_U \hat V_U, \qquad\qquad  V_D=R_D \hat V_D
\ee
with
\be
R_{U,D}\simeq \left (\begin{array}{ccc}
1&\a_{U,D} \, \tilde\eps^2&0\\
-\a_{U,D}\, \tilde \eps^2&1&0\\
0&0&1\end{array}\right )
\ee
and where $\a_U$, $\a_D$ are $\CO(1)$ unknown rotation angles. These extra rotations will modify the value of several CKM matrices elements, but leave untouched the mixing between the top and bottom quarks. To the degree of approximation that we are working, we find that such entry reads 
\be
|V_{tb}| \simeq 1-\frac{|\tilde\kappa|^2(\g^Q_{10,2})^2}{2|\rho_\mu|^2(\g^Q_{10,3})^2}=1-\frac{\tilde\kappa^2q_{R,Q}}{2\rho^2_\mu}.
\ee
Since the experimental value for this quantity is
\be
|V_{tb}|_{\rm exp.}=0.9991
\label{expVtb}
\ee
we find a typical value $|\tilde \kappa| \sim 10^{-2}-10^{-3}$. In particular using the values (\ref{values2}) we find
\be
|\tilde\kappa|=2.7\cdot10^{-3}.
\ee
That justifies the approximation $\tilde \kappa \ll m$ made in appendix  \ref{ap:wave}. 
Interestingly, this result has a direct geometrical interpretation. We have that the mixing between the second and third family is roughly given by
\be
\sqrt{1 - |V_{tb}|} \simeq \left| \frac{\tilde\kappa \sqrt{q_{R,Q}}}{\rho_\mu \sqrt{2}}\right|  \propto m_*| b x_0 - y_0|
\ee
where $(x_0,y_0)$ are the coordinates of separation of the Yukawa point $p_{\rm down}$ with respect to $p_{\rm up}$, see (\ref{pdown}). Hence we find that the mixing between the second and third family is proportional to the separation of the two Yukawa points along the particular direction $bx - y$. This combination of $x$ and $y$ is nothing but the complex variable that the holomorphic piece of the matter wavefunctions depend on, c.f. (\ref{haches}). Hence, this mixing effect can be understood as a change of wavefunction basis when moving from one Yukawa point to the other, in agreement with the results of \cite{Randall:2009dw}. Notice that the experimental value (\ref{expVtb}) then translates into a separation between Yukawa points which is roughly $10^{-2}V_{GUT}^{1/4}$, so $p_{\rm up}$ and  $p_{\rm down}$ need to be relatively close to each other, as anticipated in \cite{hv08}.

The other two mixing angles are more difficult to estimate, as this would involve obtaining explicit expressions for $R_U$ and $R_D$. In any case, it is unlikely that a rotation of this order in $\tilde \eps$ will generate the large experimental value for the Cabibbo angle. At this point one should however recall that in our analysis we have made certain approximations related to the fact that we are describing our system in a small neighbourhood of the Yukawa points. These approximations include taking a flat metric (c.f. (\ref{omega})), the limit $\mu \ll m$ that neglects the curvature of the matter curves, taking $\lam_1$ and $\lam_2$ in (\ref{prephi}) as simple functions, and taking the holomorphic piece of the matter wavefunctions as monomials (c.f. (\ref{haches})). These approximations are justified for analysing the physical couplings that involve the two heaviest families, as their real wavefunctions are by construction localised near the Yukawa points. But this need not be so for the first family, which in specific constructions may not even be described as a local chiral mode in the sense of \cite{Palti:2012aa}. Hence, we expect that the mixing angles involving the first family will be particularly sensitive to curvature corrections of the matter curves and $S_{\rm GUT}$. Finally, following the arguments in \cite{hv08}, one may estimate that corrections effects of the order $V_{GUT}^{1/4} m_*$ are precisely those necessary to generate a realistic Cabibbo angle. 

\section{Conclusions}
\label{s:conclu}

In this paper we have considered the structure of both up and down-type Yukawa couplings in local F-theory $SU(5)$ GUTs that are generated at a region of $E_8$ enhancement. In particular, we have analysed several ultra-local models that realise the breaking to $SU(5)$ via reconstructible T-branes and have rank one Yukawa matrices. We have then included the effect of non-perturbative dynamics (Euclidean D3-branes or gaugino condensates on 7-branes) and studied its impact on the flavour structure of our local models. In particular, we have seen that all families of quarks and charged leptons become massive once that these non-perturbative corrections are taken into account. 

The hierarchical structure that results from combining tree-level and non-perturbative effects is already manifest at the level of the holomorphic Yukawas, in terms of the strength of the non-perturbative effect $\eps$ which we use as an expansion parameter in our computations. This allows to classify the whole set of $E_8$ models under study by the eigenvalue hierarchy present in the up and down-type holomorphic Yukawa couplings. We have identified as promising models those whose eigenvalue hierarchy is of the form $(\CO(1), \CO(\eps), \CO(\eps^2))$ for both types of holomorphic Yukawas, as this structure has proven to be successful in reproducing empirical fermion masses for reasonable values of $\eps$ in previous work \cite{fimr12,fmrz13}.

We have carried such classification of models in section \ref{s:models}, dubbing each of the models in terms of the block-diagonal structure of the T-brane profile for the 7-brane field $\Phi$. We have seen that the \textbf{4+1} model has the appropriate hierarchy of Yukawa couplings, but has the less attractive feature of having both Higgses in the same matter curve so one would quite likely find a large $\mu$-term in a global completion. This problem can be easily solved in any of the four \textbf{3+2} models analysed subsequently, as they contain several {\bf 5}-matter curves. However, we find that two of these models have vanishing up-type Yukawas at tree level, in a similar fashion to the vanishing result of \cite{Chiou:2011js}. It would be interesting to acquire a better understanding of these vanishing results. The two remaining \textbf{3+2} models have rank-one tree-level Yukawas but fail to generate an appropriate hierarchy when including the non-perturbative effects. Finally, we have analysed four different $\textbf{2+2+1}$ and found that two of them show the desired pattern of fermion masses, and one in particular exhibits interesting mechanisms to generate realistic neutrino masses and $\mu$-term.

Consequently, we have performed a more detailed analysis this last $\textbf{2+2+1}$ model in order to compute its physical Yukawas. We have specified the whole set of worldvolume fluxes that account for chirality, $SU(5)$ breaking and doublet-triplet splitting. We have also included the non-primitive fluxes that solve for the T-brane background equations of motion. Finally, we have computed the wavefunctions for the matter and Higgs sectors in real gauge at leading order in the non-perturbative parameter and extracted the normalisation factors for the 4d matter fields.

In section \ref{s:physical} we have combined the holomorphic Yukawas with the matter field normalisation factors, obtaining the physical Yukawa couplings. While the former does not depend on the fluxes, the latter does introduce a dependence on the fluxes which has drastic phenomenological implications. Indeed, as we have seen these are crucial to fit the observed fermion masses and in particular to generate a difference between the D-type quarks and leptons mass ratios. Typical values of flux densities allow to obtain a large top quark Yukawa, with $\eps \sim 10^{-4}$ and tan $\beta \sim 10-20$. Finally, we have considered the possibility in which both Yukawas are realised at different points within the region of $E_8$ enhancement. We have seen how this separation translates into the CKM matrix and checked explicitly that the distance between the points generates quark kinetic mixing. Thus, as expected, the fact that the CKM is mostly diagonal in encoded in the fact that the two Yukawa points are very close to each other. 

The above results are very promising for the F-theory GUT programme, and it would be interesting to extend the present analysis in several directions. One obvious direction is to perform our analysis beyond the leading order in perturbation theory in $\eps$.  This would allow to gain some control over the lightest family of chiral matter, and in particular to analyse in more detail the CKM matrix. Nevertheless, as argued in the last section full control over the lightest family cannot be achieved until we extend the description of our model to a region containing the matter curves that host the SU(5) multiplets ${\bf 5}_M$ and ${\bf 10}_M$, and include curvature corrections to our computations of the wavefunction normalisation factors. In general, it would be extremely interesting to promote our ultra-local model to a local one where $S_{\rm GUT}$ is a compact four-cycle. This would allow to understand the plethora of flux densities that enter as free parameters in our model in terms of a few K\"ahler moduli, and see if the flux relations that we have obtained  are compatible with the geometry of $S_{\rm GUT}$. From a broader perspective, it would be important to embed our model within a class of fully-fledged F-theory compactifications, and interpret each of the local parameters in terms of complex and K\"ahler moduli of the compactification. One may then see whether the region of parameter values that are needed to reproduce the flavour structure observed experimentally can indeed be reached when scanning through the F-theory landscape. 

\newpage

\bigskip

\centerline{\bf \large Acknowledgments}

\bigskip

We would like to thank Anamar\'ia Font and Luis E. Ib\'a\~nez for useful discussions. G.Z. is grateful to the Max-Plack-Institut f\"ur Physik, Munich, for kind hospitality during the course of this work.
This work has been partially supported by the grant FPA2012-32828 from the MINECO, the REA grant agreement PCIG10-GA-2011-304023 from the People Programme of FP7 (Marie Curie Action), the ERC Advanced Grant SPLE under contract ERC-2012-ADG-20120216-320421 and the grant SEV-2012-0249 of the ``Centro de Excelencia Severo Ochoa" Programme. F.M. is supported by the Ram\'on y Cajal programme through the grant RYC-2009-05096. G.Z. is supported through a grant from ``Campus Excelencia Internacional UAM+CSIC" and partially by the COST Action MP1210. D.R. is supported by a grant of the Max Planck Society.

%\clearpage

\appendix

%%% Ap. A

\section{$E_8$ machinery}\label{ap:e8}
%\subsection{$E_8$ Lie algebra}

The Lie algebra of $E_8$ consists of 248 generators $Q_\alpha$. We will work in the Cartan-Weyl basis $\{H_i,E_\rho\}$ of $\mathfrak{e}_8$ and where the generators $H_i$
with $i=1,\dots, 8 $ form
a basis of the Cartan subalgebra and the remaining 240 roots are chosen to satisfy the following commutation relations
\be
[H_i , E_\rho] = \rho_i E_\rho\,.
\ee
This allows to represent the roots with a vector of charges under the Cartan subalgebra and for the case of $\mathfrak{e}_8$ the roots are
\be
(\underline{\pm 1 ,\pm 1,0,0,0,0,0,0})\,, \quad \left( \pm\frac{1}{2},\pm\frac{1}{2},\pm\frac{1}{2},\pm\frac{1}{2},\pm\frac{1}{2},\pm\frac{1}{2},\pm\frac{1}{2},\pm\frac{1}{2}\right) \, \text{with even} +\,.
\ee
For our purposes we need to decompose the $E_8$ Lie algebra as $E_8 \rightarrow SU(5)_{GUT} \times SU(5)_\perp$. In particular the branching rule for the adjoint representation of 
$E_8$ is the following
\be
\textbf{248}\rightarrow (\textbf{24},\textbf{1}) \oplus (\textbf{1},\textbf{24}) \oplus ((\textbf{10},\textbf{5}) \oplus c.c.)\oplus ((\overline{\textbf{5}},\textbf{10})\oplus c.c.)\,.
\ee
We identify the roots in the adjoint representation of $SU(5)_{GUT}$ as
\be
(0,0,0,\underline{1,-1,0,0,0})\,,
\ee
which together with the Cartan generators:
\be
\tilde{H}_1= H_4-H_5\,, \quad\tilde{H}_2= H_5-H_6\,, \quad\tilde{H}_3= H_6-H_7\,, \quad\tilde{H}_4= H_7-H_8\,.
\ee
give the adjoint representation of $SU(5)_{GUT}$.
The adjoint of $SU(5)_\perp$ is consists of the following roots
\be\begin{split}
&(\underline{\pm 1,\pm 1,0},0,0,0,0,0)+\left(\frac{1}{2},\frac{1}{2},\frac{1}{2},\frac{1}{2},\frac{1}{2},\frac{1}{2},\frac{1}{2},\frac{1}{2}\right)+
\left(\underline{\frac{1}{2},-\frac{1}{2},-\frac{1}{2}},\frac{1}{2},
\frac{1}{2},\frac{1}{2},\frac{1}{2},\frac{1}{2}\right)\\&+\left(-\frac{1}{2},-\frac{1}{2},-\frac{1}{2},-\frac{1}{2},-\frac{1}{2},-\frac{1}{2},-\frac{1}{2},-\frac{1}{2}\right)+
\left(\underline{\frac{1}{2},\frac{1}{2},-\frac{1}{2}},-\frac{1}{2},
-\frac{1}{2},-\frac{1}{2},-\frac{1}{2},-\frac{1}{2}\right)\,,
\end{split}\ee
and Cartan generators
\be
\hat{H}_1=H_2 - H_3\,, \quad\hat{H}_2=\frac{1}{2}(H_1-H_2+H_3-H_\perp)\,, \quad\hat{H}_3=\frac{1}{2}(H_1-H_2-H_3+H_\perp)\,, \quad \hat{H}_4 = H_2+H_3\,,
\ee
where $H_\perp =\sum_{i=4}^8 H_i $. 
We will label the roots of the adjoint of $SU(5)_\perp$ as follows
\be\begin{split}
E_1^{\pm} &= \pm(0,-1,1,0,0,0,0,0)\,,\\
E_2^{\pm} &= \pm\left(-\frac{1}{2},\frac{1}{2},\frac{1}{2},-\frac{1}
   {2},-\frac{1}{2},-\frac{1}{2},-\frac{1}{2},-\frac{
   1}{2}\right)\,,\\
E_3^{\pm} &= \pm(-1,0,1,0,0,0,0,0)\,,\\
E_4^{\pm} &= \pm(-1,-1,0,0,0,0,0,0)\,,\\
E_5^{\pm} &= \pm\left(-\frac{1}{2},\frac{1}{2},-\frac{1}{2},\frac{1}
   {2},\frac{1}{2},\frac{1}{2},\frac{1}{2},\frac{1}{2
   }\right)\,,\\
E_6^{\pm} &= \pm(-1,1,0,0,0,0,0,0)\,,\\
E_7^{\pm} &= \pm(-1,0,-1,0,0,0,0,0)\,,\\
   E_8^{\pm} &= \pm\left(-\frac{1}{2},-\frac{1}{2},\frac{1}{2},\frac{1}
   {2},\frac{1}{2},\frac{1}{2},\frac{1}{2},\frac{1}{2
   }\right)\,,\\
E_9^{\pm} &= \pm\left(-\frac{1}{2},-\frac{1}{2},-\frac{1}{2},-\frac{1}{2},-\frac{1}{2},-\frac{1}{2},-\frac{1}{2},-\frac{1}{2}\right)\,,\\
E_{10}^{\pm} &= \pm(0,-1,-1,0,0,0,0,0)\,.\\
\end{split}\ee

In the main text we will need two particular linear combinations of these generators
\be
Q_1 = \hat H_1 +2 \hat H_2 +2 \hat H_3 +2 \hat H_4\,, \quad Q_2 = \hat H_3+2 \hat H_4\,.
\ee
The other representations can also be identified. The roots in the representation $(\mathbf{10},\mathbf{5})$ are the following ones 
\begin{subequations}
\begin{align}
\mu_5=& \,\left(\frac{1}{2},\frac{1}{2},-\frac{1}{2},\underline{\frac{1}{2},\frac{1}{2},-\frac{1}{2},-\frac
   {1}{2},-\frac{1}{2}}\right)\,,\\
   \mu_5-\alpha_1=& \,\left(\frac{1}{2},-\frac{1}{2},\frac{1}{2},\underline{\frac{1}{2},\frac{1}{2},-\frac{1}{2},-\frac
   {1}{2},-\frac{1}{2}}\right)\,,\\
   \mu_5-\alpha_1-\alpha_2 =& \, (0,0,0,\underline{1,1,0,0,0})\,,\\
   \mu_5-\alpha_1-\alpha_2- \alpha_3 =& \, \left(-\frac{1}{2},\frac{1}{2},\frac{1}{2},\underline{\frac{1}{2},\frac{1}{2},-\frac{1}{2},-\frac
   {1}{2},-\frac{1}{2}}\right)\,,\\
   \mu_5-\alpha_1-\alpha_2-\alpha_3-\alpha_4 =& \, \left(-\frac{1}{2},-\frac{1}{2},-\frac{1}{2},\underline{\frac{1}{2},\frac{1}{2},-\frac{1}{2},-\frac{1}{2},-\frac{1}{2}}\right)\,,
\end{align}
\end{subequations}
where we identified the five $\mathbf{10}$ representations of $SU(5)_{GUT}$ with their weight under $SU(5)_{\perp}$. We called the highest weight of $SU(5)_\perp$ in the 
fundamental representation $\mu_5$ and the simple roots $\alpha_i$ of $SU(5)_\perp$. 
We can apply the same procedure to the representation $(\mathbf{ \bar 5},\mathbf{10})$ and the result is
\begin{subequations}
\begin{align}
\mu_{10}=& \,(1,0,0,\underline{-1,0,0,0,0})\,,\\
\mu_{10}-\alpha_2=& \, \left(\frac{1}{2},\frac{1}{2},-\frac{1}{2},\underline{-\frac{1}{2},\frac{1}{2},\frac{1}{2},\frac{1}{2},\frac{1}{2}}\right)\,,\\
\mu_{10}-\alpha_1-\alpha_2=& \,\left(\frac{1}{2},-\frac{1}{2},\frac{1}{2},\underline{-\frac{1}{2},\frac{1}{2},\frac{1}{2},\frac{
   1}{2},\frac{1}{2}}\right)\,,\\
 \mu_{10}-\alpha_2-\alpha_3=& \,(0,1,0,\underline{-1,0,0,0,0})\,,\\
 \mu_{10}-\alpha_1-\alpha_2-\alpha_3=& \,(0,0,1,\underline{-1,0,0,0,0})\,,\\
 \mu_{10}-\alpha_2-\alpha_3-\alpha_4=& \,(0,0,-1,\underline{-1,0,0,0,0})\,,\\
 \mu_{10}-\alpha_1-2\alpha_2-\alpha_3=& \,\left(-\frac{1}{2},\frac{1}{2},\frac{1}{2},\underline{-\frac{1}{2},\frac{1}{2},\frac{1}{2},\frac{   1}{2},\frac{1}{2}}\right)\,,\\
 \mu_{10}-\alpha_1-\alpha_2-\alpha_3-\alpha_4=& \,(0,-1,0,\underline{-1,0,0,0,0})\,,\\
  \mu_{10}-\alpha_1-2\alpha_2-\alpha_3-\alpha_4=& \,\left(-\frac{1}{2},-\frac{1}{2},-\frac{1}{2},\underline{-\frac{1}{2},\frac{1}{2},\frac{1}{2},\frac{1}{2},\frac{1}{2}}\right)\,,\\
  \mu_{10}-\alpha_1-2\alpha_2-2\alpha_3-\alpha_4=& \,(-1,0,0,\underline{-1,0,0,0,0})\,,
\end{align}
\end{subequations}
where we called $\mu_{10}$ the highest weight of the antisymmetric representation of $SU(5)_\perp$.

\section{Details on $E_8$ models}
\label{ap:details}

In this appendix we gather some additional details regarding the models analysed in section \ref{s:models}. We start by discussing how invariance under $E_8$ transformations
constrains the possible couplings. We recall that the adjoint representation of $E_8$ has the following branching rule when decomposing $E_8$ under the maximal subgroup
$SU(5)_{GUT} \times SU(5)_\perp$
\be\label{une8}
\textbf{248}\rightarrow (\textbf{24},\textbf{1}) \oplus (\textbf{1},\textbf{24}) \oplus ((\textbf{10},\textbf{5}) \oplus c.c.)\oplus ((\overline{\textbf{5}},\textbf{10})\oplus c.c.)\,.
\ee
It is convenient to introduce a basis of vectors $e_1, \dots , e_5$ for the fundamental representation
 of $SU(5)_\perp$ and this implies that a basis for the $\mathbf{10}$
representation of $SU(5)_\perp$ is given by $e_i \wedge e_j$ with $i \neq j$.\footnote{We will denote the dual basis for the antifundamental representation
of $SU(5)_\perp$ as $e_{\bar \imath}^*$ and similarly a basis for the $\mathbf{\overline{10}}$ representation of $SU(5)_\perp$ is $e^*_{\bar \imath}\wedge e_{\bar \jmath}^*$ with $i\neq j$.} Since the
$\mathbf{10}$ and the $\mathbf{\bar 5}$ representations of $SU(5)_{GUT}$ sit in the $\mathbf{5}$ and $\mathbf{10}$ of $SU(5)_\perp$ respectively we will label them
using the basis vectors we just introduced as $\mathbf{10}_i$ and the $\mathbf{\bar 5}_{ij}$. Then invariance under $E_8$ transformations boils down to invariance
under $SU(5)_{GUT} \times SU(5)_\perp$ transformations and therefore the following couplings are possible
\be
\mathbf{10}_i \cdot \mathbf{10}_j \cdot \mathbf{5}^{ij}\,,
\ee
\be
\mathbf{10}_i \cdot \mathbf{\bar 5}_{jk} \cdot \mathbf{\bar 5}_{lm}\,, \quad \epsilon_{ijklm} \neq 0\,,
\ee
where we have raised indices with the $\delta^{i \bar \jmath}$.
It is also important to analyse the possible couplings between the fields charged under $SU(5)_{GUT}$ and the singlets that come from the adjoint representation of
$SU(5)_\perp$. Using again the basis of vectors $e_i$ of the fundamental representation of $SU(5)_\perp$ the elements in the adjoint representation will be $e_i \otimes
e^{j}$ and labelling the fields in the $(\mathbf{1},\mathbf{24})$ of $SU(5) \times SU(5)_\perp$ as $\mathbf{1}^i_j$ the following renormalisable couplings with the fields charged under $SU(5)_{GUT}$ are possible
\be
\mathbf{10}_i \cdot \mathbf{\overline{10}}^j \cdot \mathbf{1}^i_j\,,
\ee
\be
\mathbf{5}_{ij} \cdot \mathbf{\bar 5}^{jk} \cdot \mathbf{1}^i_k\,.
\ee
Now we will turn to a more detailed description of the Yukawa couplings present in the models analysed in section \ref{s:models}.

\subsection*{4+1 model}

In this model the Higgs field $\Phi$ breaks $SU(5)_\perp$ to $S(U(4) \times U(1))$. This induces the following breaking pattern for the fundamental and antisymmetric representations
\bea
SU(5)_\perp &\longrightarrow& S(U(4) \times U(1))\\\nonumber
\mathbf{5} &\longrightarrow& \mathbf{4}_{\frac{1}{5}}\oplus \mathbf{1}_{-\frac{4}{5}}  \\\nonumber
\mathbf{10} &\longrightarrow& \mathbf{6}_{\frac{2}{5}}\oplus \mathbf{4}_{-\frac{3}{5}} 
\eea
where the subscript denotes the charge under the $U(1)$ that comes from the trace of the $U(4)$ factor. Thus, given the unfolding (\ref{une8}), we have that the relevant sectors are those that appear in table \ref{41}. A particular basis for these sectors is 

\begin{table}[htb]
%\footnotesize
\renewcommand{\arraystretch}{1.1}
\setlength{\tabcolsep}{3pt}
\begin{center}
\begin{tabular}{|c|c|}
\hline
$SU(5)_{GUT}$  & $S(U(4)\times U(1))$ \\
\hline
\hline
$\mathbf{10}_a$&$\mathbf{4}_{\frac{1}{5}}$\\
\hline
$\mathbf{10}_b$&$\mathbf{1}_{-\frac{4}{5}}$\\
\hline
$\mathbf{\bar 5}_a$&$\mathbf{6}_{\frac{2}{5}}$\\
\hline
$\mathbf{\bar 5}_b$&$\mathbf{4}_{-\frac{3}{5}} $\\
\hline
\end{tabular}
\end{center}
\caption{Different sectors for the 4+1 splitting. We do not show the conjugate representations.}
\label{41}
\end{table}

\begin{itemize}
\item[-] $\mathbf{10}$ sector
\be
\mathbf{10}_a : \, \left\{\begin{array}{c}e_1 \\ e_2 \\ e_3\\e_4\end{array}\right\}\,, \qquad \mathbf{10}_b : \, \left\{\begin{array}{c}e_5\end{array}\right\}\,,
\ee
\item[-] $ \mathbf{\bar 5}$ sector
\be
\mathbf{\bar 5}_a : \, \left\{\begin{array}{c}e_1\wedge e_2 \\ e_1 \wedge e_3\\ e_1\wedge e_4\\e_2\wedge e_3\\e_2\wedge e_4\\e_3\wedge e_4\end{array}\right\}\,, \quad \mathbf{\bar 5}_b : \, 
\left\{\begin{array}{c}e_1\wedge e_5\\e_2\wedge e_5\\e_3\wedge e_5\\e_4\wedge e_5\\\end{array}\right\}\,.
\ee
\end{itemize}

It is immediate to see from the charge assignments that in order to have a $SU(5)_\perp$ invariant up-type Yukawa coupling we need to take the $\mathbf{10}_M$ to be the $\mathbf{10}_a$ and the $\mathbf{5}_U$ as the $\mathbf{5}_a$. Furthermore, the $\mathbf{\bar 5}_D$ and $\mathbf{\bar 5}_M$ need to be the $\mathbf{\bar 5}_a$ and $\mathbf{\bar 5}_b$, respectively. This comes from imposing a non-vanishing down-type Yukawa coupling as well as the possibility of having a correct chiral spectrum. Indeed, if we take the $\mathbf{\bar 5}_M$ and $\mathbf{5}_U$ to be in the same curve, we cannot have three massless copies of $\mathbf{\bar 5}_M$ and achieve doublet-triplet splitting. 
Note that using gauge invariance in the $\mathbf{\bar 5}_a$ sector we can set to zero all components of the wavefunction except for two which can be chosen to be the $e_2 \wedge e_3$ and the $e_3 \wedge e_4$. In the 
following we shall the wavefunctions for these sectors $\phi_1$ and $\phi_2$ respectively.
The Yukawa matrices for the model are the following ones
\be\begin{split}
Y_{U} &= -\frac{\pi^2 }{4 \rho_m \rho_\mu} \left(\begin{array}{c c c}0& 0 & \epsilon\phi_1\frac{\theta_x+b \th_y}{2 \rho_\mu}\\0 & \epsilon\phi_1\frac{\theta_x+b \th_y}{2 \rho_\mu}&0\\ \epsilon\phi_1\frac{\theta_x+b \th_y}{2 \rho_\mu}&0&\phi_1+ \eps \, \phi_2 \frac{\th_x+ b \th_y}{2\rho_\mu} \end{array}\right)\,,
\\
Y_{D/L} &= -\frac{\pi^2 \phi_1 }{4 \rho_m \rho_\mu} \left(\begin{array}{c c c}0& 0 & \epsilon\frac{\theta_x+b \theta_y}{2 \rho_\mu}\\0 & \epsilon\frac{\theta_x+b \theta_y}{2 \rho_\mu}&0\\ \epsilon\frac{\theta_x+\theta_y}{2 \rho_\mu}&0&1 \end{array}\right)\,.
\end{split}\ee
\subsection*{3+2 models}

In this class of models the Higgs field $\Phi$ breaks $SU(5)_\perp$ to $S(U(3) \times U(2))$. In this case, the branching of the relevant representations reads

\bea
SU(5)_\perp &\longrightarrow& S(U(3) \times U(2))\\\nonumber
\mathbf{5} &\longrightarrow& (\mathbf{3},\mathbf{1})_{\frac{2}{5}}\oplus(\mathbf{3},\mathbf{1})_{-\frac{3}{5}} \\\nonumber
\mathbf{10} &\longrightarrow& (\mathbf{\bar 3},\mathbf{1})_{\frac{4}{5}}\oplus(\mathbf{3},\mathbf{2})_{\frac{1}{5}}\oplus (\mathbf{1},\mathbf{1})_{-\frac{6}{5}},
\eea
where the subscript denotes the charge under the trace of $U(3)$, which gives the fields in table \ref{32}.

\begin{table}[htb]
%\footnotesize
\renewcommand{\arraystretch}{1.1}
\setlength{\tabcolsep}{3pt}
\begin{center}
\begin{tabular}{|c|c|}
\hline
$SU(5)_{GUT}$  & $S(U(3)\times U(2))$ \\
\hline
\hline
$\mathbf{10}_a$&$(\mathbf{3},\mathbf{1})_{\frac{2}{5}}$\\
\hline
$\mathbf{10}_b$&$(\mathbf{3},\mathbf{1})_{-\frac{3}{5}}$\\
\hline
$\mathbf{\bar 5}_a$&$(\mathbf{\bar 3},\mathbf{1})_{\frac{4}{5}}$\\
\hline
$\mathbf{\bar 5}_b$&$(\mathbf{3},\mathbf{2})_{\frac{1}{5}}$\\
\hline
$\mathbf{\bar 5}_c$&$(\mathbf{1},\mathbf{1})_{-\frac{6}{5}}$\\
\hline
\end{tabular}
\end{center}
\caption{Different sectors for the 3+2 splitting. We do not show the conjugate representations.}
\label{32}
\end{table}

We pick the following basis
\begin{itemize}
\item[-] $\mathbf{10}$ sector
\be
\mathbf{10}_a : \, \left\{\begin{array}{c}e_1 \\ e_2 \\ e_3\end{array}\right\}\,, \qquad \mathbf{10}_b : \, \left\{\begin{array}{c}e_4\\e_5\end{array}\right\}\,,
\ee
\item[-] $ \mathbf{\bar 5}$ sector
\be
\mathbf{\bar 5}_a : \, \left\{\begin{array}{c}e_1\wedge e_2 \\ e_2 \wedge e_3\\ e_1\wedge e_3\end{array}\right\}\,, \quad \mathbf{\bar 5}_b : \, 
\left\{\begin{array}{c}e_1\wedge e_4\\e_2\wedge e_4\\e_3\wedge e_4\\e_1\wedge e_5\\e_2\wedge e_5\\e_3\wedge e_5\end{array}\right\}\,,
\quad \mathbf{\bar 5}_c :\, \left\{e_4 \wedge e_5\right\}\,.
\ee
\end{itemize}
In this class of models there are two possible choices for the $\mathbf{10}_M$, namely either $\mathbf{10}_a$ or $\mathbf{10}_b$. Once the $\mathbf{10}_M$ is chosen the $\mathbf{5}_U$ is uniquely fixed by demanding the possibility of having an up-type Yukawa. Then, for each of them, we have the freedom to choose between the $\mathbf{\bar 5}_M$ and $\mathbf{\bar 5}_D$. The different possibilities are
\begin{itemize}
\item   $\mathbf{10}_M= \mathbf{10}_a,\quad \mathbf{5}_U = \mathbf{5}_a,\quad \mathbf{\bar 5}_M= \mathbf{\bar 5}_b  ,\quad \mathbf{\bar 5}_D=\mathbf{\bar 5}_c $.
\item   $\mathbf{10}_M= \mathbf{10}_a,\quad \mathbf{5}_U = \mathbf{5}_a,\quad \mathbf{\bar 5}_M= \mathbf{\bar 5}_c  ,\quad \mathbf{\bar 5}_D=\mathbf{\bar 5}_b $.
\item   $\mathbf{10}_M= \mathbf{10}_b,\quad \mathbf{5}_U = \mathbf{5}_c,\quad \mathbf{\bar 5}_M= \mathbf{\bar 5}_a  ,\quad \mathbf{\bar 5}_D=\mathbf{\bar 5}_b $.
\item   $\mathbf{10}_M= \mathbf{10}_b,\quad \mathbf{5}_U = \mathbf{5}_c,\quad \mathbf{\bar 5}_M= \mathbf{\bar 5}_b  ,\quad \mathbf{\bar 5}_D=\mathbf{\bar 5}_a $.
\end{itemize}
The Yukawa matrices for both cases in which we assign the $\mathbf{10}_M$ to the $\mathbf{10}_b$ are the following ones
\be\begin{split}
Y_U &=  \frac{\pi^2\,}{2\rho_m\rho_\mu}\left (\begin{array}{ccc}
0&0&\eps\frac{\th_x+b\th_y}{2\rho_\mu}\\
0&\eps\frac{\th_x+b\th_y}{2\rho_\mu}&0\\
\eps\frac{ \th_x+b\th_y}{2\rho_\mu}&0&1\end{array}\right )\,,\\
Y_{D/L}&=\frac{\pi^2}{\rho_m\rho_{\tilde m}}\left(\begin{array}{c c c}0&0&0\\0&0&\epsilon\frac{[(2+b) \theta_x+b\theta_y]\rho_\mu}{\rho_m^{3/2} }
\\0 & 2\epsilon\frac{( \theta_x-b\theta_y)\rho_\mu}{\rho_m^{3/2} }&1+\eps \left(\frac{6b^2\theta_x\rho_\mu^3}{\rho_m^3}-\frac{16b \theta_y \rho_\mu^2}{3\rho_{\tilde m}^2}\right)\end{array}\right)\,.
\end{split}\ee
For the cases in which we assign the $\mathbf{10}_M$ to the $\mathbf{10}_a$ the Yukawa matrix for the up quarks vanishes at tree level like it was already noticed in \cite{Chiou:2011js}, therefore we will not compute the 
Yukawa matrices for these two models.

\subsection*{2+2+1 models}

In this class of models the Higgs field $\Phi$ breaks $SU(5)_\perp$ to $S(U(2) \times U(2) \times U(1))$. Again, the representations of $SU(5)_\perp$ decompose as
\bea
SU(5)_\perp &\longrightarrow& S(U(2) \times U(2)\times U(1))_\perp\\\nonumber
\mathbf{5} &\longrightarrow& (\mathbf{2},\mathbf{1})_{\frac{3}{5},-\frac{2}{5}}\oplus(\mathbf{1},\mathbf{2})_{-\frac{2}{5},\frac{3}{5}}\oplus(\mathbf{1},\mathbf{1})_{-\frac{2}{5},-\frac{2}{5}} \\\nonumber
\mathbf{10} &\longrightarrow& (\mathbf{1},\mathbf{1})_{\frac{6}{5},-\frac{4}{5}}\oplus(\mathbf{1},\mathbf{1})_{-\frac{4}{5},\frac{6}{5}}\oplus(\mathbf{2},\mathbf{1})_{\frac{1}{5},-\frac{4}{5}}\oplus(\mathbf{1},\mathbf{2})_{-\frac{4}{5},\frac{1}{5}}\oplus(\mathbf{2},\mathbf{2})_{\frac{1}{5},\frac{1}{5}}
\eea
Here the subscripts denote the charges under the traces of the two $U(2)$ factors. The different sectors are displayed in table \ref{221}.

\begin{table}[htb]
%\footnotesize
\renewcommand{\arraystretch}{1.1}
\setlength{\tabcolsep}{3pt}
\begin{center}
\begin{tabular}{|c|c|}
\hline
$SU(5)_{GUT}$  & $S(U(2)\times U(2)\times U(1))$ \\
\hline
\hline
$\mathbf{10}_a$&$ (\mathbf{2},\mathbf{1})_{\frac{3}{5},-\frac{2}{5}}$\\
\hline
$\mathbf{10}_b$&$(\mathbf{1},\mathbf{2})_{-\frac{2}{5},\frac{3}{5}}$\\
\hline
$\mathbf{10}_c$&$(\mathbf{1},\mathbf{1})_{-\frac{2}{5},-\frac{2}{5}}$\\
\hline
$\mathbf{\bar 5}_a$&$(\mathbf{1},\mathbf{1})_{\frac{6}{5},-\frac{4}{5}}$\\
\hline
$\mathbf{\bar 5}_b$&$(\mathbf{1},\mathbf{1})_{-\frac{4}{5},\frac{6}{5}}$\\
\hline
$\mathbf{\bar 5}_c$&$(\mathbf{2},\mathbf{1})_{\frac{1}{5},-\frac{4}{5}}$\\
\hline
$\mathbf{\bar 5}_d$&$(\mathbf{1},\mathbf{2})_{-\frac{4}{5},\frac{1}{5}}$\\
\hline
$\mathbf{\bar 5}_e$&$(\mathbf{2},\mathbf{2})_{\frac{1}{5},\frac{1}{5}}$\\
\hline
\end{tabular}
\end{center}
\caption{Different sectors for the 2+2+1 splitting. We do not show the conjugate representations.}
\label{221}
\end{table}

In terms of a particular basis we have the following
\begin{itemize}
\item[-] $\mathbf{10}$ sector
\be
\mathbf{10}_a: \left\{\begin{array}{c} e_1 \\ e_2 \end{array}\right\}\,, \qquad \mathbf{10}_b:\left\{\begin{array}{c} e_3 \\ e_4 \end{array}\right\}\,, \qquad \mathbf{10}_c:\left\{e_5\right\}\,,
\ee
\item[-] $\mathbf{\bar 5}$ sector
\be\begin{split}
&\mathbf{\bar 5}_a: \left\{ e_1 \wedge e_2 \right\}\,, \quad \mathbf{\bar 5}_b:\left\{ e_3 \wedge e_4 \right\}\,, \quad \mathbf{\bar 5}_c:\left\{\begin{array}{c} e_1 \wedge e_5 \\ e_2\wedge e_5 \end{array}\right\}\,, \\&
\mathbf{\bar 5}_d:\left\{\begin{array}{c} e_3 \wedge e_5 \\ e_4\wedge e_5 
\end{array}\right\}\,, \quad \mathbf{\bar 5}_e:\left\{\begin{array}{c} e_1 \wedge e_3 \\ e_2\wedge e_3 \\ e_1\wedge e_4 \\ e_2\wedge e_4\end{array}\right\}\,.  
\end{split}\ee
\end{itemize}

In order to obtain the different models we proceed as before, namely we pick a particular sector to be the $\mathbf{10}_M$ and check whether one can choose a $\mathbf{5}_U$ such that a $SU(5)_\perp$ invariant coupling $\mathbf{10}_M\mathbf{10}_M\mathbf{5}_U$ exists. Then, we check the different assignments of $\mathbf{\bar 5}_M$ and $\mathbf{\bar 5}_D$ that allow for a coupling $\mathbf{10}_M\mathbf{\bar 5}_M\mathbf{\bar 5}_D$. The different possibilities are
\begin{itemize}
\item   $\mathbf{10}_M= \mathbf{10}_a,\quad \mathbf{5}_U = \mathbf{5}_a,\quad \mathbf{\bar 5}_M= \mathbf{\bar 5}_b  ,\quad \mathbf{\bar 5}_D=\mathbf{\bar 5}_c $.
\item   $\mathbf{10}_M= \mathbf{10}_a,\quad \mathbf{5}_U = \mathbf{5}_a,\quad \mathbf{\bar 5}_M= \mathbf{\bar 5}_c  ,\quad \mathbf{\bar 5}_D=\mathbf{\bar 5}_b $.
\item   $\mathbf{10}_M= \mathbf{10}_a,\quad \mathbf{5}_U = \mathbf{5}_a,\quad \mathbf{\bar 5}_M= \mathbf{\bar 5}_d  ,\quad \mathbf{\bar 5}_D=\mathbf{\bar 5}_e $.
\item   $\mathbf{10}_M= \mathbf{10}_a,\quad \mathbf{5}_U = \mathbf{5}_a,\quad \mathbf{\bar 5}_M= \mathbf{\bar 5}_e  ,\quad \mathbf{\bar 5}_D=\mathbf{\bar 5}_d $.
\end{itemize}
One can also consider the possibilities that arise from taking the ones above and permuting the two $U(2)$ factors. However, these models are physically equivalent.
Note that again in the $\mathbf{\bar 5}_e$ sector after using gauge invariance two components cannot be set to zero, and we will choose them to coincide with the roots $e_1 \wedge e_4$ and $e_2 \wedge e_4$ calling
them $\phi_1$ and $\phi_2$ respectively.
For the first model we find the following Yukawa matrices
\be\begin{split}
Y_U &=  \frac{\pi^2}{2\rho_m\rho_\mu}\left (\begin{array}{ccc}
0&0&\eps\frac{\th_x+b\th_y}{2\rho_\mu}\\
0&\eps\frac{\th_x+b\th_y}{2\rho_\mu}&0\\
\eps\frac{\th_x+b\th_y}{2\rho_\mu}&0&1\end{array}\right )\,,\\
Y_{D/L}&=-\frac{\pi ^2 }{\rho _m (2\rho_{\mu_1}+\rho_{\mu_2})}\left(\begin{array}{c c c} 0 & 0 &\eps \,y_{13}\\ 0 & \eps \, y_{22} & \eps  \,y_{23}\\\eps \, y_{31}& \eps\, y_{32}& 1+\eps \,y_{33}\end{array}\right)\,,
\end{split}
\ee
where
\be\begin{split}
y_{13}&= -\frac{\theta _y \left(2 b \rho _{\mu _1}+c \rho _{\mu _2}\right)+\left(2 \rho _{\mu _1}+\rho _{\mu _2}\right) \theta _x}{\left(2 \rho _{\mu _1}+\rho _{\mu
   _2}\right){}^2}\,,\\
y_{22} &=\frac{\theta _y \left(4 \left(b^2+1\right) \rho _{\mu _1}^2+4 (b c+1) \rho _{\mu _2} \rho _{\mu _1}+\left(c^2+1\right) \rho _{\mu _2}^2\right)}{\left(2 \rho
   _{\mu _1}+\rho _{\mu _2}\right){}^2}\,,\\
y_{23} & = -\frac{3 (b-c) \rho _{\mu _1}^2 \rho _{\mu _2} \left(\theta _y \left(2 b \rho _{\mu _1}+c \rho _{\mu _2}\right)+\left(2 \rho _{\mu _1}+\rho _{\mu _2}\right)
   \theta _x\right)}{\left(2 \rho _{\mu _1}+\rho _{\mu _2}\right){}^3 \rho _m^{3/2}}\,,\\
   y_{31}&=-\frac{\th_y \left(2 b \rho _{\mu _1}+c \rho _{\mu _2}\right) \left(4 \left(b^2+2\right) \rho _{\mu _1}^2+4 (b c+2) \rho _{\mu _2} \rho _{\mu
   _1}+\left(c^2+2\right) \rho _{\mu _2}^2\right)}{\left(2 \rho _{\mu _1}+\rho _{\mu _2}\right){}^2}+\\
   &-\frac{\th_x \left(2 b \rho _{\mu _1}+c \rho _{\mu _2}\right){}^2}{2 \rho _{\mu _1}+\rho _{\mu _2}}\,,\\
   y_{32}&=-\frac{\rho _{\mu _1}^2 \theta _x \left(4 \left(b^2+1\right) \rho _{\mu _1}^2+4 (b c+1) \rho _{\mu _2} \rho _{\mu _1}+\left(c^2+1\right) \rho _{\mu
   _2}^2\right)}{\left(2 \rho _{\mu _1}+\rho _{\mu _2}\right){}^2 \rho _m^{3/2}}+\\
   &-\frac{\rho _{\mu _1}^2 \theta _y \left((3 c-2 b) \rho _{\mu _2}+2 b \rho _{\mu _1}\right) \left(4 \left(b^2+1\right) \rho _{\mu _1}^2+4 (b c+1) \rho _{\mu
   _2} \rho _{\mu _1}+\left(c^2+1\right) \rho _{\mu _2}^2\right)}{\left(2 \rho _{\mu _1}+\rho _{\mu _2}\right){}^3 \rho _m^{3/2}}\,,\\
y_{33} &= -\frac{6 (b-c)^2 \rho _{\mu _1}^4 \rho _{\mu _2}^2 \left(\theta _y \left(2 b \rho _{\mu _1}+c \rho _{\mu _2}\right)+\left(2 \rho _{\mu _1}+\rho _{\mu
   _2}\right) \theta _x\right)}{\left(2 \rho _{\mu _1}+\rho _{\mu _2}\right){}^4 \rho _m^3}\,.
\end{split}\ee
The result for the second model is presented in the main text in section \ref{s:E8model}.
For the third model we find the following Yukawa matrices
\be\begin{split}
Y_U &=  \frac{\pi^2}{2\rho_m\rho_\mu}\left (\begin{array}{ccc}
0&0&\eps\frac{\th_x+b\th_y}{2\rho_\mu}\\
0&\eps\frac{\th_x+b\th_y}{2\rho_\mu}&0\\
\eps\frac{\th_x+b\th_y}{2\rho_\mu}&0&1\end{array}\right )\,,\\
Y_{D/L}&=-\frac{\pi ^2 }{\rho _m \rho _{\tilde{m}}}\left(\begin{array}{c c c} 0 & 0 & 0\\ 0 & 0 & \eps  \,y_{23}\\0 & \eps\, y_{32}& \phi_2+\eps \,y_{33}\end{array}\right)\,,
\end{split}
\ee
where
\be\begin{split}
y_{23}&=-\frac{(b-c) \rho _m^{3/2} \left(\phi _1 \theta _x \rho _{\tilde{m}}+\phi _2 \rho _{\mu _2} \left(c \theta _y-\theta _x\right)\right)-b \phi _2 \rho _{\mu }
   \rho _{\tilde{m}}^{3/2} \left(b \theta _y+3 \theta _x\right)}{\rho _m^{3/2} \rho _{\tilde{m}}^{3/2}}\,,\\
y_{32}&= -\frac{(b-c) \rho _m^{3/2} \left(\phi _2 \rho _{\mu _2} \left(3 c \theta _y+\theta _x\right)-c \phi _1 \theta _y \rho _{\tilde{m}}\right)+c \phi _2 \rho _{\mu }
   \rho _{\tilde{m}}^{3/2} \left(b \theta _y-\theta _x\right)}{\rho _m^{3/2} \rho _{\tilde{m}}^{3/2}}\,,\\
   y_{33}&=-\phi_1\frac{(b-c) \rho _m^{3/2} \rho _{\tilde{m}} \left(2 b \rho _{\mu }^2 \theta _x \rho _{\tilde{m}}^{3/2}-\rho _{\mu _2}^2 \rho _m^{3/2} \left(\theta _x-3 c
   \theta _y\right)\right)-c \rho _{\mu } \rho _{\mu _2} \rho _m^{3/2} \rho _{\tilde{m}}^{5/2} \left(\theta _x-b \theta _y\right)}{\rho _m^3 \rho
   _{\tilde{m}}^3}+\\
   &-\phi_2\frac{2 c \rho _{\mu _2} \theta _y \left(b (c-b) \rho _{\mu }^2 \rho _{\tilde{m}}^{3/2}+(b+c) \rho _{\mu _2} \rho _{\mu } \rho _{\tilde{m}}^{3/2}+3 (b-c) \rho
   _{\mu _2}^2 \rho _m^{3/2}\right)}{\rho _m^{3/2} \rho _{\tilde{m}}^3}+\\
   &-\phi_2 \frac{2 \rho _{\mu } \theta _x \left(3 b^2 \rho _{\mu }^2 \rho _{\tilde{m}}^{3/2}+(b-c) (b+c) \rho _{\mu _2} \rho _{\mu } \rho _m^{3/2}-c \rho _{\mu _2}^2 \rho
   _m^{3/2}\right)}{\rho _m^3 \rho _{\tilde{m}}^{3/2}}\,.
\end{split}\ee
Finally in the fourth model we face the problem that the matter curve where the sector $\mathbf{\bar 5}_e$ is localised is actually a singular variety and therefore it is not clear how to correctly identify the various families
for the $\mathbf{\bar 5}_M$ sector. Since this constitutes a major issue in the analysis of the model we will not discuss it further.

\section{Local chirality}\label{ap:chiral}

One important consequence of the addition of fluxes as already remarked in the previous section is the generation of chiral matter in 4d. In fact, while a non-zero vev for the Higgs background $ \Phi $ gives chiral matter localised
in some matter curves $\Sigma$, this chiral matter lives in the six dimensional space $\mathbb{R}^{1,3}\times \Sigma$ and therefore the resulting four dimensional spectrum will be non-chiral after dimensional reduction.
Non zero fluxes threading the matter curves change the situation and if for a particular matter curve $\Sigma_\rho$
\be\label{eq:chir}
\int_{\Sigma_\rho} \text{Tr} \langle F\rangle \neq 0
\ee
one of the two chiralities for the matter living in $\Sigma_\rho$ will remain in the massless spectrum in 4d. In our local analysis we cannot evaluate whether the condition (\ref{eq:chir}) is met for a particular sector of the theory
for this would require knowledge of $\Sigma_\rho$ and the fluxes $F$ globally in $S_{\rm GUT}$. Nevertheless we will be able to analyse the chiral spectrum of our model using the notion of local chirality. Local chirality was discussed
in \cite{Palti:2012aa,fimr12,fmrz13} and it amounts to asking for localisation of the matter wavefunctions in a certain sector $\rho$ around the Yukawa point. To understand concretely whether matter in particular sector is localised or not it is useful to look at 
the same problem in the T-dual setup of magnetised D9-branes. As explained in \cite{afim11} the gauge field in the $\bar z$ direction is identified with $\Phi$ and so $F_{x \bar z} = D_x \Phi$ and $F_{y \bar z} = D_y \Phi$. In the case
of T-branes we additionally have the flux $F_{z \bar z } = i [\Phi , \Phi^\dag]$. In this situation local chirality at the Yukawa point can be studied looking at the Dirac index around this particular point. For a representation
$\mathcal{R}$ the Dirac index is
\be\label{Dind}
{\rm index}_{\mathcal R}  \slashed D=\frac{1}{48(2\pi)^2}\int \left ( \tr_{\mathcal R}\,F\,\w\, F\,\w \,F-\frac{1}{8}\tr_{\mathcal R}\,F\,\w\, \tr R\,\w\, R \right )
\ee
and asking for local chirality at a particular point amounts to asking for a non vanishing integrand at this particular point. Since we take zero curvature in our model the quantity we need to evaluate is the following one
\begin{eqnarray}\label{indR}
\mathcal I_{\mathcal R}\equiv\frac{i}{6}\tr_{\mathcal R}\,\left (F\,\w\, F\,\w \,F\right )_{x\bar xy\bar yz\bar z}=i\,\tr_{\mathcal R}\big ( F_{x\bar x}\{F_{y\bar y},F_{z\bar z}\}+F_{x\bar z}\{F_{y\bar x},F_{z\bar y}\}+ \\\nonumber
F_{x\bar y}\{F_{y\bar z},F_{z\bar x}\}- \{F_{x\bar x},F_{y\bar z}\}F_{z\bar y} - \{F_{x\bar y},F_{y\bar x}\}F_{z\bar z} - \{F_{x\bar z},F_{y\bar y}\}F_{z\bar x}  \big ).
\end{eqnarray}
As shown in \cite{fmrz13}, we find that the wavefunction in a representation $\mathcal R$ is localised when $\mathcal I_{\mathcal R}<0$. This can be evaluated for the the different sectors that appear in the Yukawa couplings which yields
\begin{subequations}
\begin{align}
\mathcal I_{\mathbf{10}_M}=&\, -2m^4c^4q_R(\mathbf {10}_M)  \\
\mathcal I_{\mathbf{\bar 5}_M}=&\,\,\,  -2m^4c^4q_R(\mathbf {\bar 5}_M) \\
\mathcal I_{\mathbf{5}_U}=&\, -8\mu^4 q_S(\mathbf{5}_U)  \\
\mathcal I_{\mathbf{\bar 5}_D}=&\,   -8d^2\mu^4 q_S(\mathbf{\bar 5}_D),
\end{align}
\end{subequations}
where we took $b=1$ and $\mu\ll m$, as in section \ref{s:realwave}. Using the charges in table \ref{t:sectors} we find that we need to restrict the fluxes to
\be
M_1<0,\qquad 4N_1<-N_Y,\qquad 4N_2>-N_Y
\ee
together with either
\be
3M_1<2\tilde N_Y\le0,\qquad 4M_2+\tilde N_Y+2M_1>0
\ee
or
\be
0<\tilde N_Y<-M_1,\qquad 6M_2-\tilde N_Y+3M_1>0.
\ee

\section{Holomorphic Yukawa couplings computation}\label{ap:holoyuk}

In this appendix we present the explicit computation of the Yukawas via residues as explained in subsection \ref{sec:residuemain}. Taking into account the non-perturbative corrections, the holomorphic Yukawa coupling is
\be\label{ap:yukres}
Y =  m_*^4 \pi^2 f_{abc}\,\text{Res}_p \left[\eta^a \eta^b h_{xy}\right],
\ee
where 
\be\label{ap:eta}
\eta = - i \Phi^{-1}\left[h_{xy}+i \epsilon\p_x \theta_0 \p_y \left (\Phi^{-1} h_{xy}\right)-i \epsilon\p_y \theta_0 \p_x \left (\Phi^{-1} h_{xy}\right)\right].
\ee
Here $\eta$ is a different function for every sector. In particular, $\Phi$ is the action of $\langle \Phi\rangle$ on each sector, as in eq.(\ref{phisector}), and $h_{xy}$ are given in (\ref{haches}). Finally, $\theta_0$ is a holomorphic function on $S_{\rm GUT}$ which we take linear as it appears in (\ref{th0}). Given this information we can proceed to compute the different $\eta$ functions which read,

\subsection*{$\mathbf{10}_M$}
\bea
h_{\mathbf{10}_M} & = &   bx-y \\[2mm] \label{soleta10}
i \eta_{\mathbf{10}_M}^i/\g_{\mathbf{10}_M}^i & = & - \left[ \frac{m_*^{3-i} h_{\mathbf{10}_M}^{3-i}  }{\text{det}\Phi_{\mathbf{10}_M}}\right]   \left(\begin{array}{c}m\\\mu^2 h_{\mathbf{10}_M}  \end{array}\right) + \CO(\eps^2) \\ \nonumber
& + & \eps\,   \frac{2\mu^2 (\th_x+b\th_y)\lam + m^3\th_y}{(\text{det}\Phi_{\mathbf{10}_M})^3} m_*^{3-i} h_{\mathbf{10}_M}^{3-i}\left(\begin{array}{c} 2 m \lambda\\m^3x+\lambda^2\end{array}\right)\\ \nonumber
& + &  \eps\, \frac{ (\th_x+b\th_y)}{(\text{det}\Phi_{\mathbf{10}_M})^2}  m_*^{3-i} h_{\mathbf{10}_M}^{3-i}  \left(\begin{array}{c}2m \lambda(6-i)\\m^3x(3-i)+(4-i)\lam^2\end{array}\right)
\eea

\subsection*{$\mathbf{\bar 5}_M$}
\bea
h_{\mathbf{\bar 5}_M}&=&b(x- x_0)-(y-y_0)\\
i \eta_{\mathbf{\bar 5}_M}^i/\g_{\mathbf{\bar 5}_M}^i & = &\frac{m_*^{3-i} h_{\mathbf{\bar 5}_M}^{3-i} }{\left((2 a+1)\lambda+2 \kappa \right)^2-m^3 x}\left(
\begin{array}{c}
 m \\
 (2 a+1) \lambda+2 \kappa  \\
\end{array}\right)\\
&+&\eps \frac{m_*^i (b \theta_y+\theta_x) h_{\mathbf{\bar 5}_M}^{2-i}}{(\text{det}\Phi_{\mathbf{\bar 5}_M})^2}
\left(
\begin{array}{c}
 (2 a+1) \mu ^2 m  \left(h_{\mathbf{\bar 5}_M}+2(3-i) (bx+y)\right) \\
 (2 \kappa +(2a+1)\lambda ) (2 a+1) \mu ^2 \left[h_{\mathbf{\bar 5}_M}+(3-i) (bx+y)\right] \nonumber\\
\end{array}
\right)\\
&+&\eps \frac{m_*^i (b \theta_y+\theta_x) h_{\mathbf{\bar 5}_M}^{2-i}}{(\text{det}\Phi_{\mathbf{\bar 5}_M})^2}
\left(\begin{array}{c}4 i \kappa  m\\(3-i) m^3 x+[2 \kappa +(2a+1)\lambda ] (3-i) \kappa\end{array}\right)\nonumber\\
&+&\eps\frac{ 2m_*^{3-i} h_{\mathbf{\bar 5}_M}^{3-i} (2 a+1) \mu ^2 (b \theta_y+\theta_x) \left[(2 a+1) \lambda+2 \kappa \right]}{(\text{det}\Phi_{\mathbf{\bar 5}_M})^3}
\left(
\begin{array}{c}
 2 m \left[(2 a+1)\lambda+2 \kappa \right] \\
(\text{det}\Phi_{\mathbf{\bar 5}_M}) \\
\end{array}
\right)\nonumber\\
&+&\eps\frac{ 2m_*^{3-i} h_{\mathbf{\bar 5}_M}^{3-i}}{(\text{det}\Phi_{\mathbf{\bar 5}_M})^3}
\left(\begin{array}{c}-2 m \left((2 a+1)\lambda+2 \kappa \right)m^3\theta_y\\- (\text{det}\Phi_{\mathbf{\bar 5}_M})m^3 \theta_x\end{array}\right)+\mathcal{O}(\eps^2)\,.
\eea
\subsection*{$\mathbf{ 5}_U$}
\bea
h_{\mathbf{5}_U}/\g_{\mathbf{5}_U} & = & 1\\
i \eta_{\mathbf{5}_U}/\g_{\mathbf{5}_U} & = &  -\frac{1}{ \Phi_{\mathbf{5}_U}} + \eps\, \frac{2\mu^2(\th_{x} + b \th_{y})}{\Phi_{\mathbf{5}_U}^3} + \CO(\eps^2)
\eea
\subsection*{$\mathbf{\bar 5}_D$}
\bea
h_{\mathbf{\bar 5}_D}/\g_{\mathbf{\bar 5}_D} & = & 1\\
i \eta_{\mathbf{\bar 5}_D}/\g_{\mathbf{\bar 5}_D} & = &  -\frac{1}{ \Phi_{\mathbf{\bar 5}_D}} - \eps\, \frac{2a\mu^2(\th_{x} + b \th_{y})}{\Phi_{\mathbf{\bar 5}_D}^3} + \CO(\eps^2).
\eea

Once we have these we can simply apply (\ref{ap:yukres}) to find the Yukawas. The computation of multivariate residues can be quite involved, however, in our case the matter curves intersect transversely, which means that these can be calculated in a straightforward way. Let us illustrate how this works. Consider the following residue at a point $p\in\mathbb C^2$,
\be
R= {\rm Res}_{p}\left [ \frac{f(x,y)}{\sigma_1(x,y)\sigma_2(x,y)} \right ]
\ee
where $f$, $\sigma_1$ and $\sigma_2$ are holomorphic functions. If $\sigma_1$ and $\sigma_2$ meet transversely at $p$, meaning that the determinant of the Jacobian at $p$,
\be
J(\sigma_1,\sigma_2)|_{p}=\left |\frac{\p(\sigma_1,\sigma_2)}{\p(x,y)}\right |_{p},
\ee
is non-zero, we can perform the following change of variables $(u,v)=(\sigma_1(x,y),\sigma_2(x,y))$. Then, the residue is
\be
R= {\rm Res}_{p}\left [ J(\sigma_1,\sigma_2)|^{-1}\frac{\tilde f(u,v)}{uv} \right ]=J(\sigma_1,\sigma_2)|_{p}^{-1}\tilde f(p),
\ee
where $\tilde f(u,v)=f(x,y)$. Using this expression in our case leads to the result  (\ref{hyuk}) in the main text.

\section{Zero mode wavefunctions in real gauge}\label{ap:wave}

In this appendix we discuss the computation of the real wavefunctions, both perturbative and non-perturbative. Although the structure of the couplings can be computed at the holomorphic level via residues, the normalisation factors and kinetic mixing necessarily depend on solving the D-term equation. Thus, as explained in the main text, we mainly use these solutions to compute the kinetic terms which is then combined with the holomorphic computation to give the physical Yukawa couplings.

There are two different kinds of wavefunctions in our model, those that correspond to sectors which are charged under an Abelian subgroup of $S(U(2)\times U(2)\times U(1))_\perp$ and the ones that transform as doublets of the first $U(2)$ factor. These correspond to the Higgses and matter, respectively. The former have been computed in \cite{afim11,fimr12,fmrz13} and the latter in \cite{fmrz13} so we just sketch the computation and refer the reader to those for further details.

\subsection*{Perturbative zero modes}

\subsubsection*{Higgs sectors}

These sectors are not charged under the T-brane background as can be seen from table \ref{t:sectors}. Thus, we can compute their wavefunctions by applying the techniques in \cite{afim11,fimr12,fmrz13}. The zero mode equations (\ref{zeropert}) can be recast into a Dirac-type equation, namely
\be
\left(
\begin{array}{cccc}
0 & D_x & D_y & D_z \\
-D_x & 0 & -D_{\bar{z}} & D_{\bar{y}} \\
-D_y & D_{\bar{z}} & 0 & -D_{\bar{x}} \\
-D_z & -D_{\bar{y}} & D_{\bar{x}} & 0
\end{array}
\right)
\left(
\begin{array}{c}
0 \\ \\  \vec{\vphi}_{U} \\ \quad
\end{array}
\right)\, =\, 0  
\label{Dirac5}
\ee
with
\be
D_{x} \, =\, \p_{x} + \oh(q_R \bar{x} - q_S \bar{y}) \qquad D_{y} \, =\, \p_{y} - \oh (q_R \bar{y}  + q_S \bar{x}) \qquad D_{z}\, =\, 2i \mu^2 (\bar{x}-\bar{y}) 
\label{covar}
\ee
and $D_{\bar{m}}$ their conjugates. Here we used the following particular gauge
\be\label{gaugetot}
A=\frac{i}{2}Q_R(yd\bar y-\bar ydy-xd\bar x+\bar xdx)+\frac{i}{2}Q_S(xd\bar y-\bar ydx+yd\bar x-\bar xdy)-\frac{i}{2}m^2c^2P_1(xd\bar x-\bar xdx).
\ee
which reproduces the total flux,
\be
F=iQ_R(dy\wedge d\bar y-dx\wedge d\bar x)+iQ_S(dx\wedge d\bar y+dy\wedge d\bar x)+im^2c^2P_1dx\wedge d\bar x.
\ee
The quantities $q_R$ and $q_S$ are the constant flux densities that are shown in table \ref{t:sectors}, and for concreteness we have taken the case of the up-type Higgs. The down-type Higgs can be obtained from this one by some simple replacements, see below. 

Following \cite{afim11,fimr12} one can solve this system of equations which yields 
\be
\label{wave5p}
\vec\vphi_{U}\, =\, \gamma_{U}\left ( \begin{array}{c}
i\frac{\zeta_{U}}{2\mu^2}\\
i\frac{(\zeta_{U}-\lam_{U})}{2\mu^2}\\
1\end{array}\right )\, \chi_{U}, \quad
\qquad
\chi_{U} = 
e^{\frac{q_R}{2}(|x|^2-|y|^2)-q_S (x \bar y +y\bar x)+(x-y)(\zeta_{U}\bar x-(\lam_{U}-\zeta_{U})\bar y))}
\ee
with $\lam_{U}$ the lowest solution to 
\be\label{cubU}
\lam^3_{U}-(8\mu^4+(q_R)^2+(q_S)^2)\lam_{U}+8\mu^4q_S=0
\ee
and $\zeta_{U}=\frac{\lam_{U}(\lam_{U}-q_R-q_S)}{2(\lam_{U}-q_S)}$. As shown in \cite{fmrz13} we can multiply $\chi_{U}$ by an arbitrary holomorphic function of a particular linear combination of $x$ and $y$ and still satisfy the equations of motion. This function will only be determined once we impose boundary conditions that arise when we embed this model in a compact setup. However, the dominant contribution to the Yukawa couplings comes from the average value of such function around the Yukawa point, so we may approximate it by a constant.

%Notice that, because they depend on the hypercharge flux, $q_R$ and $q_S$ take different values for the two subsectors $\mathbf{5}_1$ and $\mathbf{5}_2$ of table \ref{t:sectors}, and so the same is true for $\lam_{\mathbf{5}}$, $\zeta_{\mathbf{5}}$. In particular, imposing (\ref{cond23}) we find that $q_S(\mathbf{5}_2)=0$ and that the wavefunction for this sector is not localized along $\Sigma_{\mathbf{5}}$, as we briefly comment below. 

Now, looking at (\ref{phisector}), we see that if we set $\kappa$ to zero, we can find the solution for the down-type Higgs by simply performing the following replacement on the solution above
\be\label{rep1}
\mu^2\rightarrow -d\mu^2.
\ee
Furthermore, the effect of $\kappa$ is simply a translation in the $(x,y)$ plane which can be taken into account by shifting the solution
\be\label{shift}
x\rightarrow x-x_0,\qquad y\rightarrow y-y_0.
\ee
This yields the wavefunction for $H_D$ in a gauge in which the vector potential is shifted with respect to the one used to solve the $H_U$ sector. Thus, we need to perform a change of gauge for the background flux, namely, 
\be\label{change}
A(x-x_0,y-y_0)=A(x,y)+d\psi
\ee
with 
\be\label{gaugepar}
\psi=\frac{i}{2}Q_R(y_0\bar y-\bar y_0y-x_0\bar x+ x_0x)+\frac{i}{2}Q_S(x_0\bar y-\bar y_0x+y_0\bar x-\bar x_0y)-\frac{i}{2}m^2c^2P_1(x_0\bar x-\bar x_0x).
\ee
Taking all this into account we find that,
\be
\vec \varphi_D = \gamma_D \left(\begin{array}{c}i \frac{\zeta_D}{2d\mu^2}\\i\frac{\zeta_D-\lambda_D}{2d\mu^2}\\1\end{array}\right) e^{-i\psi}\chi_D(x-x_0,y-y_0)
\ee
with 
\be
\chi_D(x,y)=e^{\frac{q_R}{2}(|x|^2-|y|^2)-q_S (x \bar y + y \bar x)+(x-y)(\zeta_D\bar x-(\lambda_D-\zeta_D)\bar y)}.
\ee
Finally, $\lambda_D$ is the lowest solution to
\be\label{cubD}
\lambda_D^3-(8d^2 \mu^4 +q_R^2+q_S^2)\lambda_D+8d^2 \mu^4q_S=0
\ee
and $\zeta_D= \frac{\lambda_D(\lambda_D-q_R-q_S)}{2(\lambda_D-q_S)}$.

\subsubsection*{Matter sectors}

Unlike the previous sectors, these are charged under the T-brane since they transform as doublets of $U(2)$. This makes the computation of the real wavefunction more complicated. In particular, as shown in \cite{fmrz13}, it is not possible to find a simple solution in the case in which $\mu$ is non-zero since that leads to a pair of coupled partial differential equations. However, in the limit $\mu\ll m$ these two equations decouple and can be effectively reduced to ordinary differential equations. In the following we present some details on the computation while a more detailed discussion can be found in appendix  A of \cite{fmrz13}.

Similarly to what we did with the wavefunctions of the Higgses, we start by considering the $\mathbf{10}_M$ sector and we will obtain the $\mathbf{\bar 5}_M$ by simple replacements. Furthermore, for simplicity we set $b=1$ in the Higgs background (\ref{hback}).

Since this sector is a doublet under the first $U(2)$ factor, we need to have
\be
\left(
\begin{array}{c}
a_{\bar{x}}\\ a_{\bar{y} }\\ \vphi_{xy}
\end{array}
\right)
\, =\, 
\vec{\vphi}_{{10}^+} E_1^+ + \vec{\vphi}_{{10}^-} E_1^-
\ee
for which one can write an analogous equation to (\ref{Dirac5}) where
\be
a = \bmat{c} a^+ \\
a^-
\emat \qquad  \qquad
\vphi = \bmat{c} \vphi^+ \\
\vphi^-
\emat
\label{doubs}
\ee
Then, following \cite{fimr12,fmrz13}, we first solve the F-terms equations (\ref{treeF1}) and (\ref{treeF2}) to write $a$ in terms of $\varphi$, and then impose the D-term equation (\ref{treeD}) to find an equation for $\vphi$. 

Let us start by considering the case in which the primitive fluxes $\langle F_p \rangle$ are zero. Then solution to the F-terms reads
\begin{subequations}
\label{Fsol}
\begin{align}
\label{asol}
a & =  e^{fP_1/2} \bar\p \xi\\
\label{phisol}
\vphi & =  e^{fP_1/2}\left(h - i \Psi \xi \right) 
\end{align}
\end{subequations} 
where $\xi$ and $h$ are doublets with components $\xi^\pm$ and $h^\pm$ and 
\be
P_1 = \bmat{cc} 1 & 0 \\
0 & -1
\emat \qquad  \qquad
\Psi = \bmat{cc} -\mu^2(x-y) & m \\
m^2 x & -\mu^2(x-y)
\emat
\label{pepsi}
\ee
From (\ref{phisol}) we obtain
\be
\xi = i \Psi^{-1} \left( e^{-fP/2}\vphi - h\right)
\label{xisol}
\ee
which can be used to write down the D-term equation for the fluctuations (\ref{treeD}) 
\be
\p_x \p_{\bar x} \xi + \p_y \p_{\bar y} \xi + \p_x f P \p_{\bar x} \xi - i \Lambda^\dagger \left(h - i \Psi \xi \right)  = 0   
\label{dterm2} 
\ee 
where we used that $f$ does not depend on $(y,\bar y)$ and we have defined
\be
\Lambda =  e^{fP} \Psi  e^{-fP} = 
\bmat{cc} -\mu^2(x -y) & m e^{2f} \\
m^2 x e^{-2f} & -\mu^2(x-y)
\emat.
\label{ladef}
\ee
Finally, we can make the following change of variables
\be 
U =  e^{-fP/2}\vphi  \qquad ; \qquad \xi \,=\, i \Psi^{-1} \left(U - h\right)
\label{xisol2} 
\ee
and express (\ref{dterm2}) in terms of $U$
\be
\p_x \p_{\bar x} U + \p_y \p_{\bar y} U 
-  (\p_x \Psi) \Psi^{-1}  \p_{\bar x} U  + (\p_y \Psi) \Psi^{-1} \p_{\bar y} U
+  \p_x f \Psi P \Psi^{-1} \p_{\bar x} U -  \Psi\Lambda^\dagger U = 0.
\label{uterm}
\ee
As advanced earlier, this is system of coupled partial differential equations for $U^+$ and $U^-$ that have no simple solution. However, following \cite{fmrz13}, these decouple in the limit $m \gg \mu$. Furthermore, it is possible to show that there is no localised solution for $U^+$, so we set it to zero. Then, near the Yukawa point $p_{\rm up} =\{x=y=0\}$ if we approximate $f = \log c+  c^2 m^2 x\bar{x}$ we find $U^- = {\rm exp} (\lam x \bar x) h$ where $\lam$ the lowest solution to $ c^2 \lam^3 + 4c^4 m^2 \lam^2 - m^4\lam=0$. Taking this into account one finds
\be
\vec{\vphi}_{{10}^+}^j \, =\, \gamma_{{10}}^j
\left (\begin{array}{c}
\frac{i\lam}{m^2}\\
0 \\
0 \end{array}\right )
 e^{f/2} \chi_{{10}}^j 
 \qquad \quad 
\vec{\vphi}_{{10}^-}^j \, =\, \gamma_{{10}}^j
\left (\begin{array}{c}
0 \\
0 \\
1 \end{array}\right )
e^{-f/2} \chi_{{10}}^j 
\ee
where $e^{f/2} = \sqrt{c}\, e^{m^2c^2x\bar{x}/2}$ and $\chi_{\mathbf{10}}^j \, =\, e^{\lam x \bar x}\, g_j (y)$, with  $g_j$ holomorphic functions of $y$. 

Switching on the primitive gauge fluxes amounts to replacing $\p_{x,y} \raw D_{x,y}$ in the D-term, where $D_{x,y}$ is defined in (\ref{covar}), and analogously for $\bar{\p}$ in the F-terms. Then, taking $\mu \ll m$ one finds a localised solution for $U^-$ and the wavefunction reads
\be
\label{phys10} 
\vec{\vphi}_{{10}^+}^j \, =\, \gamma_{{10}}^j
\left (\begin{array}{c}
\frac{i\lam_{10}}{m^2}\\
-\frac{i\lam\zeta_{10}}{m^2} \\
0 \end{array}\right )
 e^{f/2} \chi_{{10}}^j 
 \qquad \quad 
\vec{\vphi}_{{{10}}^-}^j \, =\, \gamma_{{10}}^j
\left (\begin{array}{c}
0 \\
0 \\
1 \end{array}\right )
e^{-f/2} \chi_{{{10}}}^j 
\ee
where $\lam_{10}$ is the lowest (negative) solution to 
\be\label{cub10}
m^4 (\lambda_{10} -q_R)+\lambda c^2 \left( c^2 m^2 (q_R-\lambda_{10} )-\lambda_{10} ^2+q_R^2+q_S^2\right)=0
\ee
and $\zeta_{10} = -q_S/(\lam_{10}-q_R)$. The scalar wavefunctions $\chi_{{10}}$ are
\be
\chi_{{10}}^j \, =\,e^{\frac{q_R}{2}(|x|^2-|y|^2)-q_S (x \bar y +y\bar x)+\lam_{{10}} x (\bar x - \zeta_{{10}} \bar y)} \, g_j (y +\zeta_{{10}}x)
\label{wave10p}
\ee
where $g_j$ holomorphic functions of $y +\zeta_{{10}}x$, and $j=1,2,3$ label the different zero mode families. Similarly to \cite{hv08} we choose such family functions as
\be
g_j\, =\, m_*^{3-j}(y +\zeta_{{10}}x)^{3-j}.
\label{holorep}
\ee

Now that we have the solution to the $\mathbf{10}_M$ sector, we can obtain the corresponding solution for the $\mathbf{\bar 5}_M$ by some replacements. Similarly to the case of the down-type Higgs, we arrive at the solution by performing the shift (\ref{shift}) where now
\be
x_0\rightarrow 0,\qquad\qquad y_0\rightarrow \frac{\nu}{a},
\ee
since we need to take the limit $\mu\ll m$ as well as $\kappa\ll m$, keeping $\kappa/\mu^2=\nu$ finite. Again, we have to perform a change of gauge with parameter
\be\label{gaugepar2}
\tilde\psi=\frac{i}{2}Q_R(\nu\bar y/a-\bar \nu y/\bar a)+\frac{i}{2}Q_S(\nu\bar x/a-\bar \nu x/\bar a)
\ee
which is the same as $\psi$ but using the fact that, when taking $\mu,\kappa\ll m$ with $\kappa/\mu^2=\nu$ finite, we have $x_0=0$ and $y_0=\nu/a$. Then, the wavefunction for the matter $\mathbf{\bar 5}_M$ sector reads
\be
\vec \varphi_{5}^i = \gamma_{5}^i \left(\begin{array}{c}\frac{i \lambda_{5}}{m^2}\\-i\frac{\lambda_{5} \zeta_{5}}{m^2}\\0\end{array}\right)e^{i\tilde\psi +f/2} \chi_{5}^i(x,y-\nu/a) E^+_1+\gamma_5^i \left(\begin{array}{c}0\\0\\1\end{array}\right)e^{i\tilde\psi-f/2} \chi_{5}^i(x,y-\nu/a) E^-_1
\ee
where $\lambda_{5}$ is the lowest solution to
\be\label{cub5}
m^4 (\lambda_{5} -q_R)+\lambda_{5} c^2 (c^2 m^2 (q_R-\lambda_{10})-\lambda_{5}^2+q_R^2+q_S^2)=0\,,
\ee
and $\zeta_{5} = -q_S / (\lambda_{5} -q_R)$. Finally the wavefunctions $\chi_{5}^i$ are 
\be
\chi_{5}^i(x,y) = e^{\frac{q_R}{2}(|x|^2-|y|^2)-q_S (x\bar y+y\bar x)+\lambda_{5} x(\bar x- \zeta_{5} \bar y)} g_{5}^i(y+\zeta_{5} x)\,,
\ee
where $g_{5}^i$ are in $y+\zeta_{5} x$ and $i=1,2,3$ is a generation index. Analogously, the family functions are
\be
g_{5}^i (y+\zeta_{5} x) = m_*^{3-i} (y+\zeta_{5} x)^{3-i}\,.
\ee

\subsection*{Non-perturbative zero modes}

Here we discuss the correction to the real wavefunctions due to the non-perturbative effects. We start by considering the Higgs sectors and then move on to the matter wavefunctions. 

\subsubsection*{Higgs sectors}

The analysis of the correction to these two sectors are identical to those that appeared in section 5.1 of \cite{fimr12} and in appendix  A of \cite{fmrz13}. The zero mode equations read
\bea
\bar\partial_{\langle A\rangle}a&=&0\\
\bar\partial_{\langle A\rangle}\varphi+i[\langle \Phi\rangle,a]+\eps\,\partial\th_0\,\w\,\p_{\langle A\rangle}a&=&0\\
\omega\,\w\,\p_{\langle A\rangle}a-\frac{1}{2}[\langle \bar\Phi\rangle,\varphi]&=&0
\eea
that reduce to the following in holomorphic gauge for the up-type Higgs
\bea
\p_{\bar{x}} a_{\bar{y}} - \p_{\bar{y}} a_{\bar{x}} & = & 0 \label{ft03}\\
\p_{\bar{m}} \vphi_{xy} - i 2\mu^2 (x-y) a_{\bar{m}} & = & i\eps  \left[\th_{y} \p_x a_{\bar{m}} -  \th_{x} \p_y a_{\bar{m}}\right] + \CO(\eps^2)
\eea
regarding the F-terms while the D-term reads
\bea
& & \hspace*{-1cm}\left\{\partial_x + \bar xq_R  - \bar y q_S\right\} a_{\bar x}  +  \left\{\partial_y  - \bar yq_R - \bar x q_S \right\} a_{\bar y} + 2i \mu^2 (\bar x-\bar y) \varphi_{xy} \qquad 
 \\
&  = & 
i \eps \bar\th_{x}\left\{y q_R+ xq_S \right\} \varphi_{xy}
-i \eps \bar\th_{y}\left\{- q_Rx + y q_S \right\} \varphi_{xy}.
\nonumber
\eea
%U
The first order correction to the wavefunction is then
\be 
\varphi_{U}^{(1)} = m_*\g_{{U}} e^{(x-y)(\zeta_{{U}}\bar x-(\lam_{{U}}-\zeta_{{U}})\bar y))}\Upsilon_{{U}}
\label{chic11}
\ee
with $\lam_{{U}}$, $\zeta_{{U}}$ defined as in (\ref{wave5p}) and
\be
\Upsilon_{{U}}=\frac{1}{4\mu^2}(\zeta_{{U}} \bar{x} - (\lam_{{U}} - \zeta_{{U}})\bar{y})^2(\th_x+\th_y)+\frac{\delta_1}{2}(x-y)^2+\frac{\delta_2}{\zeta_{{U}}}(x-y)(\zeta_{{U}} y+(\lam_{{U}}-\zeta_{{U}})x)
\ee
where the constants $\delta_1$ and $\delta_2$ are 
\bea\label{d1}
\delta_1&=&- \frac{2\mu^2}{\lam_{{U}}^2}\{ \bar \th_x(q_R(\zeta_{{U}}-\lam_{{U}}) + q_S\zeta_{{U}} ) +\bar\th_y(q_R\zeta_{{U}} - q_S(\zeta_{{U}}-\lam_{{U}}))\}\\
\delta_2&=&- \frac{2\mu^2\zeta_{{U}}}{\lam_{{U}}^2}\{ \bar \th_x(q_R+q_S) +\bar\th_y(q_R-q_S)\}.
\label{d2}
\eea
Notice that the holomorphic terms in $\Upsilon_{{{U}}}$, those that depend on $\d_1$ and $\d_2$,
are there to satisfy the corrected D-term equation.

Similarly to the tree-level case, the corrected wavefunction for the down-type sector can be obtained from this one by performing the replacements (\ref{rep1}-\ref{shift}) and taking into account the change of gauge (\ref{change}) so we do not write the explicit result.

\subsubsection*{Matter sectors}

Recall that the wavefunctions in the real gauge are used essentially to compute the normalisation factors and kinetic mixing since the structure of the Yukawa couplings can be computed via residues. For the matter sectors in our model, one can prove that the mixing and normalisation factors remain unchanged when including the non-perturbative effects just by analysing the structure of the equations. In the following we discuss this point and obtain the structure of such correction for the matter sectors.

Just as in the perturbative case, let us start by turning off the primitive fluxes. Also, we consider the $\mathbf{10}_M$ sector and obtain the $\mathbf{\bar 5}_M$ by performing replacements. First, we see that the F-terms in the real gauge are solved by 
\begin{subequations}
\label{Fsolnp}
\begin{align}
\label{asolnp}
a & =  g\, \bar\p \xi\\
\label{phisolnp}
\vphi & =  g \left(h - i \Phi \xi - \eps \p \th_0  \wedge \p \xi \right) \, =\, g\, U\, dx \wedge dy
\end{align}
\end{subequations}
with
\be
g=\left ( \begin{array}{cc}
e^{f/2}&0\\
0&e^{-f/2}\end{array}\right )
\ee
and $\Phi$ is given by $\Phi|_{\mathbf {10}_M}$ in (\ref{phisector}), where we dropped the subscript $\mathbf {10}_M$ for notational convenience. The doublet $U$ can be expanded in $\epsilon$
\be
U\, =\, U^{(0)} + \eps\, U^{(1)} + \, \CO(\eps^2)
\ee
where $U^{(0)}$ in the solution for $\eps = 0$ that appeared in the last section, namely
\be
 U^{(0)}_- \, =\,  e^{\lambda_{{10}}x\bar{x}} h (y) \qquad \qquad  U^{(0)}_+ \, =\, 0.
\ee
Then, one may solve for $\xi$ from (\ref{phisolnp}) as 
\be
\begin{array}{c}
\xi\, =\, \xi^{(0)} +  i \epsilon \Phi^{-1} \left[U^{(1)} + \partial_x\th_0 \p_y\xi^{(0)} - \p_y\th_0\p_x\xi^{(0)} \right] + \mathcal O(\epsilon^2) \\
 \xi^{(0)} \, =\, i \Phi^{-1} (U^{(0)} - h)
\end{array}
\label{solxinp}
\ee
and then solve for $U^{(1)}$ by plugging in this expression into the D-term for the fluctuations (\ref{treeD}). This yields $U^{(1)}_- =0$, in the limit $\mu\ll m$. Thus, we find the following structure
\be
\xi_+ \, =\, \xi_+^{(0)} + 0 + \CO(\eps^2) \qquad \quad \xi_- \, =\, 0 + \eps\, \xi_-^{(1)} + \CO(\eps^2).
\ee
Since $\xi_{\pm}$ determines both $a$ and $\varphi$ through (\ref{Fsolnp}), we find that this particular structure implies that the solution (\ref{phys10}) looks like
\be
\vec{\vphi}_{{10}^+} \, =\, 
\left(\begin{array}{c} \bullet \\ \bullet \\ 0\end{array}\right)
 + \eps
\left(\begin{array}{c} 0 \\ 0 \\ \bullet\end{array}\right) +\CO(\eps^2)
 \qquad 
\vec{\vphi}_{{10}^-} \, =\,
\left(\begin{array}{c} 0 \\ 0 \\ \bullet\end{array}\right)
 + \eps
 \left(\begin{array}{c} \bullet \\ \bullet \\ 0\end{array}\right)+\CO(\eps^2)
\ee
Finally, it is possible to show that this structure still holds after including the non-primitive fluxes.

\end{document}